\documentclass{CSML}

\def\dOi{12(4:7)2016}
\lmcsheading%
{\dOi}
{1--58}
{}
{}
{Apr.~\phantom02, 2016}
{Dec.~28, 2016}
{}

\ACMCCS{[{\bf Software and its engineering}]: Software organization
  and properties---Software  Software verification;  Software notations and tools---Formal
  language definitions; Software notations and tools---General
  programming languages---Language types---Concurrent programming
  languages\,/\,Distributed programming languages\,/\,Parallel
  programming languages; [{\bf Theory of computation}]:  Semantics and
  reasoning---Program semantics---Operational semantics; Semantics and
  reasoning---Program constructs---Type structure}

\pdfoutput=1
\usepackage[utf8]{inputenc}

\usepackage[usenames,dvipsnames]{xcolor} 
\usepackage{amsmath}

\newcounter{tmp}

\newenvironment{bartalign}%
        {\setcounter{tmp}{\value{equation}}\setcounter{equation}{0}\align}%
        {\endalign\setcounter{equation}{\value{tmp}}}

\usepackage{amssymb}
\usepackage{caption}                
\usepackage{cmll} 
\usepackage{colonequals}            
\usepackage{environ}                
\usepackage{etex,etoolbox}          

\usepackage{eucal}

\usepackage{graphicx}               

\usepackage{nicefrac} 

\usepackage{proof}                  
\usepackage{bussproofs}
\EnableBpAbbreviations

\usepackage{rotating}               
\usepackage{subcaption}             
\usepackage{thmtools, thm-restate, thm-autoref}  
\usepackage{xifthen}                
\usepackage{appendix}               

\usepackage{amsbsy} 
\usepackage{stmaryrd}
 \usepackage{pdfsync}
\usepackage{marvosym}

\usepackage{tikz}                   
\usetikzlibrary{positioning}

\usepackage{fixltx2e} 
\usepackage{xspace} 

\usepackage[dvipsnames]{xcolor}%

\usepackage{mathtools} 

\newlist{inlinelist}{enumerate*}{1}
 \setlist*[inlinelist,1]{%
  label=(\roman*),
}

\newlist{inlinearabiclist}{enumerate*}{1}
 \setlist*[inlinearabiclist,1]{%
  label=(\arabic*),
}

\usepackage{hyperref}

\usepackage{cleveref}

\newcommand{\ifempty}[3]{%
  \ifthenelse{\isempty{#1}}{#2}{#3}%
}

\newcommand{\bnfdef}{::=}
\newcommand{\bnfmid}{\;\;\big|\;\;}

\newcommand{\nrule}[1]{{\scriptsize \textsc{#1}}}
\newcommand{\smallnrule}[1]{{\tiny \textsc{#1}}}

\newcommand{\del}[2]{\mathrm{del}_{#1}({#2})}

\newcommand{\dom}[1]{\operatorname{dom} {#1}}

\newcommand{\Nat}{\mathbb{N}}

\newcommand{\bind}[2]{{#1} \mapsto {#2}}
\newcommand{\subs}[2]{\{\nicefrac{#1}{#2}\}}
\newcommand{\fsubst}[2]{{#1} \bullet {#2}}

\newcommand{\setenum}[1]{\{#1\}}
\newcommand{\setcomp}[2]{\{{#1} \,\mid\, {#2}\}}

\newcommand{\hidden}[1]{}

\newcommand{\ie}{\text{i.e.}\xspace}

\newcommand{\Eg}{\text{E.g.}\xspace}
\newcommand{\eg}{\text{e.g.}\xspace}
\newcommand{\aka}{\text{a.k.a.}\xspace}

\def\vec{\mathaccent"017E }
\renewcommand{\vec}[1]{\ensuremath{\textit{\textbf{#1}}}}

\newcommand{\names}{\ensuremath{\mathcal{N}}}
\newcommand{\snames}{\names}

\newcommand{\vars}{\mathcal V}

\def\colorPtp{\color{ForestGreen}}
\newcommand{\pmv}[1]{\ensuremath{\mathsf{\colorPtp{#1}}}}
\newcommand{\pmvA}[1][]{{\colorPtp{\pmv{A}_{#1}}}}
\newcommand{\pmvB}[1][]{{\colorPtp{\pmv{B}_{#1}}}}
\newcommand{\pmvC}[1][]{{\colorPtp{\pmv{C}_{#1}}}}

\let\greekgamma\gamma
\let\greekGamma\Gamma
\def\contrColor{\color{Plum}}
\newcommand{\contrFmt}[1]{{\contrColor{#1}}}

\newcommand{\codefont}{\fontsize{10}{10}\selectfont}
\newcommand{\code}[1]{{\tt\codefont {#1}}}

\newcommand{\bang}{{\contrColor{\textup{\texttt{\symbol{`\!}}}}}}
\newcommand{\qmark}{{\contrColor{\textup{\texttt{\symbol{`\?}}}}}}

\newcommand{\atom}[2][]{\contrFmt{\ifempty{#1}{{\code{#2}}}{{\code{#2}}_{#1}}}}
\newcommand{\atomIn}[2][]{\atom[#1]{#2}{\qmark}}
\newcommand{\atomOut}[2][]{\atom[#1]{#2}{\bang}}

\newcommand{\Atom}{\atom{A}}
\newcommand{\AtomIn}{{\Atom^{\qmark}}}
\newcommand{\AtomOut}{{\Atom^{\bang}}}

\newcommand{\contr}[2][]{\contrFmt{#2}_{\contrColor #1}}
\newcommand{\contrP}[1][]{\mathord{\contrFmt{C}_{\contrColor{#1}}}}
\newcommand{\contrPi}[1][]{\mathord{\contrFmt{C'}_{\!\contrColor{#1}}}}
\newcommand{\contrPii}[1][]{\mathord{\contrFmt{C''}_{\!\!\contrColor{#1}}}}

\newcommand{\contrQ}[1][]{\mathord{\contrFmt{D}_{\contrColor{#1}}}}
\newcommand{\contrQi}[1][]{\mathord{\contrQ[#1]\contrColor{'}}}

\newcommand{\contrX}[1][]{\mathord{\contrFmt{X}_{\contrColor{#1}}}}

\newcommand{\abscontr}[1]{\mathcal{\contrFmt{#1}}}
\newcommand{\abscontrP}[1][]{\mathord{\abscontr{C}_{\contrColor{#1}}}}
\newcommand{\abscontrPi}[1][]{\mathord{\abscontr{C'}_{\!\contrColor{#1}}}}

\newcommand{\abscontrQ}[1][]{\mathord{\abscontr{D}_{\contrColor{#1}}}}

\renewcommand{\gamma}[1][]{\mathord{\contrFmt{\greekgamma}_{\contrFmt{#1}}}}
\newcommand{\gammai}[1][]{\mathord{\gamma[#1]\contrColor{'}}}

\renewcommand{\Gamma}[1][]{\mathord{\contrFmt{\greekGamma}_{\contrFmt{#1}}}}
\newcommand{\Gammai}[1][]{\mathord{\contrFmt{\greekGamma'}_{\contrFmt{#1}}}}

\newcommand{\pbic}[4]{
        \ifempty{#1}{{#2}, {#4}}{{\pmv {#1}} \says{#2} \mid {\pmv {#3}} \says {#4}}}
\newcommand{\bic}[2]{\pbic{\pmv A}{#1}{\pmv B}{#2}}
\newcommand{\bicAC}[2]{{\pmv A} \says {#1} \mid {\pmv C} \says {#2}}

\newcommand{\pbicSmall}[4]{
        \ifempty{#1}{{#2}, {#4}}{{\pmv {#1}} \!\says\!{#2} \!\mid\! {\pmv {#3}} \!\says\! {#4}}}
\newcommand{\bicSmall}[2]{\pbicSmall{\pmv A}{#1}{\pmv B}{#2}}

\newcommand{\lbl}[2]{\pmv{#1} \says \, {#2}}

\newcommand{\qmv}[1][]{\mathord{\contrFmt{\beta}_{\contrColor{#1}}}}
\newcommand{\qmvA}{\qmv[\pmvA]}
\newcommand{\qmvB}{\qmv[\pmvB]}

\newcommand{\buffer}[1]{\ [#1]}

\newcommand{\coco}{CO\textsubscript{2}\xspace}

\def\cocoColor{\color{Black}}
\newcommand{\cocoFmt}[1]{{\cocoColor{\code{#1}}}}

\newcommand{\expE}{\mathit{e}}

\newcommand{\false}{\cocoFmt{false}}

\newcommand{\pref}[1][]{\cocoFmt{\pi}_{{\cocoColor{#1}}}}
\newcommand{\prefi}[1][]{\cocoFmt{\pi'}_{{\cocoColor{#1}}}}

\let\greektau\tau
\renewcommand{\tau}{\cocoFmt{\greektau}}

\newcommand{\cocodo}[2]{\cocoFmt{do}_{#1}\,{#2}}
\newcommand{\tell}[2]{\cocoFmt{tell}\,{#2}}

\newcommand{\fuse}{\cocoFmt{fuse}}

\newcommand{\cond}{\mathit{if}}
\newcommand{\then}{\mathit{then}}
\newcommand{\owise}{\mathit{else}}
\newcommand{\ifte}[3]{\cond~{#1}~\then~{#2}~\owise~{#3}}
\newcommand{\cocoifte}[3]{\cocoFmt{if}~{#1}~\cocoFmt{then}~{#2}~\cocoFmt{else}~{#3}}
\newcommand{\freeze}[2]{\downarrow_{#1}{#2}}
\newcommand{\latent}[3][]{\setenum{\downarrow_{#2}{#3}}_{\ifempty{#1}{\pmvA}{#1}}}

\newcommand{\says}{\ensuremath{:}}
\newcommand{\psays}{\ensuremath{\;\mathit{:}\;}}

\newcommand{\cocoSeq}{\mathbin{\!{\cocoColor{.}}\!}}
\newcommand{\cocoPlus}{\mathbin{{\cocoColor{+}}}}
\newcommand{\cocoSum}[2][]{\mathord{{\cocoColor{\sum_{#1}{#2}}}}}
\newcommand{\cocoPar}{\mathbin{{\cocoColor{\mid}}}}

\def\sysColor{\color{Black}}
\newcommand{\sys}[2]{{#1} [{#2}] }
\newcommand{\sysFmt}[1]{{\sysColor{#1}}}
\newcommand{\sysS}[1][]{\mathord{\sysFmt{S}_{\sysColor{#1}}}}
\newcommand{\sysSi}[1][]{\mathord{\sysColor{\sysS'_{#1}}}}
\newcommand{\sysSii}[1][]{\mathord{\sysColor{\sysS''_{#1}}}}

\newcommand{\sysNil}{\sysFmt{\mathbf{0}}}
\newcommand{\emptysys}{\sysNil}

\def\ptypeColor{\color{Black}}
\newcommand{\ptypeFmt}[1]{{\ptypeColor{#1}}}

\newcommand{\ptypeF}[1][]{\mathord{\ptypeFmt{f}_{\ptypeColor{#1}}}}
\newcommand{\ptypeFi}[1][]{\mathord{\ptypeFmt{f'}_{\!\ptypeColor{#1}}}}
\newcommand{\ptypeFii}[1][]{\mathord{\ptypeFmt{f''}_{\!\!\ptypeColor{#1}}}}

\newcommand{\ptypeG}[1][]{\mathord{\ptypeFmt{g}_{\ptypeColor{#1}}}}
\newcommand{\ptypeGi}[1][]{\mathord{\ptypeFmt{g'}_{\!\ptypeColor{#1}}}}
\newcommand{\ptypeGii}[1][]{\mathord{\ptypeFmt{g''}_{\!\!\ptypeColor{#1}}}}

\newcommand{\expand}[2]{{#1}{\ptypeColor{\uparrow}}_{{#2}\,}}

\def\ctypeColor{\color{Black}}
\newcommand{\ctypeFmt}[1]{\mathcal{\ctypeColor{#1}}}
\newcommand{\chtypes}{\ctypeFmt{\ensuremath{\mathbb{P}}}}

\newcommand{\ctypeT}[1][]{\mathord{\ctypeFmt{P}_{\ctypeColor{#1}}}}
\newcommand{\ctypeTi}[1][]{\mathord{\ctypeFmt{P'}_{\!\ctypeColor{#1}}}}
\newcommand{\ctypeTii}[1][]{\mathord{\ctypeFmt{P''}_{\!\ctypeColor{#1}}}}

\newcommand{\ctypeP}[1][]{\mathord{\ctypeFmt{P}_{\ctypeColor{#1}}}}
\newcommand{\ctypePi}[1][]{\mathord{\ctypeFmt{P'}_{\ctypeColor{#1}}}}
\newcommand{\ctypeQ}[1][]{\mathord{\ctypeFmt{Q}_{\ctypeColor{#1}}}}
\newcommand{\ctypeQi}[1][]{\mathord{\ctypeFmt{Q'}_{\ctypeColor{#1}}}}
\newcommand{\ctypeR}[1][]{\mathord{\ctypeFmt{R}_{\ctypeColor{#1}}}}

\newcommand{\ctypeX}[1][]{\mathord{\ctypeFmt{X}_{\ctypeColor{#1}}}}

\newcommand{\consttovar}[1]{{\ctypeColor{\ifempty{#1}{@}{@{#1}}}}}

\newcommand{\effempty}{\ctypeFmt{\mathbf 0}}
\newcommand{\efftauqm}{\ctypeFmt{\tau_{?}}}

\newcommand{\effcontract}[1]{{\langle{#1}\rangle}}
\newcommand{\effseq}{\ctypeFmt{\dotseq}}

\newcommand{\effPar}{\mathbin{{\ctypeFmt{\mid}}}}
\newcommand{\procseq}{\ctypeFmt{\dotseq}}
\newcommand{\effrec}[2]{\ctypeFmt{\mathit{rec}}\; {#1} \effseq {#2}}
\newcommand{\effmove}[1]{\xrightarrow{\ #1\ }_{\sharp}}
\newcommand{\effwildcard}{\ctypeFmt{*}}

\newcommand{\effentails}{\mathrel{\vdash}}

\newcommand{\tsentails}[1]{\vdash_{\pmv #1}}
\newcommand{\tscompat}{\rhd}

\newcommand{\suchthat}{\mathbin{.}}%

\newcommand{\fn}[1]{\mathrm{fn}(#1)}

\newcommand{\fv}[1]{\mathrm{fv}(#1)}

\newcommand{\fnv}[1]{\mathrm{fnv}(#1)}

\def\procColor{\color{Black}}
\newcommand{\procFmt}[1]{{\procColor{#1}}}
\newcommand{\procP}[1][]{\mathord{\procFmt{P}_{\procColor{#1}}}}
\newcommand{\procPi}[1][]{\mathord{\procP[#1]\procColor{'}}}
\newcommand{\procPii}[1][]{\mathord{\procP[#1]\procColor{''}}}
\newcommand{\procQ}[1][]{\mathord{\procFmt{Q}_{\procColor{#1}}}}
\newcommand{\procQi}[1][]{\mathord{\procQ[#1]\procColor{'}}}

\newcommand{\procR}[1][]{\mathord{\procFmt{R}_{\procColor{#1}}}}
\newcommand{\procRi}[1][]{\mathord{\procR[#1]\procColor{'}}}

\newcommand{\procRec}[1]{\operatorname{{\cocoFmt{rec}\ {#1}.\,}}}
\newcommand{\procX}[1][]{\operatorname{{\procColor{\procFmt{X}_{#1}}}}}
\newcommand{\procY}[1][]{\operatorname{{\procColor{\procFmt{Y}_{#1}}}}}
\newcommand{\procNil}{\procFmt{\mathbf{0}}}
\newcommand{\pnil}{\procNil}

\newcommand{\SumIntRaw}[1]{\mathop{\contrColor{\bigoplus_{#1}}}}
\newcommand{\SumExtRaw}[1]{\mathop{\contrColor{\bigwith_{#1}}}}

\newcommand{\sumInt}{\mathbin{\contrColor{\oplus}}}
\newcommand{\sumExt}{\mathbin{\contrColor{\&}}}

\newcommand{\contrSeq}{\mathbin{\contrColor{.}}}

\newcommand{\SumInt}[3][]{\SumIntRaw{#1} {#2} \contrSeq {#3}}
\newcommand{\SumExt}[3][]{\SumExtRaw{#1} {#2} \contrSeq {#3}}

\newcommand{\sumI}[2]{{#1} \contrSeq {#2}}
\newcommand{\sumE}[2]{{#1} \contrSeq {#2}}

\newcommand{\ready}[1]{\ifempty{#1}{\contrFmt{\code{rdy}}}{\contrFmt{\code{rdy}}\; {#1}}}
\newcommand{\ctx}{\contrFmt{\code{\colorPtp{ctx}}}}

\newcommand{\rec}[2]{\contrFmt{\operatorname{\code{rec}}}\,{\contrFmt{#1}}\! \contrSeq {\contrFmt{#2}}}
\newcommand{\contrNil}{\contrFmt{\code{1}}}
\newcommand{\cnil}{\contrNil}

\newcommand{\obbl}[4][]{{{\operatorname{O}}_{#1}^{#2 @ #3}}({#4})}
\newcommand{\RdyS}[3][]{\operatorname{Rdy}^{#2 @ #3}_{#1}}

\newcommand{\compliant}[0]{\bowtie}
\newcommand{\acompliant}[0]{\compliant_{\infty}}
\newcommand{\scompliant}[0]{\compliant_1}

\newcommand{\syncmv}{\circ}
\newcommand{\cmove}[2][]{{\xrightarrow{#2}\hspace{-1.8ex}\rightarrow_{#1}}}
\newcommand{\acmove}[1]{\cmove[\infty]{#1}}
\newcommand{\scmove}[1]{\cmove[1]{#1}}
\newcommand{\abscmove}[1]{\mathrel{\cmove{#1}_{\sharp}}}
\newcommand{\abscmovectx}[1]{\mathrel{\abscmove{\ctx \says {#1}}}}

\newcommand{\sysmove}[4][]{\ifempty{#2}
  {\rightarrow_{#1}}
  {\ifempty{#4}
    {\xrightarrow{{\pmv {#2}} \psays {#3}}_{#1}}
    {\xrightarrow{{\pmv {#2}} \psays {#3}, {#4}}_{#1}}
  }
}

\newcommand{\ecmove}[1][]{\xrightarrow{#1}_{\sharp}}

\newcommand{\cta}[1]{\alpha_{\pmv A} (#1)}
\newcommand{\atomA}[1][]{{\contrColor{a_{#1}}}}
\newcommand{\atomB}[1][]{{\contrColor{b_{#1}}}}

\newcommand{\readydosys}[3]{{#3}\!\downarrow^{#2}_{#1}}
\newcommand{\readydoweak}[4][]{{#4}\!\Downarrow^{#3 @ #2}_{#1}}

\newcommand{\passivearrow}[3][]{\xrightarrow{\neq ({#2} \colon \cocodo{#3}{\!})}_{#1}}

\newcommand{\dotseq}{\mathbin{.}}
\newcommand{\eqdef}{\triangleq}
\newcommand{\mmdef}{\eqdef}
\newcommand{\wif}[3]{\textup{\texttt{\small if }} {#1} \;\textup{\texttt{\small then}}\;{#2}\;\textup{\texttt{\small else}}\;{#3}}
\newcommand{\inference}[3][]{\infer[#1]{#3}{#2}}

\NewEnviron{restheorem}[2][]
  {\begin{restatable}[#1]{thm}{#2}%
    \label{#2}
    \BODY
   \end{restatable}%
   \begin{proof}%
     See page~\pageref{#2-proof}.%
   \end{proof}}
\NewEnviron{reslemma}[2][]
  {\begin{restatable}[#1]{lem}{#2}%
    \label{#2}
    \BODY
   \end{restatable}%
   \begin{proof}%
     See page~\pageref{#2-proof}.%
   \end{proof}}

  {}

\newenvironment{proofoflem}[2][]{%
  \ifempty{#1}
  {\subsection*{Proof of Lemma~\ref{#2}}}
  {\subsection*{Proof of Lemma~\ref{#2} ({#1})}}
  \label{#2-proof}
  }%
  {}

\newenvironment{proofofthm}[2][]{%
  \ifempty{#1}
  {\subsection*{Proof of Theorem~\ref{#2}}}
  {\subsection*{Proof of Theorem~\ref{#2} ({#1})}}
  \label{#2-proof}
  }%
  {}

\theoremstyle{plain}
\newtheorem{thm}{Theorem}[section]

\newtheorem{lem}[thm]{Lemma}

\theoremstyle{definition}
\newtheorem{rem}[thm]{Remark}

\newtheorem{exa}[thm]{Example}

\newtheorem{defi}[thm]{Definition}

\newtheorem{nota}[thm]{Notation}

\begin{document}

\title[Honesty by Typing]{Honesty by Typing\rsuper*}

\author[M. Bartoletti]{Massimo Bartoletti} 
\address{Universit\`a degli Studi di Cagliari, Italy}	
\email{bart@unica.it}  

\author[A. Scalas]{Alceste Scalas}
\address{Universit\`a degli Studi di Cagliari, Italy %
\;and\; Imperial College London, UK}	
\email{alceste.scalas@imperial.ac.uk}  

\author[E. Tuosto]{Emilio Tuosto}
\address{University of Leicester, UK}
\email{emilio@leicester.ac.uk}

\author[R. Zunino]{Roberto Zunino}
\address{Universit\`a degli Studi di Trento, Italy}
\email{roberto.zunino@unitn.it}

\ACMCCS{[{\bf Theory of computation}]: Models of computation; Concurrency --
Semantics and reasoning -- Program reasoning -- Program specifications -- Program verification;
Semantics and reasoning -- Program semantics -- Operational semantics.}
\titlecomment{{\lsuper*}Full version of an Extended Abstract presented at FORTE'13.} 

\pagestyle{headings}    

\begin{abstract}
  We propose a type system for a calculus of contracting processes. 
  Processes can establish \emph{sessions}
  by stipulating \emph{contracts}, and then can interact either
  by keeping the promises made, or not. 
  Type safety guarantees that a typeable process is \emph{honest} --- 
  that is, it abides by the contracts it has stipulated 
  in all possible contexts,
  even in presence of dishonest adversaries.
  Type inference is decidable, and it allows 
  to safely approximate the honesty of processes
  using either \emph{synchronous} or \emph{asynchronous}
  communication.
\end{abstract}


\maketitle              


\section{Introduction} \label{sec:introduction}

It is commonplace that distributed applications are not easy to develop.
Besides the intrinsic issues due \eg to physical distribution, 
and to the fragility of communication networks, 
distributed applications have to be engineered within an apparent dichotomy.
On the one hand, distributed components have to \emph{cooperate} 
in order to achieve their goals and, 
on the other hand, they may have to \emph{compete}, 
\eg to 
access resources
or other components with limited availability.
This dichotomy is well witnessed by the \emph{service-oriented}
paradigm, %
which fosters \emph{dynamic} composition of distributed applications, %
with the interaction of components, \aka ``services'',
deployed by different vendors.

Cooperation and competition hardly coexist harmoniously.
A frequent simplification is to neglect competition, 
by assuming that \emph{all} components (including third-party ones) 
always adhere to some declared specification.
For instance, such a specification could be a \emph{behavioural type}, 
abstracting the input/output behaviour of a component
--- and the simplifying assumption is that 
each component is verified against its declared specification,
and has a corresponding runtime behaviour. %
We argue that this assumption is hardly realistic
in scenarios where %
third-party components can be used but not inspected, %
and thus are not guaranteed to honour their declared
behaviour. %
This is particularly relevant in competitive scenarios,
which may incentivise ``selfish'' components %
that diverge from their declared specification.

Contract-oriented computing~\cite{BZ10lics,BTZ12sacs} addresses these concerns
by disciplining the interaction among components through \emph{contracts},
which formalise promised runtime behaviours.
In a contract-oriented setting,
the components of a distributed application %
are implemented as \emph{processes}
interacting through a \emph{contract-oriented middleware}. %
Processes may \emph{advertise} contracts; %
if the middleware finds a set of \emph{compliant} contracts, %
it establishes a new \emph{session} %
involving the advertising processes.
When involved in a session, %
a process should perform the interactions
needed to realise its contract;
however, a process could also \emph{violate} its contract,
\eg by not performing some promised input/output action: %
this may happen either maliciously,
or unintentionally, because of
some implementation bug.
To discourage contract violations, the middleware
can monitor the sessions, %
and sanction the processes which do not respect their contracts.
These sanctions can be \eg pecuniary compensations,
or marginalisation: %
if a process misbehaves, the middleware could %
decrease its reputation,
and consequently its chances of being involved in further
sessions~\cite{CO2,Mukhija2007qos}.

In this scenario, to avoid sanctions, a process must be able
to respect \emph{all} the contracts it advertises, in \emph{all} possible contexts
--- even those populated by adversaries which try to make it sanctioned.
We call this property \emph{honesty}.
A crucial problem is then how to guarantee that a process is honest.


\subsection*{An example}

Consider an online store $\pmv{S}$ taking orders from buyers. %
The store sells two items: 
apples, which are always available and costs \EURtm1,
and bananas, which costs \EURtm1 when in stock, and \EURtm3 otherwise. %
In the latter case, the store orders bananas from an external distributor, 
which makes the store pay \EURtm2 per item. %

The store advertises the following contract to potential buyers $\pmvB$:
\begin{enumerate}
\item let the buyer choose between apples and bananas;
\item if the buyer chooses apples, then receive \EURtm1, and then ship the item to him;
\item if the buyer chooses bananas, offer a quotation to the buyer (\EURtm1 or \EURtm3);
\item if the quotation is \EURtm1, then receive the payment and ship;
\item if the quotation is \EURtm3, ask the buyer to pay or cancel the order;
\item if the buyer pays \EURtm3, then either ship the item to him, or refund \EURtm3.
\end{enumerate}

We can formalise such contract in several process algebras. %
For instance, we can use the following session type~\cite{Honda98esop}
(without channel passing):
\begin{align*}
  \contrP[\pmvB] \; = \;
  & \atomIn{buyA} \contrSeq \atomIn{pay1E} \contrSeq \atomOut{shipA}   
  \;\; \sumExt \\
  & \atomIn{buyB} \contrSeq (\atomOut{quote1E} \contrSeq \atomIn{pay1E} \contrSeq \atomOut{shipB} 
    \; \sumInt \;  
    \atomOut{quote3E} \contrSeq \contrPi[\pmvB]
    \; \sumInt \;  
    \atomOut{abort}
    )
  \\                  
  \contrPi[\pmvB] 
  \; = \;
  & \atomIn{pay3E} \contrSeq (\atomOut{shipB} \sumInt \atomOut{refund}) \sumExt \atomIn{quit}
\end{align*}
where \eg, $\atomIn{buyA}$ represents a label in a \emph{branching} construct
(\ie, receiving an order for apples from the buyer), while 
$\atomOut{quote1E}$ represents a label in a \emph{selection} construct
(\ie, sending an \EURtm1 quotation to the buyer). %
The operator $\sumInt$ separates branches in an \emph{internal choice},
while $\sumExt$ separates branches in an \emph{external choice}. %

The protocol between the store and a distributor $\pmv{D}$ is the following:
\[
  \contrP[\pmv D] 
  \; = \; 
  \atomOut{buyB} \contrSeq (\atomOut{pay2E} \contrSeq \atomIn{shipB} \;\sumInt\; \atomOut{quit})
\]

Note that the contracts above do not specify the \emph{actual} behaviour of the store, 
but only the behaviour it promises towards the buyer and the distributor. %
A possible informal description of the actual behaviour of the store $\pmv{S}$ 
is the following
(see~\Cref{ex:co2:online-store} for a formal specification):
\begin{enumerate}
\item $\pmv{S}$  advertises the contract $\contrP[\pmvB]$;
\item when $\contrP[\pmvB]$ is stipulated, the buyer chooses 
apples ($\atom{buyA}$) or bananas ($\atom{buyB}$);
\item if the buyer chooses apples, $\pmv{S}$ gets the payment ($\atom{pay1E}$), and ships the item ($\atom{shipA}$);
\item otherwise, if the buyer chooses bananas, $\pmv{S}$ checks if the item is in stock;
\item if bananas are in stock, $\pmv{S}$ provides the buyer with the quotation of \EURtm1 ($\atom{quote1E}$), 
  receives the payment ($\atom{pay1E}$), and ships the item ($\atom{shipB}$); 
\item otherwise, if bananas are not in stock, $\pmv{S}$ advertises the contract $\contrP[\pmv D]$;
\item when $\contrP[\pmv D]$ is stipulated, $\pmv{S}$ pre-orders bananas from the distributor ($\atom{buyB}$);
\item $\pmv{S}$ sends a \EURtm3 quotation to the buyer ($\atom{quote3E}$) 
and waits for the buyer's reply;
\item if the buyer pays \EURtm3 ($\atom{pay3E}$), 
then $\pmv{S}$  pays the distributor ($\atom{pay2E}$), 
receives the item from the distributor ($\atom{shipB}$), 
and ships it to the buyer ($\atom{shipB}$).
\end{enumerate}

\noindent The store service terminates correctly whenever 
two conditions hold:
the buyer is honest, and at step 7 the middleware selects an honest distributor. %
Such assumptions are necessary.
For instance, in their absence we have that:
\begin{enumerate} 

\item \label{ex:store-dishonest:attack-a}
if a malicious buyer does not send \EURtm3 at step 9, 
then the store does not fulfil its obligation with the distributor, 
who is expecting a payment or a cancellation;

\item \label{ex:store-dishonest:attack-b}
if the middleware finds no distributor with a contract compliant with $\contrP[\pmv D]$, then the store is stuck at line 7, 
so it does not fulfil its obligation with the buyer, who is expecting a quotation or an abort;

\item \label{ex:store-dishonest:attack-c}
if a malicious distributor does not ship the item at line 9, then the store does not fulfil its obligation with the buyer, who is expecting to receive the item or a refund;

\item \label{ex:store-dishonest:attrack-d}
if the buyer chooses $\atom{quit}$ at line 8, the store forgets to handle it;
so, it will not fulfil the contract with the distributor, 
who is expecting $\atom{pay2E}$ or $\atom{quit}$.

\end{enumerate}

\noindent Therefore, we would classify the store process above as \emph{dishonest}
(we will formalise an honest variant of the store later on, %
in~\Cref{ex:type-proc:online-store}). %
In practice, this implies that a concrete implementation of such a store could be easily attacked. %
For instance, an attacker could simply order bananas (when not in stock), 
but always cancel the transaction. %
The middleware will detect that the store is violating the contract with the distributor, 
and consequently it will sanction the store. %
Concretely, in the middleware of~\cite{CO2} the attacker will manage to never be sanctioned, 
and to arbitrarily decrease the store reputation, 
so preventing the store from being able to establish new sessions with buyers.

The example above shows that writing honest processes is an error-prone task:
this is because one has to foresee all the possible points of failure of each partner. %


\subsection*{Contributions} \label{sec:intro-contributions}

We address the problem of honesty in \coco~\cite{BTZ12sacs}, 
a core calculus for contract-oriented computing.
The main contribution of this paper is a type discipline for
statically ensuring when a \coco process is \emph{honest}.
The need for a static approximation is motivated by the fact
that honesty is an undecidable property, as shown in~\cite{BZ15wsfm}.

To obtain this result, we first study a theory of compliance between session types,
both with a synchronous and an asynchronous semantics.
We prove that asynchronous session types are Turing-powerful 
(\Cref{th:st:async-simulates-tm}). 
In~\Cref{th:st:proper-compliance-undecidable}
we show that asynchronous compliance is undecidable,
while the synchronous one is decidable.

We introduce a type system for \coco processes,
which associates behavioural types (based on Basic Parallel Processes) 
to each session name and variable. 
For these types we define a notion of abstract honesty,
which we prove decidable (\Cref{th:abs-honesty:decidable}).
Exploiting this result, we show that
our type system has a decidable type inference (\Cref{th:types:decidability}).

We establish subject reduction, \ie types simulate processes
(\Cref{th:types:subject-reduction}),
and a strong form of progress which ensures
that processes simulate types (\Cref{th:progress}).
We then exploit these results to prove type safety, which guarantees that
typeable processes are honest (\Cref{th:type-safety}).
Further, using~\Cref{th:honest-sync-async}, we can lift our type safety result
to the case of asynchronous \coco processes.

Together with the decidability of type inference, 
we obtain an algorithm to safely approximate the honesty of \coco processes,
both synchronous and asynchronous.
Our algorithm is more precise than the model-checking technique in~\cite{BMSZ15jlamp},
which can only establish the honesty of essentially finite-state processes 
(\ie, without parallel nor delimitation under recursion).
For instance, \Cref{ex:type-proc:2} shows a process for which the analysis technique proposed
in this paper is more precise than the one in~\cite{BMSZ15jlamp}.


\section{Session types as contracts} 
\label{sec:session-types}
\label{sec:contracts}

We use first-order binary session types~\cite{Barbanera10ppdp}
as contracts.
These are terms of a process algebra
featuring internal/external choice, and recursion.
We consider two semantics:
an asynchronous one, using unbounded queues (\Cref{def:st:async-semantics}),
and a synchronous one (\Cref{def:st:sync-semantics}).
We then study a \emph{compliance} relation between session types,
which is common for both semantics: 
roughly, the compliance relation formalizes when two contracts 
do not yield communication errors.
In particular, we show that compliance between session types 
is not decidable (\Cref{th:st:proper-compliance-undecidable}) 
under the asynchronous semantics,
while it is decidable under the synchronous one.
We then show that synchronous compliance 
is a sound approximation of the asynchronous one
(\Cref{th:st:compliance:sync-implies-async}).

\subsection{Syntax} \label{sec:session-types:syntax}

We assume a set of \emph{participants} 
(ranged over by ${\pmv A}, {\pmv B}, \ldots$),
and a set of \emph{actions},
partitioned into %
\emph{input actions} $\atomIn{a}, \atomIn{b}, \ldots \in \AtomIn$, %
and
\emph{output actions} $\atomOut{a}, \atomOut{b}, \ldots \in \AtomOut$. %
We let $\atomA, \atomB, \ldots$ range over $\AtomIn \cup \AtomOut$. %

\begin{defi}[Session types] \label{def:st:syntax}
  Session types are terms of the following grammar:
  \begin{align*}
    \contrP,\contrQ \;\; & ::= \;\;
    \SumInt[i \in \mathcal{I}]{\atomOut[i]{a}}{\contrP[i]} \ \bnfmid \ 
    \SumExt[i \in \mathcal{I}]{\atomIn[i]{a}}{\contrP[i]} \ \bnfmid \
    \rec{\contrX}{\contrP}
    \bnfmid \ \contrX
  \end{align*}
  where 
  \begin{inlinelist} 
  \item the index set $\mathcal{I}$ is finite,
  \item \label{item:def:contracts:syntax:pairwise-distinct}
    in a summation,
    $\atom[i]{a} = \atom[j]{a}$ implies $i = j$, and
  \item recursion variables $\contrX$ are prefix-guarded.
  \end{inlinelist}
\end{defi}

An internal sum $\SumInt[i]{\atomOut[i]{a}}{\contrP[i]}$ 
allows a participant to choose one of the labels $\atom[i]{a}$,
and then to behave according to the branch $\contrP[i]$.
Dually, an external sum  
$\SumExt[i]{\atomIn[i]{a}}{\contrP[i]}$ 
allows to wait for the other participant to choose one of the labels
$\atom[i]{a}$, 
and then behave according to the branch $\contrP[i]$.
Empty internal/external sums are identified, 
and they are denoted with $\cnil$, 
which represents a \emph{success state} wherein the interaction
has terminated correctly.

We use the binary operators to isolate a branch in a sum: 
\eg, $\contrP = (\sumI{\atomOut{a}}{\contrPi}) \sumInt \contrPii$ 
means that $\contrP$ has the form 
$\SumInt[i \in \mathcal{I}]{\atomOut[i]{a}}{\contrP[i]}$
and there exist some $i \in \mathcal{I}$ such that 
$\sumI{\atomOut{a}}{\contrPi} = \sumI{\atomOut[i]{a}}{\contrP[i]}$.
Hereafter, we will omit 
the trailing occurrences of $\cnil$,
and we will only consider session types without free
occurrences of recursion variables~$\contrX$.


\subsection{Semantics} \label{sec:session-types:semantics}

While a session type models the intended behaviour of \emph{one} 
of the two participants involved in a session,
the interaction of \emph{two}  participants {\pmv A} and {\pmv B} 
(say, with session types $\contrP$ and $\contrQ$, respectively)
is modelled by \emph{configurations} $\gamma, \gammai, \ldots$ of the form
$\bic{\contrP\buffer{\qmvA}}{\contrQ\buffer{\qmvB}}$,
where $\qmvA$ (resp.\ $\qmvB$) is an unbounded queue that stores 
the messages sent by $\pmvB$ (resp.\ by $\pmvA$) and not read yet.

\begin{defi}[Asynchronous semantics of session types] 
  \label{def:st:async-semantics} 
  A \emph{contract configuration} is a term of the form
  $\bic{\contrP\buffer{\qmvA}}{\contrQ\buffer{\qmvB}}$, 
  where $\pmv A \!\neq\! \pmv B$ and 
  $\qmvA, \qmvB \in (\AtomOut)^*$.
  We define the relation $\equiv$
  between session types as the least equivalence including
  $\alpha$-conversion of recursion variables
  and unfolding of recursion
  (that is, $\rec{\contrX}{\contrP} \equiv \contrP \subs{\rec{\contrX}{\contrP}}{\contrX}$, %
  as in the usual equi-recursive approach). %
  The labelled transition relation $\acmove{}$
  is the smallest relation between contract configurations 
  induced by the rules in~\Cref{fig:st:async-semantics} 
  up to~$\equiv$. %
\end{defi}

\begin{figure}[t]
  \[
  \begin{array}{rcll}
    \bic
    {(\sumI{\atomOut{a}}{\contrP} \sumInt \contrPi) \buffer{\qmvA}}
    {\contrQ \buffer{\qmvB}}
    & \acmove{\pmvA \says \atomOut{a}} &
    \bic{\contrP \buffer{\qmvA}}{\contrQ {\buffer{\qmvB \ \atomOut{a}}}}
    &
    \nrule{[Out]}
    \\[8pt]
    \bic
    {\contrP \buffer{\qmvA}}
    {(\sumE{\atomIn{a}}{\contrQ} \sumExt \contrQi) {\buffer{\atomOut{a} \ \qmvB}}}
    & \acmove{{\pmv B} \says \atomIn{a}} &
    \bic
    {\contrP \buffer{\qmvA}}
    {\contrQ \buffer{\qmvB}}
    & 
    \nrule{[In]}
  \end{array} 
  \]
  \caption{Asynchronous semantics of session types (symmetric rules omitted).}
  \label{fig:st:async-semantics}
\end{figure}

In rule~\nrule{[Out]}, participant {\pmv A}
chooses the branch $\atomOut{a}$ in an internal sum,
and sends $\atomOut{a}$ to the other participants’ queue.
The other participant {\pmv B} can read the message from the queue through rule~\nrule{[In]},
and then proceed as in the corresponding branch in its external choice.

\begin{defi}[Synchronous semantics of session types]
  \label{def:st:sync-semantics} 
  We define the relation~$\scmove{}$ as the subset of $\acmove{}$ where:
  \begin{inlinelist} 
  \item each queue in $\gamma$ contains at most one message;
  \item there exists at most one non-empty queue in $\gamma$.
  \end{inlinelist}
  Hereafter, we use the symbol $\syncmv \in \setenum{1,\infty}$ 
  to parameterise various notions over the synchronous/asynchronous semantics.
\end{defi}

\begin{exa}
  \label{ex:st-semantics-sync-async}
  Let \;$\contrP = \atomOut{a}.\atomIn{b}$,
  \;$\contrQ = \atomOut{b}.\atomIn{a}$\; %
  \;and\; %
  $\gamma = \bic{\contrP\!\!\buffer{}}{\contrQ\!\!\buffer{}}$. %
  Under the $\scmove{}$ semantics, we have the following traces, %
  where no participant reaches a success state\footnote{%
    We will see that both final configurations %
    are deadlocks under $\scmove{}$, %
    by \Cref{def:deadlock-message-obliviousness}.%
  }:
  \[%
  \begin{array}{c}
    \gamma%
    \;\;\scmove{\pmvA \says \atomOut{a}}\;\;
    \bic{\atomIn{b}\buffer{}}{\contrQ\buffer{\atomOut{a}}}
    \;\;\not\!\!\scmove{}%
    \qquad\text{and}\qquad%
    \gamma%
    \;\;\scmove{\pmvB \says \atomOut{b}}\;\;
    \bic{\contrP\buffer{\atomOut{b}}}{\atomIn{a}\buffer{}}
    \;\;\not\!\!\scmove{}%
  \end{array}  
  \]
  
  Under the $\acmove{}$ semantics, instead, we have the following
  traces, all leading to the success state for both participants:
  {\small%
  \[
  \begin{array}{@{\hskip -0.5mm}c}
    \gamma%
    \;\acmove{\pmvA \says \atomOut{a}}\;
    \bicSmall{\atomIn{b}\!\!\buffer{}}{\contrQ\!\!\buffer{\atomOut{a}}}
    \;\acmove{\pmvB \says \atomOut{b}}\;
    \bicSmall{\atomIn{b}\!\!\buffer{\atomOut{b}}}{\atomIn{a}\!\!\buffer{\atomOut{a}}}
    \;\acmove{\pmvA \says \atomIn{b}}\;
    \bicSmall{\contrNil\!\!\buffer{}}{\atomIn{a}\!\!\buffer{\atomOut{a}}}
    \;\acmove{\pmvB \says \atomIn{a}}\;
    \bicSmall{\contrNil\!\!\buffer{}}{\contrNil\!\!\buffer{}}
    \\%
    \gamma%
    \;\acmove{\pmvA \says \atomOut{a}}\;
    \bicSmall{\atomIn{b}\!\!\buffer{}}{\contrQ\!\!\buffer{\atomOut{a}}}
    \;\acmove{\pmvB \says \atomOut{b}}\;
    \bicSmall{\atomIn{b}\!\!\buffer{\atomOut{b}}}{\atomIn{a}\!\!\buffer{\atomOut{a}}}
    \;\acmove{\pmvB \says \atomIn{a}}\;
    \bicSmall{\atomIn{b}\!\!\buffer{\atomOut{b}}}{\contrNil\!\!\buffer{}}    
    \;\acmove{\pmvA \says \atomIn{b}}\;
    \bicSmall{\contrNil\!\!\buffer{}}{\contrNil\!\!\buffer{}}
    \\%
    \gamma%
    \;\acmove{\pmvB \says \atomOut{b}}\;
    \bicSmall{\contrP\!\!\buffer{\atomOut{b}}}{\atomIn{a}\!\!\buffer{}}
    \;\acmove{\pmvA \says \atomOut{a}}\;
    \bicSmall{\atomIn{b}\!\!\buffer{\atomOut{b}}}{\atomIn{a}\!\!\buffer{\atomOut{a}}}
    \;\acmove{\pmvA \says \atomIn{b}}\;
    \bicSmall{\contrNil\!\!\buffer{}}{\atomIn{a}\!\!\buffer{\atomOut{a}}}
    \;\acmove{\pmvB \says \atomIn{a}}\;
    \bicSmall{\contrNil\!\!\buffer{}}{\contrNil\!\!\buffer{}}
    \\%
    \gamma%
    \;\acmove{\pmvB \says \atomOut{b}}\;
    \bicSmall{\contrP\!\!\buffer{\atomOut{b}}}{\atomIn{a}\!\!\buffer{}}
    \;\acmove{\pmvA \says \atomOut{a}}\;
    \bicSmall{\atomIn{b}\!\!\buffer{\atomOut{b}}}{\atomIn{a}\!\!\buffer{\atomOut{a}}}
\;\acmove{\pmvB \says \atomIn{a}}\;
    \bicSmall{\atomIn{b}\!\!\buffer{\atomOut{b}}}{\contrNil\!\!\buffer{}}    
    \;\acmove{\pmvA \says \atomIn{b}}\;
    \bicSmall{\contrNil\!\!\buffer{}}{\contrNil\!\!\buffer{}}
  \end{array}
  \]
  }

\end{exa}

The following~\namecref{th:st:async-simulates-tm}
states that session types with the asynchronous semantics
simulate Turing machines.
Our result is closely related to analogous expressiveness results 
for Communicating Finite State Machines (CFSMs)~\cite{Brand81tr,Finkel97tcs}.
The construction in~\cite{Brand81tr} is used to show that a system of two CFSMs 
with \emph{mixed choices} can simulate Turing machines;
the one in~\cite{Finkel97tcs} is technically different, 
as it uses two copies of the same \emph{non-deterministic} CFSM with no mixed choices.
Compared to~\cite{Brand81tr,Finkel97tcs}, our result is slightly stronger: 
indeed, session types are deterministic 
(in the sense of constraint~\ref{item:def:contracts:syntax:pairwise-distinct}
in~\Cref{def:st:syntax})
and without mixed choices, 
so our encoding has to work in a restricted model compared to CFSMs.
Essentially, our proof constructs a
system of two deterministic CFSMs without mixed choice.

\newcommand{\readTM}[1]{\mathit{read}({#1})}
\newcommand{\writeTM}[1]{\mathit{write}\langle{#1}\rangle}

\begin{thm}
  \label{th:st:async-simulates-tm}
  Session types (with the $\acmove{}$ semantics) can simulate Turing machines.
\end{thm}
\begin{proof}\emph{(Sketch)}.
  Assume as given a deterministic Turing machine $M$ 
  with states $Q = \setenum{q_1, \ldots,  q_n}$,
  initial/halting states $q_0,q_h \in Q$,
  tape alphabet $\Sigma$ (with blank symbol $\#$),
  and transition function~$\delta$.
  Let $s_0 \ldots s_n \in \Sigma^*$ be the initial tape.
  Each configuration of $M$ can be represented as a string
  $s_0 s_1 \ldots s_{i-1} q_k s_{i} \ldots s_m \, {\it end}$, which is obtained from the tape
  $s_0 s_1 \ldots s_m$ by inserting the state/symbol $q_k$ at the position of the head ($i$).

  We now define two session types $\contrP$, $\contrQ$ such that
  $\bic{\contrP \buffer{}}{\contrQ \buffer{}}$ simulates $M$ stepwise.
  Intuitively, $\contrP$ acts as a transducer that
  inputs the symbols of a configuration of $M$, and
  outputs the symbols of the next configuration. 
  Instead, $\contrQ$ simply outputs the initial configuration of $M$
  and then echoes every symbol back, until a ${\it stop}$ message is received.
  To perform its task, $\contrP$ works in a streaming fashion, emitting the output
  configuration while reading the input one. Indeed, to determine the next symbol to emit,
  we need only a bounded lookahead. In this way, the procedure requires a finite amount
  of memory, and is amenable to be implemented in a session type.
  After reaching the $\it end$ marker, $\contrP$ restarts, so to handle the 
  new configuration received from $\contrQ$. 
  If $\contrP$ eventually finds a configuration in the halting state,
  it signals ${\it stop}$ to $\contrQ$.

  More precisely, we define the session type $\contrQ$ as follows:
  \begin{align*}
    \contrQ
    & \; = \; \atomOut{q_0} \, . \, \atomOut{s_0} \ldots \atomOut{s_n} \, . \, \atomOut{end} \, . \, 
      (\rec{\contrX}{
      \atomIn{stop} \;\sumExt\; \textstyle
      \SumExt[\atom{x} \in \Sigma \cup Q \cup \setenum{{\it end}}]{\atomIn{x}}{\atomOut{x} \, . \, \contrX}}
      )
  \end{align*}

  In order to define the session type $\contrP$, 
  it is convenient to first give a set of (recursive) defining equations,
  and then apply Beki\'c theorem~\cite{Bekic84} %
  (see Theorem 10.1 in~\cite{Winskel93book})
  to obtain the (recursive) term $\contrP$.
  To this purpose, we introduce some auxiliary notation.
  The shortcut $\writeTM{x} . \contr{T}$ 
  stands for $\atomOut{x} . \contr{T}$, 
  or for $\contr{T}$ if $x$ is the special value $\bot$.
  Dually, $\readTM{x} . \contr{T}$ stands for 
  $\SumExt[x \in \Sigma\cup Q \cup \setenum{{\it end}}]{\atomIn{x}}{\contr{T}}$; 
  here the index $x$ may appear in $\contr{T}$, 
  hence it is considered a binder. 
  The defining equations are as follows:
  \begin{align*}
    \contr[0]{T}(p) 
    & = \readTM{s} . \contr[1]{T}(p,s) 
    \\
    \contr[1]{T}(p,s) 
    & = \begin{cases}
      \writeTM{p} . \contr[0]{T}(s) 
      & \text{if $s \in \Sigma$} \\
      \readTM{s_2}. \contr[2]{T}(p,s,s_2) & \text{if $s \in Q$} \\
      \writeTM{p} . \writeTM{\#} . \writeTM{\it end} . \contr[0]{T}(\bot) 
      & \text{if $s = {\it end}$}
    \end{cases}
    \\
    \contr[2]{T}(p,q,s_2) &= \begin{cases}
      \writeTM{p} . \writeTM{q_h} . \writeTM{s_2} . \writeTM{\it stop}
      & \text{if $q = q_h$} \\
      \writeTM{q'} . \writeTM{p} . \contr[0]{T}(s_3) & \text{if $dir = L$}  \\
      \writeTM{p} . \writeTM{q'} . \contr[0]{T}(s_3) & \text{if $dir = -$} \\
      \writeTM{p} . \writeTM{s_3} . \writeTM{q'} . \readTM{s_4} . \contr[0]{T}(s_4) & \text{if $dir = R$}
    \end{cases}
    \\
    & \hspace{20pt}
      \text{where $(q', s_3, dir) = \delta(q,s_2)$}
  \end{align*}
  and we obtain $\contrP$ by applying Beki\'c theorem
  on $\contr[0]{T}(\bot)$.

  The contract configuration 
  $\bic{\contrP \buffer{}}{\contrQ \buffer{}}$ simulates $M$ on the initial tape. 
  This is because $\contrP$ receives its own outputs back, and so repeatedly performs the steps of $M$.
\end{proof}


\subsection{Compliance} \label{sec:session-types:compliance}

We define compliance between session types by taking inspiration 
from the notion of safety on Communicating Finite State Machines~\cite{Finkel05infoco,Lange15popl}.
We start with some auxiliary definitions.

\begin{defi}[Deadlock \& message-obliviousness]
  \label{def:deadlock-message-obliviousness}%
  We say that a configuration $\gamma$ is:
  \begin{itemize}

    \item a \emph{deadlock under the $\cmove[\syncmv]{}$ semantics} \;iff:
    $\,
    \gamma \not\!\!\cmove[\syncmv]{}
    \;\;\text{ and }\;\;
    \gamma \neq \bic{\contrNil\buffer{}}{\contrNil\buffer{}}
    $

  \item \emph{message-oblivious under the $\cmove[\syncmv]{}$ semantics} \;iff\; 
    $\gamma = \bic{\contrP\buffer{\atomOut{a} \qmvA}}{\contrQ\buffer{\qmvB}}$
    \;and
    \[
    \gamma
    \;\; \cmove[\syncmv]{\pmvA \says \atomOut[1]{a}} \;\;
    \cdots
    \;\; \cmove[\syncmv]{\pmvA \says \atomOut[k]{a}} \;\;
    \gammai 
    \;\; \text{ implies } \;\;
    \gammai
    \not\!\!\cmove[\syncmv]{\;\; \pmvA \says \atomIn{a} \;\;}
    \]
    or the symmetric condition holds when the roles of $\pmvA$ and $\pmvB$ are swapped.
  \end{itemize}
\end{defi}

\noindent The notion of deadlock is standard;
\emph{message-obliviousness} characterizes the configurations
where an enqueued message is left unread forever.

Our compliance relation requires that no reachable configuration
is a deadlock or is message-oblivious.

\begin{defi}[Compliance]
  \label{def:st:compliance}
  We say that $\contrP$ and $\contrQ$ are \emph{compliant 
    under the $\cmove[\syncmv]{}$ semantics}
  (in symbols, $\contrP \compliant_{\syncmv} \contrQ$) whenever:
  $
  \bic{\contrP\buffer{}}{\contrQ\buffer{}} 
  \;\cmove[\syncmv]{}^*\; 
  \gamma
  $
  implies that $\gamma$ is neither deadlock %
  nor a message-oblivious configuration %
  under the $\cmove[\syncmv]{}$ semantics.
\end{defi}

Note that when applying this notion to the synchronous semantics,
deadlock-freedom implies absence of message-oblivious configurations,
and so our compliance $\scompliant$ coincides exactly 
with the usual (symmetric) progress-based notion in~\cite{Barbanera10ppdp,BSZ14concur,BCZ15plabs}.

\begin{thm}
  \label{th:st:proper-compliance-undecidable}
  $\scompliant$ is decidable; $\acompliant$ is undecidable.
\end{thm}
\begin{proof}
  Decidability of $\scompliant$ follows because
  the state space of $\bic{\contrP\buffer{}}{\contrQ\buffer{}}$
  under $\scmove{}$ is finite.
  Undecidability of $\acompliant$ follows by~\Cref{th:st:async-simulates-tm}.
\end{proof}

The following~\namecref{th:st:compliance:sync-implies-async},
which is a corollary of results in~\cite{BSZ14concur,Scalas15phd}, 
shows that synchronous compliance implies asynchronous compliance.
We can then use a decision procedure for synchronous compliance
(which is decidable)
to safely approximate asynchronous compliance 
(which is undecidable).

\begin{thm}
  \label{th:st:compliance:sync-implies-async}
  $\scompliant \;\subsetneq\; \acompliant$.
\end{thm}
\begin{proof}\emph{(Sketch)}.
  The inclusion $\subseteq$ is proved in two parts:
  first we show that deadlock-freedom holds in the asynchronous semantics;
  second, we show that absence of message-oblivious configurations holds as well.

  The first part follows from Proposition 3 in~\cite{BSZ14concur},
  showing that, for session types,
  client-biased progress (denoted by $\dashv$ in~\cite{BSZ14concur}),
  can be lifted from synchronous to asynchronous semantics.
  Our deadlock-freedom is equivalent to the intersection between
  client-biased progress $\dashv$ and the symmetric server-biased progress $\vdash$.
  Applying Proposition~3 of~\cite{BSZ14concur} twice, we get that
  the relation $\dashv\vdash \, = \, \dashv \cap \vdash$
  (\ie, our deadlock freedom)
  can be lifted to the asynchronous semantics as well.

  For the second part 
  we exploit Theorems 4.9, 4.13 and 5.22 in~\cite{Scalas15phd}.
  These results establish relations between progress,
  the \emph{I/O compliance} relation of~\cite{BSZ14concur}
  (a notion of compliance which is stricter than progress on arbitrary LTSs,
  and coincides with it on the LTSs of session types),
  and the notion \emph{orphan message configuration} of~\cite{Scalas15phd}
  (\ie, configurations containing messages which are sent but never received).
  More precisely, assume that
  $\contrP \scompliant \contrQ$, 
  \ie that $\contrP \dashv\vdash \contrQ$ using the notation of~\cite{Scalas15phd}.
  Then:
  \begin{enumerate}

  \item \label{proof:st:compliance:sync-implies-async:1}
    by Theorem 4.9 in~\cite{Scalas15phd}, 
    $\contrP \dashv\vdash \contrQ$ iff 
    $\contrP$ and $\contrQ$ are I/O compliant;

  \item \label{proof:st:compliance:sync-implies-async:2}
    by Theorem 4.13 in~\cite{Scalas15phd},
    I/O compliance between session types 
    is preserved when passing from the synchronous to the asynchronous semantics;

  \item \label{proof:st:compliance:sync-implies-async:3}
    by Theorem 5.22 in~\cite{Scalas15phd},
    if two session types are I/O compliant in the asynchronous semantics,
    their asynchronous execution cannot have orphan message configurations;

  \item \label{proof:st:compliance:sync-implies-async:4}
    by Definition 5.5 in~\cite{Scalas15phd} 
    and by~\Cref{def:deadlock-message-obliviousness},
    orphan message and message-oblivious configurations
    coincide for session types.

  \end{enumerate}  

\noindent   To obtain the thesis, assume that
  $\contrP \scompliant \contrQ$
  and
  $
  \bic{\contrP\buffer{}}{\contrQ\buffer{}} 
  \;\cmove[\infty]{}^*\; 
  \gamma
  $.
  By items~\eqref{proof:st:compliance:sync-implies-async:1}--\eqref{proof:st:compliance:sync-implies-async:4}
  above, $\gamma$ is \emph{not} a message-oblivious configuration.

  \medskip
  The strict inclusion it witnessed by
  $\contrP = \atomOut{a}.\atomIn{b}$ and
  $\contrQ = \atomOut{b}.\atomIn{a}$, %
  since
  $\contrP \not\scompliant \contrQ$ %
  and $\contrP \acompliant \contrQ$ %
  (see the traces in \Cref{ex:st-semantics-sync-async}).
\end{proof}


\section{The \coco calculus} \label{sec:co2}

We now present an instance 
the process calculus \coco~\cite{BTZ12coordination}
with the contracts of~\Cref{sec:contracts}.


\subsection{Syntax} \label{sec:co2-syntax}

Let $\vars$ and $\snames$ be disjoint sets of,
respectively, \emph{session variables} (ranged over by $x,y,\ldots$) and 
\emph{session names} (ranged over by $s,t,\ldots$),
and let $u,v,\ldots$ range over \emph{channels} $\vars \cup \snames$.
Finite sequences are in bold, 
\eg $\vec{u},\vec{v},\ldots$ denote finite sequences of channels.

\begin{table}[t]
  \small
  \[
  \begin{array}{ll}

    \begin{array}{ll}
      \pmv{A}, \pmv{B}, \ldots & \text{Participant names}
      \\
      \atomA, \atomB, \ldots & \text{Actions}
      \\
      \contrP, \contrQ, \ldots & \text{Contracts}
      \\
      \gamma, \gammai, \ldots & \text{Contract configurations} 
      \\
      \gamma \cmove{} \gammai & \text{Transition of contract configurations}
    \end{array}

                             & \hspace{0pt}

                               \begin{array}{ll}
                                 u, v,\ldots \mbox{\hspace{50pt}} & \text{Channels, comprising:} 
                                 \\
                                 \hspace{12pt} s,t, \ldots \in \snames 
                                                                  & \hspace{12pt} \text{Session names} 
                                 \\
                                 \hspace{12pt} x, y, \ldots \in \vars
                                                                  & \hspace{12pt} \text{Variables} 
                                 \\
                                 \procP,\procQ,\ldots & \text{Processes} 
                                 \\
                                 \sysS, \sysSi,\ldots & \text{Systems}
                                 \\
                                 \sysS \sysmove{}{}{} \sysSi & \text{Transition of systems}
                               \end{array}

  \end{array}
  \]
  \caption{Summary of notation.} \label{def:notation}
\end{table}

\begin{defi}[\coco syntax]
  \label{def:co2:syntax}
  The syntax of \coco is defined as follows:
  \[\begin{array}{rcll}
      \procP  
      & ::= & 
              \cocoSum[i]{\pref[i] \cocoSeq \procP[\!i]}
              \bnfmid
              \procP \cocoPar \procP
              \bnfmid
              (u)\procP
              \bnfmid
              (\procRec{\procX(\vec y)} \procP)(\vec u)
              \bnfmid
              \procX(\vec u)
              \hspace{51pt} & \emph{(Processes)}
      \\[.5pc]
      \pref
      & ::= & 
              \tau
              \bnfmid  
              \tell {} {\freeze{u}{\contrP}}
              \bnfmid  
              \cocodo u {\atomA}
                            & \emph{(Prefixes)}
      \\[.5pc]
      \sysS  
      & ::= &  
              \emptysys 
              \bnfmid
              \sys {\pmvA} \procP 
              \bnfmid 
              \sys s \gamma 
              \bnfmid
              \setenum{\freeze u \contrP}_{\pmv A} 
              \bnfmid 
              (u) \sysS
              \bnfmid 
              \sysS \cocoPar \sysS 
                            & \emph{(Systems)}
    \end{array}
    \]
    We also assume the following syntactic constraints on processes and systems:
    \begin{enumerate}
      
    \item recursion is prefix-guarded;
      
    \item in $(\vec u)(\sys {\pmv A} \procP \cocoPar \sys{\pmv B} \procQ \cocoPar \cdots)$,
      it must be $\pmvA \neq \pmvB$;
      
    \item in $(\vec u)(\sys s \gamma \cocoPar \sys t \gammai \cocoPar \cdots)$,
      it must be $s \neq t$;
      
    \item we denote with $\fuse$ and $\efftauqm$ two special prefixes
      which cannot occur in processes,
      and with $\pmv K$ a special participant name which cannot occur in systems.
      
    \end{enumerate}
  \end{defi}

  Processes specify the behaviour of participants;
  they can be prefix-guarded finite sums 
  $\textstyle \cocoSum[i]{\pref[i].\procP[i]}$,
  parallel compositions $\procP \cocoPar \procQ$,
  delimited processes $(u) \procP$,
  recursive calls $\procX(\vec u)$, or %
  recursive processes $(\procRec{\procX(\vec y)} \procP)(\vec u)$.
  Prefixes include the silent action $\tau$, 
  \emph{contract advertisement} $\tell{}{\freeze u \contrP}$, 
  and \emph{action execution} $\cocodo{u}{\atomA}$.
  In a prefix $\pref \in \setenum{\tell{}{\freeze u \contrP}, \cocodo{u}{\atomA}}$, 
  the identifier $u$ refers to the target session
  involved in the execution of $\pref$.
  The special prefixes $\fuse$ and $\efftauqm$ are technical, 
  and are only used in labels of system transitions (see~\Cref{fig:co2:semantics}).
  Intuitively, $\efftauqm$ labels $\cocodo{x}{}$ actions under a delimitation $(x)$,
  while $\fuse$ labels session creations.
  Systems are parallel compositions of 
  \emph{agents} $\sys{\pmvA}{\procP}$,
  \emph{sessions} $\sys s \gamma$,
  \emph{latent contracts} $\latent u \contrP$,
  and delimited systems $(u) \sysS$. %
  A \emph{latent contract} $\latent[\pmvA]{x}{\contrP}$ 
  represents a contract $\contrP$ signed by participant $\pmvA$
  but not stipulated yet; %
  the variable $x$ will be instantiated to a fresh session name
  upon stipulation. %

  Delimitation $(u)$ binds channels, both in processes and systems.
  In the recursive process $(\procRec{\procX(\vec y)} \procP)(\vec u)$,
  the variables in $\vec y$ bind their occurrences in~$\procP$.
  Variables and names which are not bound by such binders are
  {\em free}; their sets are denoted by $\fv\_$ and $\fn\_$, respectively.
  For a system/process $Z$,
  we denote with $\fnv{Z}$ its free channels,
  \ie $\fnv{Z} = \fn{Z} \cup \fv{Z}$,
  and we say that $Z$
  is \emph{closed} when $\fnv{Z} = \emptyset$. %

  \begin{nota}
    We write $\pnil$ for $\cocoSum[\emptyset]{\procP}$, and 
    $\pref[1].\procQ[1] \cocoPlus \procP$
    for $\cocoSum[i \in I \cup \setenum{1}]{\pref[i].\procQ[i]}$ provided that 
    $\procP = \cocoSum[i \in I]{\pref[i].\procQ[i]}$ and $1 \not\in I$. 
    As usual, we omit trailing occurrences of $\pnil$.
    When the sequence of arguments $\vec{u}$ is empty, 
    we will write $\procX()$ instead of $\procX(\vec{u})$.
  \end{nota}



\subsection{Semantics} \label{sec:co2-semantics}

\begin{figure}[t]
{\[
  \begin{array}{c}
    ({\vec u}) \sys {\pmv A} {(\vec v)\procP}
    \equiv {({\vec u},{\vec v}) \sys{\pmv A} \procP}
    \hspace{25pt}
    Z \cocoPar \emptysys \equiv Z
    \hspace{15pt}
    Z \cocoPar Z' \equiv Z' \cocoPar Z
    \hspace{15pt}
    (Z \cocoPar Z') \cocoPar Z'' \equiv Z \cocoPar (Z' \cocoPar Z'')
    \\[5pt]
    \text{$\alpha$-conversion of delimited channels}
    \hspace{15pt}
    Z \cocoPar (u)Z' \equiv (u)(Z \cocoPar Z') 
    \;\;\text{if}\ u \not\in \fnv Z
    \\[5pt]
    (u)(v)Z \equiv (v)(u)Z
    \hspace{25pt}
    (u)Z \equiv Z
    \;\;\text{if}\ u \not\in \fnv Z
    \hspace{25pt}    
    \latent{s}{\contrP} \equiv \pnil
  \end{array}
  \]}
\caption{Structural congruence for \coco ($Z$ ranges over processes/systems).} \label{fig:co2:equiv}
\end{figure}

We formalise the semantics of \coco as a reduction relation on
systems (\Cref{def:co2:semantics}). %
This uses a structural congruence $\equiv$, 
which is the smallest congruence relation satisfying the
equations in~\Cref{fig:co2:equiv}. %
The axioms in~\Cref{fig:co2:equiv} are mostly standard;
note that
$({\vec u}) \sys {\pmvA} {(\vec v)\procP} \equiv {({\vec u},{\vec v}) \sys{\pmv A} \procP}$ allows
to move delimitations between systems and processes; %
we use the last axiom $\latent{s}{\contrP} = \pnil$ 
to collect garbage terms possibly arising from variable substitutions. %
In order to define honesty in~\Cref{sec:honesty},
we decorate transitions with labels,
by writing $\sysmove A \pref {}$ for a reduction where participant 
$\pmv{A}$ fires prefix $\pref$.

\begin{defi}[\coco semantics] \label{def:co2:semantics}
  The relation $\sysmove{A}{\pref}{}$ between systems
  (considered up-to structural congruence $\equiv$,
  and parametric w.r.t.\ synchronous/asynchronous semantics of session types)
  is the smallest relation
  closed under the rules of~\Cref{fig:co2:semantics}.
\end{defi}

\begin{figure}[t]
\small
\[
  \begin{array}{c}
    {\sys {\pmv A} {\tau \cocoSeq \procP \cocoPlus \procPi \cocoPar \procQ}
      \; {\sysmove{A}{\tau}{}}_{\syncmv} \;
      \sys {\pmv A} {\procP \cocoPar \procQ}
    } \quad \smallnrule{[Tau]}
    \\[14pt]
    \begin{array}{l}
    \sys {\pmv A} {\tell {} {\freeze u \contrP} \cocoSeq \procP \cocoPlus \procPi \cocoPar \procQ}
      \; {\sysmove{A}{\tell {} {\freeze u \contrP}}{}}_{\syncmv} \;
      \sys {\pmv A} {\procP \cocoPar \procQ} \ \cocoPar\ 
      \latent u \contrP
    \end{array}
    \quad \smallnrule{[Tell]}
    \\[10pt]
    \inference[\smallnrule{[Do]}]
    {\gamma \cmove{{\pmv A} \says {\atomA}}_{\syncmv} \gammai}
    {\sys {\pmv A} {\cocodo s {\atomA} \cocoSeq \procP \cocoPlus \procPi \cocoPar \procQ}
      \cocoPar  
      \sys s {\gamma} 
      \; {\sysmove{A}{\cocodo s {\atomA}}{}}_{\syncmv} \; 
      \sys {\pmv A} {\procP \cocoPar \procQ}
      \cocoPar 
      \sys s {\gammai} 
    } 
    \\[15pt]
    \inference[\smallnrule{[Rec]}]
    {\sys {\pmv A} {\procP\subs{\procRec{\procX(\vec y)} \procP}{\procX}\subs{\vec u}{\vec y}
       \cocoPar \procQ} \cocoPar \sysS 
     \; {\sysmove{A}{\pref}{}}_{\syncmv} \; \sysSi}
    {\sys {\pmv A} {(\procRec{\procX(\vec y)} \procP)(\vec u) \cocoPar \procQ} \cocoPar \sysS 
      \; {\sysmove{A}{\pref}{}}_{\syncmv} \; 
      \sysSi}
    \\[12pt]
    \inference[\smallnrule{[Fuse]}]
    {\contrP \compliant_{\syncmv} \contrQ &
    \gamma = \bic{\contrP\buffer{}}{\contrQ\buffer{}} &
    \sigma = \subs s {x,y} &
    s \not\in \fnv{\sysS}
    }
    {(x,y)(\sysS\ \mid\ 
    \setenum{\freeze x \contrP}_{\pmv A}\ \mid\ 
    \setenum{\freeze y \contrQ}_{\pmv B})
    \; {\sysmove{K}{\fuse{}}{}}_{\syncmv} \;
    (s)(\sysS\sigma \ \mid\ \sys s {\gamma})
    }
    \\[10pt]
    \inference[\smallnrule{[Par]}]
    {\sysS \; {\sysmove{A}{\pref}{}}_{\syncmv} \; \sysSi}
    {\sysS \mid \sysSii \; {\sysmove{A}{\pref}{}}_{\syncmv} \; \sysSi \mid \sysSii}
    \\[10pt]
    \inference[\smallnrule{[Del]}]
    {\sysS \; {\sysmove{A}{\pref}{}}_{\syncmv} \; \sysSi}
    {(u)\sysS \; {\sysmove{A}{\del u {\pref}}{}}_{\syncmv} (u)\sysSi}
    \hspace{20pt}
    \begin{array}{c}
      \\[-37pt]
      \text{where } \del u {\pref} =
      \begin{cases}
        \tau & \text{if } \pref = \tell{}{\freeze u \contrP} \\
        \efftauqm & \text{if } \pref = \cocodo{u}{\atomA} \\
        \pref & \text{otherwise}
      \end{cases}
    \end{array}
  \end{array}
  \]
  \caption{Reduction semantics of \coco.} \label{fig:co2:semantics}
\end{figure}

We now comment on the rules in~\Cref{fig:co2:semantics}.
Rule \nrule{[Tau]} is standard.
Rule \nrule{[Tell]} advertises the latent contract 
$\latent x {\contrP}$.
Rule \nrule{[Fuse]} stipulates contract: 
if there are two compliant contracts, 
a fresh session $s$ is created;
the latent contracts are consumed, and
the substitution $\sigma$ is applied to the system,
to instantiate the variables $x, y$ to the session name $s$.
The participant $\pmv{K}$ performing this move models 
a contract broker, similar to the one in~\cite{CO2}.
Rule \nrule{[Do]} allows a participant $\pmvA$ to 
perform some action $\atomA$ in a session $s$,
whose state $\gamma$ evolves accordingly to $\gamma'$.
Rule \nrule{[Del]} allows a system to evolve under a delimitation.
Note that the label $\pref$ fired in the premise
becomes $\tau$ or $\efftauqm$ in the consequence, 
when $\pref$ contains the delimited channel.
This transformation is defined by the function $\del{u}{\pref}$:
for instance, 
$
  (x) \, \sys {\pmv A} {\tell {} {\freeze{x} \contrP}. \procP} 
  \sysmove{A}{\tau}{} 
  (x) \, (\sys {\pmv A} \procP \,\mid\, \setenum{\freeze{x\!}{\contrP}}_{\pmv A})
$. 
Here, it would make little sense to have 
the label $\lbl {A} {\tell {} {\freeze{x\!}{\contrP}}}$, 
as $x$ (being delimited) may be $\alpha$-converted.
Rule \nrule{[Par]} is standard.
Rule \nrule{[Rec]} is used to unfold recursions.



\subsection{Examples} \label{sec:co2-examples}

\begin{exa}[Honest/dishonest choice] \label{ex:co2:honest-choice}
Consider the following processes:
\begin{align*}
  \procP & \; = \;
  (x) \;
  \tell{} {\freeze x {(\atomOut{a} \sumInt \atomOut{b})}}
  \cocoSeq 
  \cocodo x {\atomOut{a}}
  \\
  \procQ & \; = \;
  (y) \;
  \tell{} {\freeze y {(\atomIn{a} \sumExt \atomIn{b})}}
  \cocoSeq 
  \cocodo y {\atomIn{a}}
\end{align*}
A possible computation of the system
$\sysS[0] = \sys {\pmv A} {\procP} \mid \sys {\pmv B} {\procQ}$,
both under the synchronous and the asynchronous semantics,
is the following:
\begin{align}
    \label{eq:co2:honest-choice1}
    \sysS[0] 
    & {\sysmove B {\tau}{}}_{\syncmv}
    && \sys{\pmv A} {\procP} 
    \ \cocoPar\ 
    (y) \, \left(\sys{\pmv B} {\cocodo y {\atomIn a}}
    \ \cocoPar\ 
    \setenum{\freeze y {\atomIn{a} \sumExt \atomIn{b}}}_{\pmv B}
    \right)
    \\
    \label{eq:co2:honest-choice2}
    & {\sysmove A {\tau}{}}_{\syncmv} &&
    (x,y)\, \left( \sys{\pmv A} {\cocodo x {\atomOut a}} 
    \ \cocoPar \ 
    \sys{\pmv B} {\cocodo y {\atomOut a}}
    \ \cocoPar \
    \setenum{\freeze x {\atomOut{a} \sumInt \atomOut{b}}}_{\pmv A}
    \ \cocoPar \ 
    \setenum{\freeze y {\atomIn{a} \sumExt \atomIn{b}}}_{\pmv B}
    \right)
    \\
    \label{eq:co2:honest-choice3}
    & {\sysmove K \fuse {}}_{\syncmv} &&
    (s) \, \left( \sys{\pmv A} {\cocodo s {\atomOut a}} \ \cocoPar\ 
    \sys{\pmv B} {\cocodo s {\atomIn a}} \ \cocoPar\ 
    \sys s {\bic{\atomOut{a} \sumInt \atomOut{b} \buffer{}}{\atomIn{a} \sumExt \atomIn{b} \buffer{}}}
    \right) 
    \\
    \label{eq:co2:honest-choice4}
    & {\sysmove{A}{\efftauqm}{}}_{\syncmv} &&
    (s) \, \left( \sys{\pmv A}{\pnil} 
    \ \cocoPar \ 
    \sys{\pmv B} {\cocodo s {\atomIn a}} 
    \ \cocoPar \ 
    \sys s {\bic{\cnil \buffer{}}{\atomIn{a} \sumExt \atomIn{b} \buffer{ \atomOut a}}}
    \right)
    \\
    \label{eq:co2:honest-choice5}
    & {\sysmove{B}{\efftauqm}{}}_{\syncmv} &&
    (s) \, \left( \sys{\pmv A}{\pnil} 
    \ \cocoPar \ 
    \sys{\pmv B} {\pnil} 
    \ \cocoPar \ 
    \sys s {\bic{\cnil \buffer{}}{\cnil \buffer{}}}
    \right)
\end{align}
  Transitions~\eqref{eq:co2:honest-choice1} and~\eqref{eq:co2:honest-choice2}
  are obtained by applying rules \nrule{[Tell]},
  \nrule{[Par]}, and \nrule{[Del]},
  and by using structural congruence to move delimitations.
  Transition~\eqref{eq:co2:honest-choice3}
  is obtained by rule~\nrule{[Fuse]},
  since $\atomOut{a} \sumInt \atomOut{b}$ and $\atomIn{a} \sumExt \atomIn{b}$ are compliant
  (both under $\scompliant$ and $\acompliant$).
  Finally, transitions~\eqref{eq:co2:honest-choice4} and~\eqref{eq:co2:honest-choice5}
  are obtained by rule \nrule{[Do]}.

  We anticipate that, under both semantics of \coco, 
  $\procP$ is honest while $\procQ$ is dishonest.
  Intuitively, $\procP$ is honest because,
  in all possible contexts,
  it always fulfils its contract $\atomOut{a} \sumInt \atomOut{b}$
  by performing $\cocodo x {\atomOut{a}}$.
  The process $\procQ$ is not honest
  because \eg the system
  $\sys {\pmv A} {(x) \;
  \tell{} {\freeze x {\atomOut{b}}}
  \cocoSeq 
  \cocodo x {\atomOut{b}}
  } \mid \sys {\pmv B} {\procQ}$
  admits a computation leading to:
  \[
  (s) \, \left( \sys{\pmv A}{\pnil} 
    \ \cocoPar \ 
    \sys{\pmv B} {\cocodo s {\atomIn a}} 
    \ \cocoPar \ 
    \sys s {\bic{\cnil \buffer{}}{\atomIn{a} \sumExt \atomIn{b} \buffer{\atomOut b}}}
  \right)
  \]
  where $\pmvB$ is not fulfilling its contract at session $s$.
  Note in fact that $\pmvB$ declares in his contract to be able to read $\atom{a}$ or $\atom{b}$
  (chosen \emph{externally}, by the other endpoint of the session),
  while the only action in the process of $\pmvB$ is to read $\atom{a}$.
  We will formally establish the honesty of $\procP$ 
  in~\Cref{ex:type-proc:honest-choice},
  and the dishonesty of $\procQ$ in~\Cref{ex:honesty:honest-choice}.
  \qed
\end{exa}

\begin{exa}[Dishonest interleaving]
  \label{ex:co2:dishonest-interleaving}
  Let:
  \[
    \procP \; = \;
    (x,y) \; 
    \tell{}{\freeze x {\atomIn a}} \, \cocoSeq \, 
    \tell{}{\freeze y {\atomOut b}} \, \cocoSeq \,
    \cocodo{x}{\atomIn a} \, \cocoSeq \,
    \cocodo{y}{\atomOut b}
  \]
  We anticipate that $\procP$ is \emph{not} honest 
  (neither under the synchronous nor the asynchronous semantics of \coco).
  Indeed, in both semantics we can reduce the system 
  \[
  \sys {\pmv A} {\procP} 
  \; \cocoPar \;
  \sys {\pmv B}{(z) \;
    \tell{}{\freeze z {\atomIn b}} \, \cocoSeq \, 
    \cocodo{z}{\atomIn b}} \ 
  \; \cocoPar \;
  \sys {\pmv C}{(w) \;
    \tell{}{\freeze w {\atomOut a}} \, \cocoSeq \,
    \pnil} 
  \]
  into the system:
  \[
    \sysS \; = \; (s,t) \, (
    \sys {\pmvA} {\cocodo t {\atomIn a} \procseq {\cocodo s {\atomOut b}}} 
    \, \cocoPar \, 
    \sys {\pmvB} {\cocodo{z}{\atomIn b}} 
    \, \cocoPar \, 
    \sys {\pmvC} {\pnil} 
    \, \cocoPar \,
    {\sys t {\bicAC{\atomIn{a}\buffer{}}{\atomOut{a}\buffer{}}}} 
    \, \cocoPar\, 
    {\sys s {\bic{\atomOut{b}\buffer{}}{\atomIn{b}\buffer{}}}} 
    )
  \]
  The system $\sysS$ cannot reduce further. %
  Indeed, {\pmv C} (dishonestly) avoids to perform the internal choice $\atomOut{a}$
  required by his contract, 
  and so {\pmv A} is stuck, waiting for $\atomIn{a}$ from {\pmv C}. %
  Intuitively, $\procP$ is dishonest
  because $\pmvA$ does not perform
  the obligation $\atomOut b$ at session~$s$. 
  This intuitive argument will be made formal in~\Cref{ex:honesty:dishonest-interleaving}.

  We anticipate that an honest variant of the process $\procP$ would be the following:
  \[
    \procQ \; = \;
    (x,y) \; 
    \tell{}{\freeze x {\atomIn a}} \, \cocoSeq \, 
    \tell{}{\freeze y {\atomOut b}} \, \cocoSeq \,
    (\cocodo{x}{\atomIn a} \, \cocoPar \,
    \cocodo{y}{\atomOut b})
  \]
  Note that in $\procQ$ the causal dependency between
  $\cocodo{x}{\atomIn a}$ and $\cocodo{y}{\atomIn b}$ is lost.
  Another honest variant of $\procP$, 
  preserving such causal dependency (but slightly changing the contract at $y$)
  will be presented in~\Cref{ex:type-proc:honest-interleaving}.
  \qed
\end{exa}

\begin{exa}[Online store]
  \label{ex:co2:online-store}
  We formalise the online store $\pmv{S}$ from~\Cref{sec:introduction}
  as the \coco process $\procP$ below,
  where we assume that $\procR[i] = \pnil$, for $i \in 1..4$.
  \begin{align*}
    \procP 
    & \; = \; (x) \;
      \tell{}{\freeze x {\contrP[\pmvB]}} \cocoSeq 
      \left(
      \cocodo x {\atomIn{buyA}} \cocoSeq 
      \cocodo x {\atomIn{pay1E}} \cocoSeq 
      \cocodo x {\atomOut{shipA}}
      \; \cocoPlus \;
      \cocodo x {\atomIn{buyB}} \cocoSeq 
      \procPi(x)
      \right)
    \\
    \procPi(x) 
    & \; = \;
      \cocodo x {\atomOut{quote1E}} \cocoSeq 
      \cocodo x {\atomIn{pay1E}} \cocoSeq 
      \cocodo x {\atomOut{shipB}}
      \; \cocoPlus \;
      (y) \, \tell{} {\freeze y {\contrP[\pmv{D}]}} \cocoSeq 
      \procPii(x,y)
    \\
    \procPii(x,y) 
    & \; = \;
      \cocodo y {\atomOut{buyB}} \; \cocoSeq \; \cocodo x {\atomOut{quote3E}} \; \cocoSeq \; (
    \\
    & \hspace{50pt}
      \cocodo x {\atomIn{pay3E}} \; \cocoSeq \; (
    \\
    & \hspace{80pt}
      \cocodo y {\atomOut{pay2E}} \; \cocoSeq \; (
    \\
    & \hspace{110pt}
      \cocodo y {\atomIn{shipB}} \; \cocoSeq \; \cocodo x {\atomOut{shipB}}
    \\
    & \hspace{110pt}
      \cocoPlus \; \procR[1]
    \\
    & \hspace{80pt}
    ) \; \cocoPlus \; \procR[2]
    \\
    & \hspace{50pt}
    \; ) \; \cocoPlus \; \procR[3]
    \\
    & \phantom{=}
    \hspace{15pt}  ) \; \cocoPlus \; \procR[4]
  \end{align*}
  As anticipated in~\Cref{sec:introduction}, 
  this process is not honest;
  we will show later on in~\Cref{ex:type-proc:online-store}
  an honest version of the online store 
  (advertising the same contracts).
  This version will be obtained by suitably instantiating
  the processes $\procR[i]$ within $\procPii$.
  \qed
\end{exa}


\section{Honesty} \label{sec:honesty}

We now define when participants are {\em honest}, %
\ie when they fulfil their contracts, in \emph{all} execution contexts. %
We start by introducing some auxiliary notions, %
which are parameterised 
over the synchronous/asynchronous semantics of session types
(\ie, over $\syncmv \in \setenum{1,\infty}$).
The definition of honesty given in this~\namecref{sec:honesty} 
extends to the asynchronous case the one in~\cite{BZ15wsfm}. 


The set of \emph{obligations} $\obbl[\syncmv] {\pmvA} s \sysS$ yields the actions 
(at a session $s$ in $\sysS$) participant $\pmvA$ must choose from 
in order to respect her contract.

\begin{defi}[Obligations] 
  \label{def:obligations}
  We define the set of actions $\obbl[\syncmv] {\pmv A} {s} {\sysS}$ as follows:
  \[
  \obbl[\syncmv] {\pmv A} {s} {\sysS} 
  \; = \;
  \setcomp {\atomA} {\exists \gamma,\sysSi \; : \;
    \sysS 
    \equiv \sys s \gamma \cocoPar \sysSi \text{ and } 
    \gamma \cmove[\syncmv]{{\pmv A} \says {\atomA}}}
  \]
\end{defi}

The set $\readydosys{u}{\pmv A}{\sysS}$
(called \emph{ready-do} set)
collects the actions $\atomA$
that the process of $\pmvA$ would perform, if enabled in session $u$.
The set $\readydoweak[\syncmv]{u}{\pmv A}{\sysS}$ 
is a weak variant of $\readydosys{u}{\pmvA}{\sysS}$
(parameterised over the $\sysmove[\syncmv]{}{}{}$ semantics),
which contains the next reachable ready actions of $\pmvA$.

\begin{defi}[Ready-do] 
  \label{def:readydo}
  \label{def:readydo-weak} 
  We define the sets of actions
  $\readydosys{u}{\pmv A}{\sysS}$ and $\readydoweak[\syncmv]{u}{\pmv A}{\sysS}$ as:
  \[
  \begin{array}{rcl}
    \readydosys{u}{\pmv A}{\sysS} 
    & = &
    \left\{
    \atomA \cocoPar\ 
    \exists \vec{v},\procP,\procPi,\procQ,\sysSi \suchthat\;
    \sysS \equiv (\vec{v})\left( 
    \sys {\pmv A} {\cocodo{u}{\atomA} \cocoSeq \procP \cocoPlus \procPi \cocoPar \procQ} 
    \cocoPar \sysSi
    \right) \;\land\; u \not \in \vec{v}
    \right\}
    \\
    \readydoweak[\syncmv]{u}{\pmv A}{\sysS}
    & = &
    \Big\{
    {\atomA}
    \; \cocoPar \;
    {\exists \sysSi : \sysS \passivearrow[\syncmv]{\pmv A}{u}{\!\!\!\!}^* \; \sysSi
    \text{ and } \atomA \in \readydosys{u}{\pmv A}{\sysSi}}
  \Big\}
  \end{array}
  \]
  where $\sysS \passivearrow[\syncmv]{\pmv A}{u} \sysSi$ iff
  $
  \;
  \exists {\pmvB}, \pref \suchthat\;\;
  \sysS \sysmove[\syncmv]{\pmvB}{\pref}{} \sysSi 
  \;\land\;
  (
  {\pmv A} \neq \pmvB
  \;\lor\;
  \forall \atomA \suchthat\;
  \pref \neq \cocodo{u}{\atomA}
  )
  $.
\end{defi}

A participant $\pmvA$ is \emph{ready} in session $s$ when
either $\pmv A$ has no obligations at $s$, 
or $\pmv A$ is weakly ready to perform \emph{some} output action in her obligations, 
or $\pmv A$ is weakly ready to perform \emph{all} the input actions in her obligations.
This reflects the fact that to respect an internal choice it is enough to perform one of its outputs,
while to respect an external choice one has to be able to perform all of its inputs.

\begin{defi}[Readiness] \label{def:ready}
  $\RdyS[\syncmv] {\pmv A}{s}$ is the set of systems $\sysS$ such that:%
  \[
  \obbl[\syncmv] {\pmv A}{s}{\sysS} = \emptyset
  \quad \lor \quad
  \obbl[\syncmv] {\pmv A}{s}{\sysS} 
  \,\cap\, 
  \AtomOut
  \,\cap\, 
  \readydoweak[\syncmv]{s}{\pmv A}{\sysS} \; \neq \; \emptyset
  \quad \lor \quad
  \emptyset \; \neq \; (\obbl[\syncmv] {\pmv A}{s}{\sysS} \cap \AtomIn) \; \subseteq \;
  \readydoweak[\syncmv]{s}{\pmv A}{\sysS}
  \]
  Then, we say that {\pmv A} is \emph{$\syncmv$-ready in $\sysS$} iff \;
  for all $s,\sysSi,\vec u$, %
  $\sysS \equiv (\vec u) \sysSi$ implies $\sysSi \in \RdyS[\syncmv]{\pmv A}{s}$.
\end{defi}

\begin{rem}
  \label{footnote:simplified-ready}%
  \Cref{def:ready} could be simplified as
  $
  \obbl[\syncmv]{\pmv A}{s}{\sysS} = \emptyset%
  \;\lor\;
  \obbl[\syncmv]{\pmv A}{s}{\sysS}
  \,\cap\,
  \readydoweak[\syncmv]{s}{\pmv A}{\sysS} \; \neq \; \emptyset
  $,
  because $(\obbl[\syncmv]{\pmv A}{s}{\sysS} \cap \AtomIn)$ contains at most one element.
  However, we prefer to use the same definition of~\cite{BZ15wsfm}
  to inherit its undecidability results.
\end{rem}

A process $\procP$ is honest when, 
for all contexts where $\sys{\pmv A}{\procP}$ may be engaged in, 
{\pmv A} is persistently ready in all the reducts of that context.

\begin{defi}[Honesty] \label{def:honest}
  We say that:
  \begin{enumerate}
    
  \item \emph{$\sysS$ is $\pmv{A}$-free} 
    iff it has no latent/stipulated
    contracts of $\pmv{A}$, nor processes of $\pmv{A}$
    
  \item $\procP$ is \emph{$\syncmv$-honest in $\sysS$} iff\;\  
    $\forall {\pmv A} : \left(
      \sysS \text{ is {\pmv A}-free} %
      \,\land\, \sys {\pmv A} \procP \cocoPar \sysS \sysmove[\syncmv]{}{}{}^* \sysSi %
    \right)
    \implies {\pmv A} \text{ is $\syncmv$-ready in } \sysSi$
    
  \item \label{it:honesty:honest} $\procP$ is \emph{$\syncmv$-honest} iff\;\ 
    $\forall \sysS : \procP \text{ is $\syncmv$-honest in } \sysS$.
    
  \end{enumerate}
\end{defi}

The {\pmv A}-freeness requirement in~\Cref{def:honest}
is used just to rule out those systems which already carry stipulated 
or latent contracts of {\pmv A} outside $\sys{\pmv A}{\procP}$, 
\eg, $\latent{x}{\atomOut{pay}}$.
In the absence of {\pmv A}-freeness, the context could trivially 
make a process dishonest.
Note that $\sys{\pmvA}{\procP}$ is vacuously honest when $\procP$
advertises no contracts.



We prove below the dishonesty of the process $\procQ$ from~\Cref{ex:co2:honest-choice}
and of the process $\procP$ from~\Cref{ex:co2:dishonest-interleaving}.
Note that processes $\procP$ from~\Cref{ex:co2:honest-choice} and
$\procQ$ from~\Cref{ex:co2:dishonest-interleaving} are honest.
However, at this point of the paper
we do not have a convenient proof technique to cope 
with the universal quantification over contexts required in~\Cref{def:honest}.
We will establish the honesty of these processes using the type system 
in~\Cref{sec:type-system}.

\begin{exa}[Dishonest choice] \label{ex:honesty:honest-choice}
  Recall from~\Cref{ex:co2:honest-choice} the process:
  \[
  \procQ \; = \;
  (y) \;
  \tell{} {\freeze y {(\atomIn{a} \sumExt \atomIn{b})}}
  \cocoSeq 
  \cocodo y {\atomIn{a}}
  \]
  We prove that $\procQ$ is \emph{not} $\infty$-honest
  (by~\Cref{th:honest-sync-async}, $\procQ$ is not even $1$-honest).
  Recall from~\Cref{ex:co2:honest-choice} 
  that the system
  $\sys {\pmv A} {
    (x) \;
    \tell{} {\freeze x {\atomOut{b}}}
    \cocoSeq 
    \cocodo x {\atomOut{b}}
  } \mid \sys {\pmv B} {\procQ}$
  may evolve (under the asynchronous semantics) to:
  \[
  \sysS \; = \; (s) \, \sysSi
  \qquad
  \text{where}
  \quad
  \sysSi \; = \;
  \sys{\pmv A}{\pnil} 
  \ \cocoPar \ 
  \sys{\pmv B} {\cocodo s {\atomIn a}} 
  \ \cocoPar \ 
  \sys s {\bic{\cnil \buffer{}}{\atomIn{a} \sumExt \atomIn{b} \buffer{\atomOut b}}}
  \]
  We have that $\sysSi \not\in \RdyS[\infty]{\pmvB}{s}$,
  because none of the three disjunctive clauses in~\Cref{def:ready} are safisfied:
  indeed, the first two clauses are trivially false,
  while the third one is false because
  $\obbl[\infty] {\pmvB} s {\sysSi} = \setenum{\atomIn b} \not\subseteq \setenum{\atomIn a} = \readydoweak[\infty]{s}{\pmvB}{\sysSi}$.
  Therefore, $\pmvA$ is \emph{not} $\infty$-ready in $\sysS$. %
  \qed
\end{exa}

\begin{exa}[Dishonest interleaving] \label{ex:honesty:dishonest-interleaving}
  Recall from~\Cref{ex:co2:dishonest-interleaving} the process
  \[
    \procP \; = \;
    (x,y) \; 
    \tell{}{\freeze x {\atomIn a}} \, . \, 
    \tell{}{\freeze y {\atomOut b}} \, . \,
    \cocodo{x}{\atomIn a} \, . \,
    \cocodo{y}{\atomOut b}
  \]
  Recall from~\Cref{ex:co2:dishonest-interleaving} 
  that $\sys {\pmv A} {\procP}$ can be put in a system which evolves 
  (under both the synchronous and asynchronous semantics) 
  to $\sysS \equiv (s) \, \sysSi$, where:
  \[
    \sysSi \; = \; (t) \, (
    \sys {\pmvA} {\cocodo t {\atomIn a} \procseq {\cocodo s {\atomOut b}}} 
    \, \cocoPar \, 
    \sys {\pmvB} {\cocodo{z}{\atomIn b}} 
    \, \cocoPar \, 
    \sys {\pmvC} {\pnil} 
    \, \cocoPar \,
    {\sys t {\bicAC{\atomIn{a}\buffer{}}{\atomOut{a}\buffer{}}}} 
    \, \cocoPar\, 
    {\sys s {\bic{\atomOut{b}\buffer{}}{\atomIn{b}\buffer{}}}} 
    )
  \]
  We have that $\sysSi \not\in \RdyS[\infty]{\pmvA}{s}$,
  because none of the clauses in~\Cref{def:ready} are safisfied:
  in particular, 
  $\obbl[\infty] {\pmvA} s {\sysSi} = \setenum{\atomOut b}$ and 
  $\readydoweak[\infty]{s}{\pmv A}{\sysSi} = \emptyset$,
  so their intersection is empty. %
  Therefore, $\pmvA$ is \emph{not} ready in $\sysSi$, 
  and so $\procP$ is \emph{not} $\infty$-honest
  (by~\Cref{th:honest-sync-async}, $\procP$ is not even $1$-honest). %
  \qed
\end{exa}


The following theorem states the undecidability of honesty
(both under the synchronous and the asynchronous semantics).

\begin{thm} 
  \label{th:honesty-undecidable}
  The problem of deciding whether $\procP$ 
  is $\syncmv$-honest is not recursive. 
\end{thm}
\begin{proof}
  For $\syncmv = 1$, we can 
  reduce the halting problem to checking dishonesty, 
  similarly to~\cite{BZ15wsfm}.
  The construction in~\cite{BZ15wsfm} 
  can be easily adapted also to the case $\syncmv = \infty$.
\end{proof}

Honesty under the synchronous semantics implies honesty under the asynchronous one (\Cref{th:honest-sync-async}).
As a consequence, any static analysis over-approximating synchronous honesty 
(like \eg, the type system in~\Cref{sec:type-system}) 
also over-approximates asynchronous honesty.

\begin{thm} 
  \label{th:honest-sync-async}
  If $\procP$ is $1$-honest, then $\procP$ is $\infty$-honest.
\end{thm}
\begin{proof}\emph{(Sketch)}.
  We prove the contrapositive:
  if $\procP$ is \mbox{$\infty$-dishonest} then it is also \mbox{$1$-dishonest}.
  Let $\sysS[\infty]$ be such that
  $\procP$ is $\infty$-dishonest in $\sysS[\infty]$.
  By~\Cref{def:honest}, there exists a trace
  $\sys {\pmv A} \procP \cocoPar \sysS[\infty] \sysmove[\infty]{}{}{}^* \sysSi[\infty]$
  such that $\pmvA$ is \emph{not} ready in $\sysSi[\infty]$.
  We can then craft a new context $\sysS[1]$ for which $\procP$ is $1$-dishonest. 
  Dishonesty is achieved by allowing $\procP$ to perform the same transitions as in the previous trace, so reaching a state where $\pmvA$ is not ready.
  To make this possible, 
  we need to ensure that the prefixes fired in the asynchronous semantics do not become stuck in the synchronous one.
  In particular, a $\cocodo{x}{\atomA}$ prefix can become stuck for two reasons:
  \begin{inlinearabiclist}
  \item when $\atomA$ is an output and the message queue is full, or 
  \item the session $x$ cannot be established because
    the contracts fused in the original trace are compliant under $\compliant_{\infty}$
    but not under $\compliant_{1}$.
  \end{inlinearabiclist}
  To address these issues, we craft $\sysS[1]$ so that:
  \begin{inlinearabiclist}
  \item it always keeps its (1-bounded) queues empty;
  \item it advertises all the \emph{syntactic duals}~\cite{Barbanera10ppdp}
    of the contracts in $\procP$, instead of the contracts used in the asynchronous trace.
  \end{inlinearabiclist}
  Consequently, the context $\sysS[1]$ allows $\pmvA$ to:
  \begin{inlinearabiclist}
  \item fire all the prefixes as in the asynchronous trace, and 
  \item reach a state $\sysSi[1]$ where 
    $\pmvA$ has the same process as in $\sysSi[\infty]$, and 
    the contracts of $\pmvA$ in all sessions are the same as in $\sysSi[\infty]$.
  \end{inlinearabiclist}
  To make $\pmvA$ not ready in $\sysSi[1]$ (under the synchronous semantics), 
  the context simply avoids to do anything (but keeping its queues empty).
  Since $\pmvA$ is not ready in the asynchronous trace, 
  it has some unfulfilled obligation in a session of $\sysSi[\infty]$;
  hence, the same obligation is unfulfilled in $\sysSi[1]$,
  because $\sysSi[1]$ allows $\pmvA$ to fire no more prefixes than $\sysSi[\infty]$.
\end{proof}

\begin{exa} \label{ex:honest-sync-async}
  We now provide an example of the context construction
  sketched in the proof of~\Cref{th:honest-sync-async}.
  Consider the process:
  \[
  \procP \; = \;
  (x) \; \tell {} {\freeze x {(\atomOut{a}.\atomOut{b}.\atomOut{c}.\atomIn{d})}}
  \cocoSeq 
  \cocodo x {\atomOut{a}} 
  \cocoSeq 
  \cocodo x {\atomOut{b}} 
  \cocoSeq 
  \cocodo x {\atomOut{c}}
  \]
  We have that $\procP$ is $\infty$-dishonest. 
  To show that, consider \eg the context:
  \[
  \procQ \; = \;
  (y) \; \tell {} {\freeze y {(\atomOut{d}.\atomIn{a}.\atomIn{b}.\atomIn{c})}}
  \cocoSeq 
  \cocodo y {\atomOut{d}} 
  \]
  Under the asynchronous semantics, 
  the system $\sys{\pmvA}{\procP} \cocoPar \sys{\pmvB}{\procQ}$
  may evolve to a system wherein $\pmvA$ is not ready. %
  In this computation,
  first a session between $\pmvA$ and $\pmvB$ is created, since their contracts
  are compliant under~$\acompliant$.
  Then, $\pmvA$ enqueues all the outputs $\atomOut{a},\atomOut{b},\atomOut{c}$. 
  Finally, $\pmvB$ enqueues $\atomOut{d}$. %
  This enables the input action $\atomIn{d}$ of $\pmvA$, 
  but since $\pmvA$ does not perform such obligation, we conclude that $\pmvA$ is not ready.

  Under the synchronous semantics, however,
  $\sys{\pmvA}{\procP} \cocoPar \sys{\pmvB}{\procQ}$ 
  does \emph{not} evolve to a system where $\pmvA$ is not ready. 
  Indeed, in that case no session is created 
  since the two contracts are no longer compliant: 
  intuitively, making two contracts interact requires queues 
  to be longer than one message. 
  However, following the proof of~\Cref{th:honest-sync-async}, we can craft a
  context where $\pmvA$ is not $1$-honest, changing the process of $\pmvB$ as follows:
  \[
  \procQi \; = \;
  (y) \; \tell {} {\freeze y {(\atomIn{a}.\atomIn{b}.\atomIn{c}.\atomOut{d})}}
  \cocoSeq 
  \cocodo y {\atomIn{a}} 
  \cocoSeq 
  \cocodo y {\atomIn{b}} 
  \cocoSeq 
  \cocodo y {\atomIn{c}} 
  \cocoSeq 
  \cocodo y {\atomOut{d}}
  \]
  Now the contract of $\pmvA$ and the one of $\pmvB$ are compliant under $\scompliant$,
  and the system $\sys{\pmvA}{\procP} \cocoPar \sys{\pmvB}{\procQi}$
  may evolve to a system wherein $\pmvA$ is not ready.
  In this computation, after the session is created
  $\pmvA$ sends all the outputs to $\pmvB$, which are immediately received by $\pmvB$
  (who always empties his queue).
  When $\pmvB$ enqueues $\atomOut{d}$, $\pmvA$ is not ready to receive.
  \qed
\end{exa}

\begin{exa}
  \label{ex:honesty:sync-async}
  The converse of~\Cref{th:honest-sync-async} does not hold,
  in general.
  \Eg, the process:
  \[
    (x,y) \;
    \tell{}{\freeze x {(\atomOut{a} \contrSeq \atomOut{a})}} \cocoSeq 
    \cocodo{x}{\atomOut{a}} \cocoSeq 
    \tell{}{\freeze y {\atomIn{b}}} \cocoSeq 
    \cocodo{x}{\atomOut{a}} \cocoSeq
    \cocodo{y}{\atomIn{b}}
  \]
  is $\infty$-honest but not \mbox{$1$-honest}.
  Indeed, under the asynchronous semantics all the outputs can always be fired,
  while this is not the case in the synchronous case. 
  In particular, under the synchronous semantics, $\cocodo{y}{\atomIn{b}}$
  is reachable only if the process at the other endpoint of session $x$
  has read the first $\atomOut{a}$, allowing the second output to be performed.
  \qed
\end{exa}


\section{Types} \label{sec:types}

In this~\namecref{sec:types} we introduce types for \coco processes.
We will exploit them later on in~\Cref{sec:type-system} to devise a type system 
for honesty.
Note that, by~\Cref{th:honest-sync-async},  
we can focus on the synchronous semantics of contracts and systems, only:
therefore, hereafter we will implicitly assume that 
the synchronicity parameter $\syncmv$ is always~$1$.
If a process is typeable, then we will guarantee its honesty
both under the synchronous and the asynchronous semantics.

Assume we want to verify the honesty of a participant $\pmvA$ in a system $\sysS$.
We start by fixing an arbitrary channel $u$, and transforming $\sysS$ so to gather all the information involving $u$, while abstracting away the rest of the system. 
The result is a {\em pointed abstract system} (\Cref{sec:abs-sys}),
\ie, a pair $(\Gamma, \ctypeP)$,
whose first component describes the possible sessions which might be established on $u$,
while the component $\ctypeP$ 
(a \emph{pointed abstract process}, see~\Cref{def:types:abs-proc})
abstracts the behavior of $\pmv A$ on channel~$u$.
In particular, the component $\Gamma$ is either 
a set of contracts (when no session has been established yet),
or otherwise an \emph{abstract contract configuration} $\abscontrP$ %
(\Cref{sec:type-cabs}).
Our type system will abstract processes and systems on \emph{all} channels simultaneously, inferring a \emph{type} $\ptypeF$, which is a function mapping each channel $u$ to a pointed abstract process $\ptypeF(u) = \ctypeP$.


\subsection{Abstract contract configurations} \label{sec:type-cabs}

Let $\pmvA$ be a participant, and let $\gamma$ be a contract configuration
(as in~\Cref{def:st:async-semantics}):
in the \emph{abstract contract configuration} $\cta{\gamma}$
we maintain only the contract of $\pmvA$,
while recording the message sent by $\pmvA$ (if any),
and abstracting the context. %
An abstract contract configuration can be either a contract $\contrP$,
a term $\ctx~{\atomIn{a}}.\contrP$ representing that $\pmvA$ has sent a message and the context has not read it yet, 
or a term $\ready{\atomIn{a}}.\contrP$ representing that $\pmvA$ has to read a message sent by the context.

\begin{defi}[Abstract contract configurations]
  \label{def:abs-contracts}
  The syntax of \emph{abstract contract configurations}
  $\abscontrP, \abscontrQ,\ldots$
  is defined as follows, where $\contrP$ is a (concrete) contract %
  from \Cref{def:st:syntax}:
  \[
  \abscontrP, \abscontrQ 
  \;\; \bnfdef \;\;
  \contrP 
  \bnfmid 
  \ctx~{\atomIn{a}}.\, \contrP 
  \bnfmid
  \ready{\atomIn{a}}.\, \contrP
  \]
  For all participants $\pmvA$ and contract configurations $\gamma$
  involving $\pmvA$, we define %
  the \emph{abstraction of $\gamma$} w.r.t.~$\pmvA$, %
  in symbols $\cta{\gamma}$, as follows (symmetric cases omitted):
  \[
  \begin{array}{l}
    \cta{\bic{\contrP \buffer{}}{\contrQ \buffer{}}} 
    \; = \; 
    \contrP
    \\[4pt]
    \cta{\bic{\contrP \buffer{}}{\contrQ \buffer{\atomOut{a}}}}
    \; = \;
    \ctx~\atomIn{a}.\, \contrP
    \\[4pt]
    \cta{\bic{\atomIn{a}.\contrP \sumExt \contrPi \buffer{\atomOut{a}}}{\contrQ \buffer{}}} 
    \; = \;
    \ready{\atomIn{a}}.\, \contrP
  \end{array}
  \]
  The LTS $\abscmove{}$ on abstract contract configurations
  is defined by the rules in~\Cref{fig:abs-contr:semantics}. 
\end{defi}

In an internal sum, $\pmvA$ chooses a branch; 
in an external sum, the choice is made by the context
(the $\ctx$ in the label indicates that the action is performed 
by the participant at the other endpoint of the session); %
in a $\ready{\atomIn{a}}.\contrP$ the atom $\atomIn{a}$ is fired. %

\def\absContractsSemantics{
  \begin{array}{ll}
    \sumI{\atomOut{a}}{\contrP} \sumInt \contrPi
    \abscmove{\atomOut{a}}
    \ctx~{\atomIn{a}}.\, \contrP
    \hspace{25pt}
    &
    \ready{\atomIn{a}}.\, \contrP \abscmove{\atomIn{a}} \contrP 
    \\[7pt]
    \sumE{\atomIn{a}}{\contrP} \sumExt \contrPi 
    \abscmovectx{\atomOut{a}} 
    \ready{\atomIn{a}}.\, \contrP
    \hspace{25pt}
    &
    \ctx~\atomIn{a}.\, \contrP \abscmovectx{\atomIn{a}} \contrP
  \end{array}
}

\begin{figure}[t]
  \[
  \absContractsSemantics
  \vspace{-5pt}
  \]
  \caption{Semantics of abstract contract configurations.}
  \label{fig:abs-contr:semantics}
\end{figure}

The following~\namecref{lem:cabs:conc-to-abs} 
states that each transition of a contract configuration $\gamma$
can be simulated by its abstraction $\cta{\gamma}$.
This result was already established as Theorem~4.5 in~\cite{BMSZ15jlamp},
so we omit its proof here.

\newcommand{\absatomA}{\atomA}

\newcommand{\itemlemabscontractconctoabsi}{
  \gamma
  \; \cmove{\pmv A \says {\absatomA}} \;
  \gammai
  \;\;\implies\;\;
  \cta{\gamma} \abscmove{\absatomA} \cta{\gammai}}

\newcommand{\itemlemabscontractconctoabsii}{
  \gamma
  \; \cmove{\pmv B \says \absatomA} \;
  \gammai
  \;\;\implies\;\;
  \cta{\gamma} \abscmovectx{\absatomA} \cta{\gammai}
  \hspace{30pt} (\pmv B \neq \pmv A)
}

\begin{lem} \label{lem:cabs:conc-to-abs}
  For all contract configurations $\gamma$, $\gammai$
  such that $\cta{\gamma}$ is defined:
  \begin{bartalign} %
    \label{item:lem:abs-contract:conc-to-abs-i} 
    & \itemlemabscontractconctoabsi
    \\
    \label{item:lem:abs-contract:conc-to-abs-ii} 
    & \itemlemabscontractconctoabsii
  \end{bartalign}
\end{lem}


\subsection{Pointed abstract systems} \label{sec:abs-sys}

Pointed abstract processes are Basic Parallel Processes 
(BPPs)~\cite{Mayr98phd}
where prefixes are of the following kinds: 
atoms ${\atomOut a}, {\atomIn b}, \ldots$, 
nonblocking silent actions $\tau$, 
possibly blocking silent actions $\efftauqm$, 
and contract advertisement actions $\effcontract{\contrP}$.

\begin{defi}[Pointed abstract processes and systems] 
  \label{def:types:abs-proc}
  \label{def:types:abstract-proc-pair}
  \label{def:types:abstract-proc-semantics}
  The syntax of \emph{pointed abstract processes} $\ctypeT$ 
  and \emph{prefixes} $\alpha$
  is defined as follows:
  \begin{align*}
    \ctypeT 
    \;\coloncolonequals\;\; 
    & \effempty 
    \bnfmid 
    \alpha \effseq \ctypeT 
    \bnfmid 
    \ctypeT + \ctypeT
    \bnfmid 
    \ctypeT \mid \ctypeT 
    \bnfmid 
    \effrec{\ctypeX}{\ctypeT} 
    \bnfmid 
    \ctypeX \\
    \alpha \;\coloncolonequals\;\; & 
    \atomOut a
    \bnfmid
    \atomIn a
    \bnfmid
    \tau 
    \bnfmid 
    \efftauqm
    \bnfmid 
    \effcontract{\contrP}
  \end{align*}
  where sums and recursions are prefix-guarded.
  We denote with $\chtypes$ the set of all pointed abstract processes.
  A \emph{pointed abstract system} is a pair $(\Gamma, \ctypeT)$,
  where $\Gamma$ is either a set of (concrete) contracts %
  (\Cref{def:st:syntax}) %
  or an abstract contract configuration %
  (\Cref{def:abs-contracts}).
  The semantics of pointed abstract processes and systems 
  is given
  in~\Cref{fig:abs-sys:semantics}.
\end{defi}

The set $\Gamma$ grows when the process $\ctypeT$ in $(\Gamma,\ctypeT)$
advertises a contract $\contrQ$ (rule \nrule{[A-Tell1]}).
After one of the contracts in $\Gamma$ has been stipulated, 
the set is reduced to a single contract $\contrP[i]$ (rule \nrule{[A-Fuse]}),
and further advertisements are neglected (rule \nrule{[A-Tell2]}).
Rule \nrule{[A-Do]} models a $\cocodo{}{\atomA}$ action performed by $\ctypeT$,
while rule \nrule{[A-Ctx]} models an action performed by the context 
(\ie, the participant at the other endpoint of the session). %
Some pointed abstract system transitions %
will be shown in \Cref{ex:type-realization}.

\begin{figure}[t]
  \[
  \begin{array}{c}
    \inference[\smallnrule{[C-Pref]}]
    {}
    {\alpha \effseq \ctypeT \effmove{\alpha} \ctypeT}
    \hspace{20pt} 
    \inference[\smallnrule{[C-SumL]}]
    {\ctypeT \effmove{\alpha} \ctypeTi}
    {\ctypeT + \ctypeTii \effmove{\alpha} \ctypeTi} 
    \hspace{20pt}
    \inference[\smallnrule{[C-ParL]}]
    {\ctypeT \effmove{\alpha} \ctypeTi}
    {\ctypeT \mid \ctypeTii \effmove{\alpha} \ctypeTi \mid \ctypeTii}
    \\[12pt]
    \inference[\smallnrule{[C-Rec]}]
    {\ctypeT\subs{\effrec{\ctypeX}{\ctypeT}}{\ctypeX} \effmove{\alpha} \ctypeTi}
    {\effrec{\ctypeX}{\ctypeT} \effmove{\alpha} \ctypeTi}
    \hspace{16pt}
    \begin{array}{l}
      \text{commutative monoidal laws for $\mid$ and $+$}
      \\[10pt]
    \end{array}
  \end{array}
  \]
  \[
  \begin{array}{c}
    \inference[\smallnrule{[A-Tell1]}]
    {\ctypeT \effmove{\effcontract{\contrQ}} \ctypeTi}
    {(\setenum{\contrP[1],\ldots,\contrP[n]},\ctypeT) \ecmove[\tau] 
    (\setenum{\contrP[1],\ldots,\contrP[n],\contrQ},\ctypeTi)}
    \hspace{20pt}
    \inference[\smallnrule{[A-Tell2]}]
    {\ctypeT \effmove{\effcontract{\contrQ}} \ctypeTi}
    {(\abscontrP,\ctypeT) \ecmove[\tau] (\abscontrP,\ctypeTi)}
    \\[10pt]
    \inference[\smallnrule{[A-Fuse]}]
    {i \in \setenum{1,\ldots,n}}
    {(\setenum{\contrP[1],\ldots,\contrP[n]},\ctypeT) \ecmove[\efftauqm] (\contrP[i],\ctypeT)} 
    \hspace{20pt}
    \inference[\smallnrule{[A-Tau]}]
    {\ctypeT \effmove{\alpha} \ctypeTi & \alpha \in
      \setenum{\tau, \efftauqm}}
    {(\Gamma,\ctypeT) \ecmove[\alpha] (\Gamma,\ctypeTi)}
    \\[10pt]
    \inference[\smallnrule{[A-Do]}]
    {\abscontrP \;\abscmove{\atomA}\; \abscontrPi & \ctypeT \effmove{\atomA} \ctypeTi}
    {(\abscontrP,\ctypeT) \ecmove[\atomA] (\abscontrPi,\ctypeTi)}
    \hspace{20pt}
    \inference[\smallnrule{[A-Ctx]}]
    {\abscontrP \abscmove{\ctx: \atomA} \abscontrPi}
    {(\abscontrP,\ctypeT) \ecmove[\efftauqm] (\abscontrPi,\ctypeT)}
  \end{array}
  \]
  \caption{Semantics of pointed abstract processes and systems.}
  \label{fig:abs-sys:semantics}
\end{figure}


\subsection{Types} \label{sec:proc-type}

We will abstract the behavior of processes and systems as
a \emph{type}~$\ptypeF$,
which describes the abstract behaviour on \emph{all} channels
as a pointed abstract process. %
Intuitively, we abstract the behaviour of $\procP$ on a \emph{free} 
channel $u$ as $\ptypeF(u)$,
a pointed abstract process in~$\chtypes$.
On all other channels (\ie, those not free in $\procP$), 
the process $\procP$ has the same behaviour:
therefore, we cumulatively abstract the behaviour of $\procP$ on these channels
as $\ptypeF(\effwildcard)$, 
where $\effwildcard$ is a ``dummy'' element not in $\names \cup \vars$.
This makes it possible to limit $\ptypeF$ to a finite domain, 
namely the free channels of $\procP$ and $\effwildcard$ 
(as we will establish in~\Cref{lem:types:dom-fnv}). %

\begin{defi}[Type]\label{def:types:f}
  A \emph{type} is a partial function $\ptypeF \colon \names \cup \vars
  \cup \setenum{\effwildcard} \rightarrow \chtypes$
  from names/variables to pointed abstract processes,
  such that $\dom{\ptypeF}$ is finite and comprises $\effwildcard$. %
\end{defi}

Recall that we consider pointed abstract processes and systems up-to structural equivalence:
consequently, also types are up-to structural equivalence, \ie
$\ptypeF = \ptypeFi$ whenever $\dom{\ptypeF} = \dom{\ptypeFi}$ and
$\ptypeF(u) \equiv \ptypeFi(u)$ for all $u \in \dom{\ptypeF}$.

We define below an operator which expands the domain of types.
This will be exploited later on in~\Cref{sec:type-proc} in our type system.
For instance, to obtain the type of $\procP \cocoPar \procQ$
we need to expand the domains of the types of $\procP$ and $\procQ$
to all the free channels of $\procP \cocoPar \procQ$.

\begin{defi}[Domain expansion]
  \label{def:types:expand}
  For all types $\ptypeF$ and for all $A \subseteq \names \cup \vars$, 
  we define the type $\expand{\ptypeF}{A}$ as: 
  $
  \expand{\ptypeF}{A} \; = \; \ptypeF \setenum{\bind{A \setminus \dom \ptypeF}{\ptypeF(\effwildcard)}}
  $.
\end{defi}


\section{Abstract honesty} \label{sec:abs-honesty}

We now introduce a notion of \emph{abstract} honesty for pointed abstract processes and systems. Unlike honesty for concrete processes, the abstract notion does not require a universal quantification over all contexts, which is key to prove its decidability 
(\Cref{th:abs-honesty:decidable}).
We will exploit abstract honesty in our type system, 
when typing delimited processes.
Intuitively, the typing of $\procP$ constructs a pointed abstract process for each name/variable in $\procP$.
The typing also checks the abstract honesty of these pointed abstract processes:
the proof of type safety exploits these checks to guarantee that typeability implies honesty.

As done for concrete processes, we build abstract honesty over readiness.
Intuitively, a pointed abstract system $(\abscontrP,\ctypeP)$ is ready
if it can weakly perform some action 
whenever $\abscontrP$ has enabled actions of $\pmvA$.
When checking these weak transitions, we only consider those representing non-blocking steps, \ie $\tau$ actions. 
By contrast, $\efftauqm$ transitions represent potentially blocking actions, and so they are not followed, since there is no guarantee that they are enabled in the concrete system.

In order to be honest, a pointed abstract process must keep itself ready 
upon transitions, including the potentially blocking ones. %
Readiness must be checked against all the contracts
that may be stipulated along the reductions of the abstract process, 
starting from the empty set of contracts.
\Cref{def:abs-honesty} below formalises the abstract honesty for types 
(\Cref{def:abs-honesty:honest-type}). 
This involves an auxiliary definition, 
\ie the abstract honesty of pointed abstract processes 
(\Cref{def:abs-honesty:honest-proc}), which in turn involves defining 
the readiness and abstract honesty
(respectively, \Cref{def:abs-honesty:honest-sys,def:abs-honesty:ready}) 
for pointed abstract systems.
Note that, unlike the corresponding condition in~\Cref{def:honest},
\Cref{def:abs-honesty:honest-proc} in~\Cref{def:abs-honesty}
does not universally quantify over all contexts.

\begin{defi}[Abstract honesty]
  \label{def:abs-honesty}
  We say that:
  \begin{enumerate}

  \item \label{def:abs-honesty:ready}
    \quad
    $(\Gamma,\ctypeT)$ is \emph{ready} iff:
    \hspace{10pt}
    $
    \Gamma \abscmove{\atomA}
    \;\; \implies \;\;
    \exists b \; \suchthat \;
    (\Gamma,\ctypeT) \; {\ecmove[\tau]}{}^* \; \ecmove[\atomB]
    $
    
  \item \label{def:abs-honesty:honest-sys}
    \quad
    $(\Gamma,\ctypeT)$ is \emph{honest} iff: \;
    $
    (\Gamma,\ctypeT) \;\ecmove^*\; (\abscontrP,\ctypeTi) 
    \;\; \implies \;\;
    (\abscontrP,\ctypeTi) \;\text{is ready}
    $
    
  \item \label{def:abs-honesty:honest-proc}
    \quad
    $\ctypeT$ is \emph{honest} iff: 
    \hspace{25pt}
    $(\contrFmt{\emptyset},\ctypeT)$ is honest

  \item \label{def:abs-honesty:honest-type}
    \quad
    $\ptypeF$ is \emph{honest} iff:
    \hspace{25pt}
    $\ptypeF(u)$ is honest, 
    for all $u \in \dom \ptypeF$.

  \end{enumerate}
\end{defi}

\begin{exa}[Honest/dishonest choice] \label{ex:abs-honesty:honest-choice}
  Let $\ctypeT = \effcontract{\atomOut{a} \sumInt \atomOut{b}} . \, \atomOut{a}$.
  The LTS of the pointed abstract system $\left(\contrFmt{\emptyset}, \ctypeT\right)$ is:
  \begin{center}
    \scalebox{0.9}{%
      \begin{tikzpicture}
        \def\absTag{{\tiny{$\sharp$}}}%
        \node               (P0) at (0,0) {$(\contrFmt{\emptyset}, \ctypeT)$};
        \node[right=of P0]  (P1) {$(\setenum{\atomOut{a} \sumInt \atomOut{b}}, \atomOut{a})$};
        \node[right=of P1] (P2) {$(\atomOut{a} \sumInt \atomOut{b}, \atomOut{a})$};
        \node[right=of P2]  (P4)  {$(\ctx~{\atomIn{a}}, \effempty)$};
        \node[right=of P4]  (P5)  {$(\cnil, \effempty)$};

        \draw[->] (P0)--(P1.west)  node [midway,above] {\scriptsize$\tau$} node[right,below] {\absTag};
        \draw[->] (P1)--(P2.west)  node [midway,above] {\scriptsize$\efftauqm$} node[right,below] {\absTag};
        \draw[->] (P2)--(P4.west)   node [midway,above] {\scriptsize$\atomOut{a}$} node[right,below] {\absTag};
        \draw[->] (P4)--(P5.west)   node [midway,above] {\scriptsize$\efftauqm$} node[right,below] {\absTag};
      \end{tikzpicture}
    } 
  \end{center}

  \noindent
  To prove that $\ctypeT$ is honest, we check for readiness all the reducts in the LTS:
  \begin{enumerate}
    
  \item $(\contrFmt{\emptyset}, \ctypeT)$:
    nothing to check (no contracts advertised yet).
    
  \item $(\setenum{\atomOut{a} \sumInt \atomOut{b}}, \atomOut{a})$ 
    nothing to check (no contracts stipulated yet).
    
  \item $(\atomOut{a} \sumInt \atomOut{b}, \atomOut{a})$
    is ready, because
    $(\atomOut{a} \sumInt \atomOut{b},\atomOut{a}) \ecmove[\atomOut{a}]$.
    
  \item $(\ctx~{\atomIn{a}}.\,\cnil, \effempty)$ and $(\cnil,\effempty)$
    are vacuously ready,
    since $\cnil$ and $\ctx~{\atomIn{a}}.\,\cnil$ cannot take %
    $\abscmove{\atomA}$-transitions (for any $\atomA$).
    
  \end{enumerate}
  
  \smallskip\noindent
  Now, let $\ctypeQ = \effcontract{\atomIn{a} \sumExt \atomIn{b}} . \, \atomIn{a}$.
  The LTS of $(\contrFmt{\emptyset}, \ctypeQ)$ is the following:
  \begin{center}
    \scalebox{0.9}{%
      \begin{tikzpicture}
        \def\absTag{{\tiny{$\sharp$}}}%
        \node               (P0) at (0,0) {$(\contrFmt{\emptyset}, \ctypeQ)$};
        \node[right=of P0]  (P1) {$(\setenum{\atomIn{a} \sumExt \atomIn{b}}, \atomIn{a})$};
        \node[right=of P1]  (P2) {$(\atomIn{a} \sumExt \atomIn{b}, \atomIn{a})$};
        \node[right=of P2]  (P4a)  {$(\ready~{\atomIn{a}}, \atomIn{a})$};
        \node[below=of P4a]  (P4b)  {$(\ready~{\atomIn{b}}, \atomIn{a})$};
        \node[right=of P4a]  (P5)  {$(\cnil, \effempty)$};

        \draw[->] (P0)--(P1.west)  node [midway,above] {\scriptsize$\tau$} node[right,below] {\absTag};
        \draw[->] (P1)--(P2.west)  node [midway,above] {\scriptsize$\efftauqm$} node[right,below] {\absTag};
        \draw[->] (P2)--(P4a.west) node [midway,above] {\scriptsize$\ctx:\atomOut{a}$} node[right,below] {\absTag};
        \draw[->] (P2)--(P4b.west) node [midway,right] {\scriptsize$\ctx:\atomOut{b}$} node[right,below] {\absTag};
        \draw[->] (P4a)--(P5.west) node [midway,above] {\scriptsize$\atomIn{a}$} node[right,below] {\absTag};
      \end{tikzpicture}
    } 
  \end{center}
  
  \noindent
  In this case we have that the reduct $(\ready~{\atomIn{b}}, \atomIn{a})$ is \emph{not} ready:
  indeed, $\ready{\atomIn{b}} \abscmove{\atomIn{b}}$,
  while $(\ready{\atomIn{b}},\atomIn{a})$ cannot take %
  $\ecmove[\atomA]$-transitions (for any $\atomA$).
  Therefore, $\ctypeQ$ is \emph{not} abstractly honest.

  We anticipate that $\ctypeT$ and $\ctypeQ$ are the pointed abstract processes inferred by our type system,
  under the delimitations of the processes $\procP$ and $\procQ$
  discussed in~\Cref{ex:co2:honest-choice}.
  Using the abstract honesty of $\ctypeT$
  we will show in~\Cref{ex:type-proc:honest-choice} that $\procP$ is typeable, 
  hence honest.
  Instead, the abstract dishonesty of $\ctypeQ$ will prevent us from typing
  $\procQ$ --- and rightly so, 
  because we know from~\Cref{ex:honesty:honest-choice}
  that $\procQ$ is dishonest.
  \qed
\end{exa}

\begin{exa} \label{ex:type-realization}
  Let $\contrP = \atomOut{a} \sumInt \atomOut{b}$,
  and let 
  $\ctypeT = \effcontract{\contrP} \mid \tau \effseq \atomOut{a}$.
  To determine whether $\ctypeT$ is honest, 
  we check for readiness
  all the reducts of the pointed abstract system $\left(\contrFmt{\emptyset}, \ctypeT\right)$: %
  \begin{center}
    \scalebox{0.9}{%
      \begin{tikzpicture}
        \def\absTag{{\tiny{$\sharp$}}}%
        \node               (P0) at (0,0) {$(\contrFmt{\emptyset}, \ctypeT)$};
        \node[right=of P0]  (P1a) {$(\contrFmt{\emptyset}, \effcontract{\contrP} \mid \atomOut{a})$};
        \node[right=of P1a] (P2a) {$(\setenum{\contrP}, \atomOut{a})$};
        \node[right=of P2a] (P3)  {$(\contrP, \atomOut{a})$};
        \node[right=of P3]  (P4)  {$(\ctx~{\atomIn{a}}, \effempty)$};
        \node[right=of P4]  (P5)  {$(\cnil, \effempty)$};

        \node[below=of P1a] (P1b) {$(\setenum{\contrP}, \tau \effseq \atomOut{a})$};
        \node[right=of P1b] (P2b) {$(\contrP, \tau \effseq \atomOut{a})$};

        \draw[->] (P0)--(P1a.west)  node [midway,above] {\scriptsize$\tau$} node[right,below] {\absTag};
        \draw[->] (P1a)--(P2a.west) node [midway,above] {\scriptsize$\tau$} node[right,below] {\absTag};
        \draw[->] (P2a)--(P3.west)  node [midway,above] {\scriptsize$\efftauqm$} node[right,below] {\absTag};
        \draw[->] (P3)--(P4.west)   node [midway,above] {\scriptsize$\atomOut{a}$} node[right,below] {\absTag};
        \draw[->] (P4)--(P5.west)   node [midway,above] {\scriptsize$\efftauqm$} node[right,below] {\absTag};
        
        \draw[->] (P0)--(P1b.north west) node [midway,above] {\scriptsize$\tau$} node[right,below] {\!\!\!\!\!\absTag};
        \draw[->] (P1b.north east)--(P2a.south) node [midway,above] {\scriptsize$\tau$} node[right,below] {\absTag};
        
        \draw[->] (P1b)--(P2b.west) node [midway,above] {\scriptsize$\efftauqm$} node[right,below] {\absTag};
        \draw[->] (P2b.north east)--(P3.south)  node [midway,above] {\scriptsize$\tau$} node[right,below] {\absTag};
      \end{tikzpicture}
    } 
  \end{center}
  
  \noindent
  We have that:
  \begin{enumerate}
    
  \item $(\contrFmt{\emptyset}, \ctypeT)$ 
    and $(\contrFmt{\emptyset}, \effcontract{\contrP} \mid \atomOut{a})$:
    nothing to check (no contracts advertised yet).
    
  \item $(\setenum{\contrP},\tau \procseq \atomOut{a})$ 
    and $(\setenum{\contrP}, \atomOut{a})$:
    nothing to check (no contracts stipulated yet).
    
  \item $(\contrP,\tau \procseq \atomOut{a})$
    is ready, because
    $(\contrP,\tau \procseq \atomOut{a}) \ecmove[\tau] \ecmove[\atomOut{a}]$.
    
  \item $(\contrP, \atomOut{a})$
    is ready, because
    $(\contrP,\atomOut{a}) \ecmove[\atomOut{a}]$.
    
  \item $(\cnil,\effempty)$ and $(\ctx~{\atomIn{a}}, \effempty)$
    are vacuously ready,
    because $\cnil$ and $\ctx~{\atomIn{a}}.\,\cnil$ cannot take %
    $\abscmove{\atomA}$-tran\-sitions (for all $\atomA$).
    
  \end{enumerate}
  Summing up, we conclude that $\ctypeT$ is honest.
  \qed
\end{exa}

Silent moves and contract advertisements
of pointed abstract processes preserve honesty,
while input/output moves may break honesty: for instance, 
in
$
\ctypeP = \atomOut{a}.\effcontract{\atomOut{b}}
\effmove{\atomOut{a}}
\effcontract{\atomOut{b}}
= \ctypePi
$
we have that $\ctypeP$ is honest 
(because $(\emptyset,\ctypeP)$ is stuck),
while $\ctypePi$ is dishonest.

\begin{reslemma}{lem:types:chan-types-moves-honesty}
  For all 
  $\alpha \in \setenum{\tau, \efftauqm, \effcontract{\contrPi}}$: \;
  $
  \ctypeT \effmove{\alpha} \ctypeTi
  \;\land\; 
  \ctypeT\text{ honest} 
  \;\implies\; \ctypeTi\text{ honest}
  $
\end{reslemma}

The following~\namecref{lem:types:multiplicity} gives a compositional
criterion to check the abstract readiness of a pointed abstract system $(\Gamma,\ctypeP)$:
indeed, it is enough to check the parallel components of $\ctypeP$ independently.

\begin{reslemma}[Abstract readiness and parallel composition]
  {lem:types:multiplicity}
  For all $\Gamma$, $\ctypeP$, and $\ctypeQ$:
  \[
  (\Gamma,\ctypeP \mid \ctypeQ) \text{ ready}
  \;\; \iff \;\;
  (\Gamma,\ctypeP) \text { ready}
  \;\lor\;
  (\Gamma,\ctypeQ) \text { ready}
  \]
\end{reslemma}

Note that, unlike readiness, honesty is \emph{not} compositional.
\Eg, the direction $\Rightarrow$ of \Cref{lem:types:multiplicity} 
would be false since
$(\atomOut{a} \contrSeq \atomOut{b},\; \atomOut{a} \mid \atomOut{b})$ is honest,
while neither $(\atomOut{a} \contrSeq \atomOut{b},\; \atomOut{a})$ 
nor $(\atomOut{a} \contrSeq \atomOut{b},\; \atomOut{b})$ are such.
The direction $\Leftarrow$ would be false since
$(\atomOut{a} \contrSeq \atomOut{b},\;
\atomOut{a} \effseq \atomOut{b})$ is honest,
while $(\atomOut{a} \contrSeq \atomOut{b},\;
\atomOut{a} \mid \atomOut{a} \effseq \atomOut{b})$
is not.

\medskip
\Cref{th:abs-honesty:decidable} below establishes
one of our main results: checking the honesty of a type $\ptypeF$ is decidable.
Since abstract honesty will be used as a side condition in our typing rules for \coco processes,  
this result is crucial to obtain decidability for both type checking and inference (\Cref{{th:types:decidability}}).
Our proof reduces abstract honesty to submarking reachability in Petri nets,
which is decidable~\cite{Hack76phd,Mayr84siam}.
To define the reduction, we first map a pointed abstract system $(\emptyset, \ctypeP)$ 
into a Petri Net which preserves its semantics. 
Roughly, all the reducts of $\ctypeP$ are parallel compositions of processes taken, possibly more than once, from a finite set of subterms of $\ctypeP$. 
Hence, we can associate a place to each such subterm, and use the tokens to count their multiplicity. 
Further, there are only finitely many states for the $\Gamma$ component, so we can associate a place to each of them, and use a single token to represent the current $\Gamma$.
The correctness of our reduction relies on~\Cref{lem:types:multiplicity},
which implies that readiness can be established by inspecting at most
one token for each place.

\begin{thm}[Decidability of abstract honesty]
  \label{th:abs-honesty:decidable}
  Abstract honesty of pointed abstract processes is decidable.
\end{thm}
\begin{proof}
  To decide whether $\ctypeP$ is (abstractly) honest,
  by~\Cref{def:abs-honesty:honest-proc} of~\Cref{def:abs-honesty}
  we need to decide whether the pointed abstract system 
  $(\emptyset, \ctypeP)$ is honest. 
  We define the sets: 
  \begin{itemize}
  \item $\setenum{\ctypeP[j]}_{j}$, 
    comprising the closed nonempty sums which 
    are subterms of some unfolding of $\ctypeP$.
    We consider each $\ctypeP[j]$ up-to $\equiv$ and unfolding of recursion.

  \item $\Gamma(\ctypeP)$, comprising elements of two kinds:
    \begin{inlinelist}
    \item sets $\setenum{\contrP[1],\ldots,\contrP[n]}$ 
      of contracts occurring in $\ctypeP$, and 
    \item the reducts of each $\contrP[j]$ occurring in $\ctypeP$,
      according to the semantics in~\Cref{fig:abs-contr:semantics}.
    \end{inlinelist}
  \end{itemize}
  Every pointed abstract system reachable from $(\emptyset, \ctypeP)$
  has the form $(\Gamma[i], \ctypeQ[i])$, where
  $\Gamma[i] \in \Gamma(\ctypeP)$, 
  and $\ctypeQ[i]$ can be uniquely written (up to $\equiv$ and unfolding) 
  as the parallel composition of some terms in $\setenum{\ctypeP[j]}_{j}$,
  possibly taken multiple times.

  We now define a Petri net $N=(P,T)$. 
  The places $P$ comprise all the elements in $\Gamma(\ctypeP)$ 
  (called \emph{$\Gamma$-places}), 
  and all the terms in $\setenum{\ctypeP[j]}_{j}$ 
  (called \emph{$\ctypeP$-places}). 
  Intuitively, we want the reachable markings of $N$ 
  to have exactly one token in a $\Gamma$-place 
  (while all the other $\Gamma$-places have none);
  instead, $\ctypeP$-places can contain any number of tokens. 
  The idea is that the single token in the $\Gamma$-places corresponds to 
  the first component in $(\Gamma[i],\ctypeQ[i])$,
  while the tokens in $\ctypeP$-places determine the component $\ctypeQ[i]$.
  More precisely, the number of tokens in place $\ctypeP[j]$ 
  is the number of terms $\ctypeP[j]$ 
  in the parallel decomposition of $\ctypeQ[i]$
  (see~\cite{Aceto09tocl}).
  The initial marking of $N$ is the one corresponding to $(\emptyset, \ctypeP)$.
  The transitions of $N$ reflect the semantics of pointed abstract systems 
  in~\Cref{fig:abs-sys:semantics}. 
  For instance, we encode rule~\nrule{[A-Do]} in the Petri net
  by moving the single token in the $\Gamma$-places 
  from the place $\abscontrP$ to $\abscontrPi$,
  and simulating the firing of action $a$ as follows.
  First, we find the parallel component $\sum_k a_k . \ctypeR[k] $ of $\ctypeP$
  in the rule with the prefix $a_k = a$ to be fired. 
  To rewrite one copy of that sum with $\ctypeR[k]$,
  we consume one token in the $\ctypeP$-place associated to the sum, 
  and we produce tokens for $\ctypeR[k]$ in all the places corresponding to its parallel decomposition.
  This construction extends the one in~\cite{Esparza97fi},
  which shows an isomorphism between the transition system of a BPP 
  and the reachability graph of its Petri net.
  Summing up, the firing sequences of $N$ correspond to the computations
  of $(\emptyset,\ctypeP)$.

  Now, note that the set $\Gamma(\ctypeP)$ is finite, 
  because its elements of the form $\setenum{\contrP[1],\ldots,\contrP[n]}$ 
  contain only contracts syntactically occurring in $\ctypeP$, 
  and its elements of the form $\abscontrPi$ are reducts of some  $\contrP[j]$, %
  so they are finite because the semantics in~\Cref{fig:abs-contr:semantics}
  is finite-state.
  Further, also the set $\setenum{\ctypeP[j]}_{j}$ is finite,
  because its elements are considered up-to.
  Therefore, $N$ is a finite Petri net.

  We reduce the problem of checking dishonesty of $(\emptyset,\ctypeP)$
  to the \emph{submarking reachability problem} in $N$, 
  which is decidable for finite Petri nets~\cite{Hack76phd,Mayr84siam}. 
  A submarking is a mapping from a \emph{subset} of the places $P'$ to $\Nat$,
  which partially specifies a marking of the whole net.
  The submarking reachability problem asks, 
  given a submarking $m': P' \rightarrow \Nat$, 
  to establish whether or not some marking $m$ is reachable such that
  $m(p) = m'(p)$ for all $p \in P'$.

  Say that a marking is \emph{ready} if it corresponds to 
  a ready pointed abstract system.
  For every multiset $X$ of places, we denote with $\chi_{X}$
  the marking which associates each place to the corresponding number of
  occurrences in $X$.
  Then, for all $\Gamma$ we define the submarking $M_{\Gamma}$ as follows:
  \begin{equation}
    \label{eq:th:abs-honesty:decidable:defM}
    M_{\Gamma}(p) = \begin{cases}
      1 & \text{if $p = \Gamma$} \\
      0 & \text{if $p$ is a $\ctypeP$-place and $\chi_{\setenum{\Gamma,p}}$ is ready} \\
      \text{undefined} & \text{otherwise}
    \end{cases}
    \tag{\ref{th:abs-honesty:decidable}a}
  \end{equation}
  We now prove that, for any reachable marking $m$:
  \begin{equation}
    \label{eq:th:abs-honesty:decidable:fact1}
    m \text{ non ready}
    \;\; \iff \;\;
    \exists \atomA, \Gamma : 
    \Gamma \abscmove{\atomA}
    \;\land\;
    M_{\Gamma} \text{ submarking of } m 
    \tag{\ref{th:abs-honesty:decidable}b}
  \end{equation}
  For the $\Rightarrow$ direction, 
  let $\Gamma[i]$ be the single $\Gamma$-place with one token. 
  Since $m$ is not ready, we must have $\Gamma[i] \abscmove{\atomA}$
  for some $\atomA$. 
  To obtain the thesis, choose $\Gamma = \Gamma[i]$, and
  assume by contradiction that $M_{\Gamma}$ is 
  \emph{not} a submarking of $m$, \ie
  $M_{\Gamma}(p) \neq m(p)$ for some $p$ in the domain of $M_{\Gamma}$. 
  We cannot have that $p = \Gamma$, since in that case 
  by~\Cref{eq:th:abs-honesty:decidable:defM} it must be $M_{\Gamma}(p) = m(p) = 1$. 
  Hence, $p$ must be a $\ctypeP$-place for which $\chi_{\setenum{\Gamma,p}}$ is ready. 
  Further, since $M_{\Gamma}(p) = 0 \neq m(p)$, 
  we have that $m(p)$ has at least as many $\ctypeP$-tokens as the ready marking $\chi_{\setenum{\Gamma,p}}$. 
  By~\Cref{lem:types:multiplicity} (adapted to Petri nets in the natural way), 
  adding more $\ctypeP$-tokens to a ready marking cannot make it non ready,
  hence $m$ is ready --- contradiction.
  
  For the $\Leftarrow$ direction of~\Cref{eq:th:abs-honesty:decidable:fact1}, 
  assume that $\Gamma \abscmove{\atomA}$ and 
  $M_{\Gamma}$ is a submarking of~$m$. 
  By~\Cref{eq:th:abs-honesty:decidable:defM}, $m(\Gamma)=1$. 
  By contradiction, assume that $m$ is ready.
  Then, by (the adaptation of)~\Cref{lem:types:multiplicity}, 
  one can remove all the $\ctypeP$-tokens from $m$ but one, 
  while preserving its readiness; 
  so, assume that such token is in place $p$ (this implies that $m(p)>0$).
  The marking obtained in this way is $\chi_{\setenum{\Gamma,p}}$, 
  and since it is ready, 
  \Cref{eq:th:abs-honesty:decidable:defM} gives $M_{\Gamma}(p)=0$.
  Since $M_{\Gamma}$ is a submarking of $m$, 
  this implies that $m(p)=0$
  --- contradiction.

  To conclude, we now exploit the decidability of 
  the submarking reachability problem on Petri nets 
  to decide abstract dishonesty.
  First, we equivalently rephrase dishonesty in terms of submarking reachability:
  \begin{align}
    \nonumber
    (\emptyset,\ctypeP) \text{ dishonest}
    & \iff 
      \exists (\abscontrPi,\ctypePi) \;\text{non ready} :
      (\emptyset,\ctypeP) \;\ecmove^*\; (\abscontrPi,\ctypePi)
    && \text{(by~\Cref{def:abs-honesty})}
    \\
    \nonumber
    & \iff
      \exists m \text{ reachable in $N$ and non ready}
    && \text{(by construction of $N$)}
    \\
    \nonumber
    & \iff
      \exists m \text{ reachable in $N$ and }
    && 
    \\
    \nonumber
    & \phantom{\iff \exists m \;} 
      \exists \atomA, \Gamma : 
      \Gamma \abscmove{\atomA}
      \;\land\;
      M_{\Gamma} \text{ submarking of } m 
    && \text{(by~\Cref{eq:th:abs-honesty:decidable:fact1})}
    \\
    \label{eq:th:abs-honesty:decidable:fact2}
    & \iff
      \exists \atomA, \Gamma : 
      \Gamma \abscmove{\atomA}
      \;\land\;
      M_{\Gamma} \text{ reachable submarking}
    &&
       \tag{\ref{th:abs-honesty:decidable}c}
  \end{align}

  To conclude, we show how to verify the last formulation 
  of dishonesty (\Cref{eq:th:abs-honesty:decidable:fact2}).
  To this purpose, note that we can effectively and finitely enumerate 
  all the possible $\Gamma$ and~$\atomA$.
  In each case we can easily check whether $\Gamma \abscmove{\atomA}$.
  Further, we can effectively construct the submarking $M_{\Gamma}$ 
  following~\Cref{eq:th:abs-honesty:decidable:defM}.
  The only non-trivial task is checking whether $\chi_{\setenum{\Gamma,p}}$ is ready.
  According to~\Cref{def:abs-honesty} (adapted to Petri nets),
  this just requires to check whether such marking has some weakly ready action,
  \ie, if it can fire some action $\atomB$ after a finite sequence of $\tau$ actions.
  This can be reduced once again to a submarking reachability problem:
  more precisely, we start by labelling the transitions of $N$ 
  as for the moves of pointed abstract systems; 
  then we remove from $N$ all the $\efftauqm$ transitions,
  while making all the non-$\tau$ transitions 
  (\ie, the ready actions $\atomB$) fill a special place.
  At this point, it suffices to check whether the special place can eventually become nonempty, which is a submarking reachability problem.
\end{proof}



\section{A type system for honesty} 
\label{sec:type-system}

We now introduce a type system for \coco.  
Type inference is decidable (\Cref{th:types:decidability}),
and type safety guarantees that typeable processes are honest
(\Cref{th:type-safety}).


\subsection{Process typing} \label{sec:type-proc}

Our type system associates types to \coco processes.
Basically, in \Cref{def:types:typing-prefixes} %
we abstract the \coco prefixes
as actions of pointed abstract processes (\Cref{def:types:abs-proc}). %
To give some intuition, assume 
we want to abstract the behaviour of a process $\procP$ over a channel $u$, 
and $\procP$ has a prefix acting on some channel $v$.  
We have two cases.
\begin{itemize}

\item If $v \neq u$, we abstract the prefix as a $\tau$ when it is statically known to be unblocking, otherwise we abstract it as $\efftauqm$.
For instance, if $\procP$ has a prefix $\tell{}{\freeze{v}{\contrP}}$, 
which requires no synchronisation with the context,
then we abstract it with a $\tau$ action (unblocking).
Instead, if $\procP$ has a prefix $\cocodo{v}{\atomA}$,
which can only be fired with a suitable configuration in session $v$,
we abstract it as $\efftauqm$ (potentially blocking).

\item If $v = u$, the abstraction is more precise, recording the effect of the prefix.
For instance, we abstract 
the prefix $\tell{}{\freeze{v}{\contrP}}$ as $\effcontract{\contrP}$,
while we abstract $\cocodo{v}{\atomA}$ as $\atomA$.

\end{itemize}

\begin{defi}[Prefix abstraction] \label{def:types:typing-prefixes}
  For all $u \in \names \cup \vars \cup \setenum{\effwildcard}$, we
  define the mapping $[\cdot]_u$ from \coco prefixes to prefixes of 
  pointed abstract processes as follows:
  \[
  \begin{array}{c}
    [\tau]_u \; = \; \tau 
    \qquad\quad%
    [\efftauqm]_u \; = \; \efftauqm
    \qquad\quad%
    [\fuse]_u \; = \; \efftauqm
    \\[5pt]
    [\tell {} {\freeze v \contrP}]_u \; = \;
    \begin{cases}
      \effcontract{\contrP}%
      &\text{if $v = u$}
      \\
      \tau%
      &\text{otherwise}
    \end{cases}
    \qquad\qquad%
    [\cocodo v {\atomA}]_u 
    \; = \; 
    \begin{cases}
      \atomA%
      &\text{if $v=u$}
      \\%
      \efftauqm%
      &\text{otherwise}
    \end{cases}
  \end{array}
  \]
\end{defi}

Our type system extends prefix abstraction to the whole process,
on \emph{all} channels.
To properly deal with delimited channels, 
we additionally check honesty of 
the pointed abstract processes associated to them.
Typing judgments for processes have the form
$\effentails \procP \colon \ptypeF$.

\begin{defi}[Process typing]
  \label{def:types:effentails}
  Typing rules for processes are given in~\Cref{fig:type-proc}. %
  We assume an injective function $\consttovar{}$
  which associates a recursion variable $\ctypeX$
  to each constant $\procX()$.
  We say that $\ptypeF$ is \emph{inhabited}
  whenever $\effentails \procP \colon \ptypeF$, for some $\procP$. %
\end{defi}

\begin{figure}[t]
  \[
  \begin{array}{c}
    \inference[\smallnrule{[T-Nil]}]
    {}
    {\effentails \procNil \colon \lambda u \dotseq \ifte{u=\effwildcard}{\effempty}{\bot}}
    \quad
    \inference[\smallnrule{[T-Par]}]
    {\effentails \procP \colon \ptypeF & \effentails \procQ \colon \ptypeG & A = \dom \ptypeF \cup \dom \ptypeG}
    {\effentails \procP \cocoPar \procQ \colon
      \lambda u\dotseq \expand{\ptypeF}{A}\!(u) \mid \expand{\ptypeG}{A}\!(u)} 
    \\[15pt]
    \inference[\smallnrule{[T-Sum]}]
    {\forall i \in I \neq \emptyset \dotseq \effentails \procP[i] \colon \ptypeF[i] & 
      A = \bigcup_{i \in I} (\dom {\ptypeF[i]} \cup \fnv{\pref[i]})
    }
    {\effentails \cocoSum[i \in I]{\pref[i] \procseq \procP[i]} \colon
      \lambda u\dotseq\sum_{i \in I} [\pref[i]]_u \effseq \expand{\ptypeF[i]}{A}\!(u)}
    \quad
    \inference[\smallnrule{[T-Del]}]
    {\effentails \procP \colon \ptypeF & \expand{\ptypeF}{\setenum{u}}\!(u)\; \mbox{honest}}
    {\effentails (u)\procP \colon
      \ptypeF\setenum{\bind{u}{\bot}}}
    \\[15pt]
    \inference[\smallnrule{[T-Rec]}]
    { \effentails \procP \colon \ptypeF}
    { \effentails (\procRec{\procX()} \procP)() \colon \lambda u . \, \effrec{\consttovar{\procX}}{\ptypeF(u)}}
    \quad
    \inference[\smallnrule{[T-Var]}] 
    {}
    {\,\effentails \procX() \colon \lambda u \dotseq \ifte{u = \effwildcard}{\consttovar{\procX}}{\bot}} 
  \end{array}
  \]
  \caption{Typing rules for processes.}
  \label{fig:type-proc}
\end{figure}

We now comment on the rules in~\Cref{fig:type-proc}.
In all the rules we take care of making the types defined only on the channels that can be observed, and on the dummy channel~$\effwildcard$, 
which represents the other channels (see~\Cref{lem:types:dom-fnv}):
technically, in each judgement $\effentails \procP \colon \ptypeF$ we ensure that
$\dom{\ptypeF} = \fnv{\procP} \cup \setenum{\effwildcard}$. 
To this purpose we often suitably extend, in the conclusions of the rules, 
the domain of the types mentioned in the premises; 
this is done through the operator  
$\expand{\cdot}{\cdot}$ introduced in~\Cref{def:types:expand}.
Rule~\nrule{[T-Nil]} types the empty process with a map assigning the type $\effempty$ 
to the dummy channel $\effwildcard$ 
(and undefined on the other channels, since $\fnv{\pnil} = \emptyset$).
The type of a parallel composition is the pointwise parallel composition
of the component types (rule~\nrule{[T-Par]}).
Rule \nrule{[T-Sum]} types non-empty summations as (abstract) summations,
by abstracting the prefixes according to~\Cref{def:types:typing-prefixes}.

Rule \mbox{\nrule{[T-Del]}} types delimited processes $(u) \procP$:
since the channel $u$ is bound in $(u)\procP$
in the conclusion we remove it from the domain of the type.
In the rule premise, the pointed abstract process $\ptypeF(u)$ 
(with the domain expanded to include $u$, if needed),
abstracts the \emph{whole} behaviour of the participant under observation 
at session $u$.
At this point, checking the (abstract) honesty of $\ptypeF(u)$
guarantees that $\procP$ respects its obligations at session $u$.
Note that omitting or delaying the honesty check of $\ptypeF(u)$ at this point
would allow a dishonest behaviour to be typeable:
for instance, we would incorrectly type the dishonest process
$(u) \tell {} {\freeze u {\atomOut{a}}} \procseq \pnil$;
if we delay the the honesty check after the delimitation we would 
be able to type the process, 
since the resulting type $\ptypeFi$ is only defined on $\effwildcard$,
and $\ptypeFi(\effwildcard) = \tau$, which is abstractly honest.
Note that verifying $\ptypeF(u)$ honest is decidable by~\Cref{th:abs-honesty:decidable}:
we exploit this fact to prove that type inference is decidable as well.

Finally, rules \nrule{[T-Rec]} and \nrule{[T-Var]} 
deal with recursive processes and process variables. 
Note that only recursive calls without parameters are typeable.

\begin{exa}[Honest choice] \label{ex:type-proc:honest-choice}
  Recall from~\Cref{ex:co2:honest-choice} the process: 
  \[
  \procP \; = \; (x) \procPi
  \qquad
  \text{where}
  \quad
  \procPi
  \; = \;
  \tell{}{\freeze x {(\atomOut{a} \sumInt \atomOut{b}})} 
  \procseq \,
  \cocodo x {\atomOut a}
  \]
  We can type $\procP$ as follows,
  where $\ptypeF[\pnil] = \lambda u \dotseq \ifte{u=\effwildcard}{\effempty}{\bot} = \setenum{\bind{\effwildcard}{\effempty}}$:
  \medskip
  \begin{center}
    \AXC{}
    \RL{\smallnrule{[T-Nil]}}
    \UIC{$\effentails \pnil \colon \ptypeF[\pnil]$}
    \RL{\smallnrule{[T-Sum]}}
    \UIC{$\effentails \cocodo x {\atomOut a} \procseq \pnil \colon \lambda u.\, [\cocodo x {\atomOut a}]_u \effseq \, \expand{\ptypeF[\pnil]}{\setenum{x,\effwildcard}}\!(u) \; = \; \ptypeFii$}
    \RL{\smallnrule{[T-Sum]}}
    \UIC{$\effentails \procPi \colon \lambda u.\, [\tell{}{\freeze x {(\atomOut{a} \sumInt \atomOut{b}})}]_u \effseq \, \ptypeFii\,(u) \; = \; \ptypeFi$}
    \AXC{$\ptypeFi(x)$ honest}
    \RL{\smallnrule{[T-Del]}}
    \BIC{$\effentails \procP \colon \ptypeFi\setenum{\bind{x}{\bot}} \; = \; \ptypeF$}
    \DisplayProof
  \end{center}
  where
  $ \,
  \ptypeFi = \setenum{
    \bind{x}{\effcontract{\atomOut{a} \sumInt \atomOut{b}} \effseq \atomOut{a} \effseq \effempty}, \;
    \bind{\effwildcard}{\tau \effseq \efftauqm \effseq \effempty}
  }
  $.
  Note that the premise of rule~\nrule{[T-Del]} holds, because
  $\effcontract{\atomOut{a} \sumInt \atomOut{b}} \effseq \atomOut{a} \effseq \effempty$
  is abstractly honest, as shown in~\Cref{ex:abs-honesty:honest-choice}.
  Since $\procP$ has no free variables,
  its type $\ptypeF$ has domain $\setenum{\effwildcard}$, and we have
  $
  \ptypeF(\effwildcard) 
  =
  \ptypeFi(\effwildcard) 
  = 
  \tau \effseq \efftauqm \effseq \effempty
  $.
  From the typeability of $\procP$, 
  type safety (\Cref{th:type-safety})
  will allow us to deduce that $\procP$ is honest.
  \qed
\end{exa}

\begin{exa}[Dishonest choice] \label{ex:type-proc:dishonest-choice}
  Recall from~\Cref{ex:co2:honest-choice} the process:
  \[
  \procQ \; = \;
  (y) \; \procQi
  \qquad
  \text{where}
  \quad
  \procQi \; = \;
  \tell{} {\freeze y {(\atomIn{a} \sumExt \atomIn{b})}}
  \cocoSeq 
  \cocodo y {\atomIn{a}}
  \]
  We show that $\procQ$ is \emph{not} typeable.
  The only possible typing derivation for $\procQ$ would have
  the following form, where
  $ \,
  \ptypeGi = \setenum{
    \bind{y}{\effcontract{\atomIn{a} \sumExt \atomIn{b}} \effseq \atomIn{a} \effseq \effempty}, \;
    \bind{\effwildcard}{\tau \effseq \efftauqm \effseq \effempty}
  }
  $:
  \medskip
  \begin{center}
    \AXC{}
    \RL{\smallnrule{[T-Nil]}}
    \UIC{$\effentails \pnil \colon \ptypeF[\pnil]$}
    \RL{\smallnrule{[T-Sum]}}
    \UIC{$\effentails \cocodo y {\atomIn a} \procseq \pnil \colon \lambda u.\, [\cocodo y {\atomIn a}]_u \effseq \, \expand{\ptypeF[\pnil]}{\setenum{y,\effwildcard}}\!(u) \; = \; \ptypeGii$}
    \RL{\smallnrule{[T-Sum]}}
    \UIC{$\effentails \procQi \colon \lambda u.\, [\tell{}{\freeze y {(\atomIn{a} \sumExt \atomIn{b}})}]_u \effseq \, \ptypeGii\,(u) \; = \; \ptypeGi$}
    \AXC{$\ptypeGi(y)$ honest}
    \RL{\smallnrule{[T-Del]}}
    \BIC{$\effentails \procQ \colon \ptypeGi\setenum{\bind{y}{\bot}} \; = \; \ptypeG$}
    \DisplayProof
  \end{center}
  The rightmost premise in rule~\nrule{[T-Del]} is false, because 
  $\ptypeGi(y) = \effcontract{\atomIn{a} \sumExt \atomIn{b}} \effseq \atomIn{a} \effseq \effempty$ 
  is \emph{not} abstractly honest,
  as shown in~\Cref{ex:abs-honesty:honest-choice}.
  Therefore, the (dishonest) process $\procQ$ is not typeable.
  \qed
\end{exa}

\begin{exa}[Dishonest interleaving] \label{ex:type-proc:dishonest-interleaving}
  Recall from~\Cref{ex:co2:dishonest-interleaving} the process:
  \[
    \procP \; = \;
    (x,y) \; 
    \tell{}{\freeze x {\atomIn a}} \, . \, 
    \tell{}{\freeze y {\atomOut b}} \, . \,
    \cocodo{x}{\atomIn a} \, . \,
    \cocodo{y}{\atomOut b}
  \]
  The only possible typing derivation for $\procP$ would require 
  the use of rule~\nrule{[T-Del]} to close the delimitation on $y$.
  The premise of such rule should verify the abstract honesty of 
  $
  \tau 
  \effseq 
  \effcontract{\atomOut{b}} 
  \effseq 
  \efftauqm
  \effseq 
  \atomOut{b} 
  $
  --- which is \emph{not} abstractly honest,
  because the required $\atomOut{b}$ action is potentially blocked
  by the prefix $\efftauqm$.
  Therefore, $\procP$ is not typeable.
  \qed
\end{exa}

The following~\namecref{ex:type-proc:honest-interleaving}
shows an honest process which interleaves its actions in two sessions,
while enjoying typeability. 
As such, it respects its obligations in both sessions.

\begin{exa}[Honest interleaving] \label{ex:type-proc:honest-interleaving}
  Consider the process:
  \[
  \procRi 
  \; = \;
  \tell{}{\freeze x {\atomIn{a}}}
  \cocoSeq \,
  \tell{}{\freeze y {(\atomOut{b} \sumInt \atomOut{c})}}
  \cocoSeq \,
  \big(
  \cocodo x {\atomIn a} \cocoSeq \cocodo y {\atomOut b}
  \cocoPlus
  \tau \cocoSeq (\cocodo y {\atomOut c} \cocoPar \cocodo x {\atomIn a})
  \big)
  \]
  The process $\procR = (x,y) \, \procRi$ can be seen as an honest variant
  of the process~$\procP$ 
  in~\Cref{ex:co2:dishonest-interleaving}.
  The key difference w.r.t.\ $\procP$ is that, 
  in the internal choice at session~$y$,
  $\procR$ adds the option $\atomOut{c}$,
  playing the role of an ``abort'' message.
  After advertising the two contracts, the implementation of $\procR$
  proceeds as follows:
  \begin{inlinelist}
  \item if $\procR$ receives $\atom{a}$ in session $x$, 
    it will send $\atom{b}$ in session $y$;
  \item otherwise, if the $\tau$ prefix is fired (modelling \eg, a timeout),
    then $\procR$ will send the abort message $\atom{c}$ in $y$, 
    while staying ready to receive $\atom{a}$ in $x$.
  \end{inlinelist}

  We can type the process $\procRi$ with the following type $\ptypeF$
  (where we omit $\ptypeF(\effwildcard)$):
  \[
  \setenum{\;
    \bind{x}{\effcontract{\atomIn a} . \tau . \big( \atomIn{a}.\efftauqm + \tau.(\efftauqm \mid \atomIn{a}) \big) },\;\;
    \bind{y}{\tau . \effcontract{\atomOut{b} \sumInt \atomOut{c}} . \big( \efftauqm . \atomOut{b} + \tau . (\atomOut{c} \mid \efftauqm) \big)}
  \; }
  \]
  Since both $\ptypeF(x)$ and $\ptypeF(y)$ are abstractly honest,
  then we can apply twice rule~\nrule{[T-Del]}, 
  obtaining that $\procR$ is typeable,
  hence honest by type safety (\Cref{th:type-safety}). 
  \qed
\end{exa}

\begin{exa}[Online store]
  \label{ex:type-proc:online-store}
  Recall from~\Cref{ex:co2:online-store} the dishonest specification $\procP$
  of the online store in~\Cref{sec:introduction}.
  By defining the processes $\procR[i]$ within $\procP$ as follows:
  \begin{align*}
    \procR[1] 
    & \; = \;
      \tau \, \cocoSeq \, (\cocodo x {\atomOut{refund}} \cocoPar \cocodo y {\atomIn{shipB}})
    \\
    \procR[2] 
    & \; = \;
    \tau \, \cocoSeq \,  (\cocodo x {\atomOut{refund}} \cocoPar \cocodo y {\atomOut{quit}})
    \\
    \procR[3] 
    & \; = \;
      \cocodo x {\atomIn{quit}} \cocoSeq \cocodo y {\atomOut{quit}}
      \; \cocoPlus \;
      \tau \cocoSeq \left(
      (\cocodo x {\atomIn{pay3E}} \cocoSeq \cocodo x {\atomOut{refund}} \cocoPlus \cocodo x {\atomIn{quit}}) 
      \cocoPar 
      \cocodo y {\atomOut{quit}}
      \right)
    \\
    \procR[4] 
    & \; = \;
      \tau \, \cocoSeq \,  \left(
      \cocodo x {\atomOut{abort}} \cocoPar \cocodo y {\atomOut{buyB}} \cocoSeq \cocodo y {\atomOut{quit}}
      \right)
  \end{align*}
  we obtain an honest (and typeable) variant of the online store.
  Intuitively, $\procP$ in~\Cref{ex:co2:online-store} is dishonest 
  in the contexts where the counterpart in one of the two sessions stops 
  to cooperate. 
  That makes $\procP$ stuck waiting for a message from that session, 
  and no longer interacting in the other session, hence becoming not ready there. 
  The processes $\procR[i]$ above deal with these situations, 
  by performing the needed
  compensations in order to make the store ready.
  For instance, $\procR[4]$ deals with the case where 
  the session $y$ with the distributor is not established (or delayed): 
  in such case, the action $\atomOut{buyB}$ at $y$ cannot be 
  fired, but still the store must carry on the interaction with the buyer at $x$.
  To this purpose, $\procR[4]$ starts with a $\tau$ prefix, 
  modelling a timeout, and then performs $\atomOut{abort}$ on $x$.
  The compensation actions at $y$ are needed in case 
  the session with the distributor is established after the timeout.
  \qed
\end{exa}

In the following~\namecref{ex:type-proc:2}
we type a process which recursively advertises contracts
and respects its obligations in all sessions.

\begin{exa} \label{ex:type-proc:2}
  Let 
  $
  \procPi
  = 
  \tell{}{\freeze x {\atomOut{a}}} \procseq \procY()
  \, \cocoPar \,
  \cocodo x {\atomOut a}
  $.
  We can type the process $\procPi$ as follows,
  where $\ptypeF[\procY] = \lambda u \dotseq \ifte{u = \effwildcard}{\consttovar{\procY}}{\bot}$,
  and $\ptypeF[\pnil]$, $\ptypeFii$
  are as in the previous~\namecref{ex:type-proc:honest-choice}:
  \medskip
  \begin{center}
    \AXC{}
    \RL{\smallnrule{[T-Var]}}
    \UIC{$\effentails \procY \colon \ptypeF[\procY]$}
    \RL{\smallnrule{[T-Sum]}}
    \UIC{$\effentails \tell{}{\freeze x {\atomOut{a}}} \procseq \procY() \colon \lambda u \dotseq [\tell{}{\freeze x {\atomOut{a}}}]_u \effseq \expand{\ptypeF[\procY]}{\setenum{x,\effwildcard}}\!(u)$}
    
    \AXC{}
    \RL{\smallnrule{[T-Nil]}}
    \UIC{$\effentails \pnil \colon \ptypeF[\pnil]$}
    \RL{\smallnrule{[T-Sum]}}
    \UIC{$\effentails \cocodo x {\atomOut a} \colon \ptypeFii$}

    \RL{\smallnrule{[T-Par]}}
    \BIC{$\effentails \procPi \colon \lambda u \dotseq \ifte{u \in \setenum{x,\effwildcard}}{([\tell{}{\freeze x {\atomOut{a}}}]_u \effseq \consttovar{\procY} \mid [\cocodo x {\atomOut a}]_u)}{\bot} \; = \; \ptypeFi$}

    \DisplayProof
  \end{center}

  \medskip\noindent
  Note that 
  $\ptypeFi(x) = \effcontract{\atomOut{a}} \effseq \consttovar{\procY} \mid \atomOut{a}$
  is abstractly honest.
  Therefore, we can type the recursive process
  $\procP = (\procRec{\procY()} (x) \procPi) ()$ as follows:

  \medskip
  \begin{center}
    \AXC{$\effentails \procPi \colon \ptypeFi$}
    \AXC{$\ptypeFi(x)$ honest}
    \RL{\smallnrule{[T-Del]}}
    \BIC{$\effentails (x) \procPi \colon \ptypeFi\setenum{\bind{x}{\bot}} \; = \; \ptypeF$}
    \RL{\smallnrule{[T-Rec]}}
    \UIC{$\effentails (\procRec{\procY()} \procP)() \colon \lambda u . \, \effrec{\consttovar{\procY}}{\ptypeF(u)}$}

    \DisplayProof
  \end{center}
  We note that the process $\procP$ is infinite-state, 
  because of the delimitation and the parallel under recursion.
  \qed
\end{exa}

\Cref{ex:type-proc:2} shows a case where the analysis technique proposed 
in this paper is more precise than the one in~\cite{BMSZ15jlamp}.
Indeed, since $\procP$ is typeable
(and type inference is decidable,~\Cref{th:types:decidability}),
then by~\Cref{th:type-safety} our analysis technique effectively
proves that $\procP$ is honest.
Instead, the model checking algorithm in~\cite{BMSZ15jlamp} would diverge
on $\procP$, because it can only handle finite-state processes. 
The technique in~\cite{BMSZ15jlamp} could be extended by exploiting standard model-checking algorithms for Petri nets (such as \cite{Esparza94decidability}), 
so to be capable of verifying the honesty of some infinite-state processes. 
However, such an extension would still fail to handle the process $\procP$ of~\Cref{ex:type-proc:2}, because the delimitation under the recursion makes $\procP$ not expressible as a Petri net.



\subsection{System typing} \label{sec:sys-type}

Observe that the type system for processes is enough to guarantee 
whether a participant is honest. %
However, in order to establish subject reduction
we have to consider system transitions 
(because the semantics of a process depends on the system wherein it is run),
and so we need to extend our type system to \coco systems.

Type judgments for systems are of two kinds,
\mbox{$\tsentails{A}\colon$} and \mbox{$\tsentails{A}\tscompat$}. %
A judgment of the form $\tsentails{A} \sysS \colon \ptypeF$ 
guarantees that a participant $\pmvA$ in $\sysS$ behaves according to $\ptypeF$. %
Instead, a judgment of the form $\tsentails{A} \sysS \tscompat \ptypeF$ 
means that $\pmvA$'s process is \emph{not} in $\sysS$, 
and $\sysS$ is guaranteed to be \emph{compatible} with a participant
$\pmvA$ which behaves as $\ptypeF$. %
Our notion of compatibility is quite liberal: 
intuitively, it just checks that 
every contract of $\pmvA$ in the context $\sysS$ 
has indeed been advertised by $\pmvA$.

\begin{defi}[System typing]
  \label{def:types:system-typing}
  The relations $\tsentails{A} \sysS \colon \ptypeF$ and
  \mbox{$\tsentails{A} \sysS \tscompat \ptypeF$} are the smallest relations
  closed under the rules in~\Cref{fig:type-sys}.
\end{defi}

\begin{figure}[t]
  \[
  \small
  \begin{array}{c}
    \inference[\smallnrule{[T-SA]}]
    {\effentails \procP \colon \ptypeF}
    {\tsentails{A} {\sys {\pmv A} {\procP}} \colon \ptypeF}
    \quad
    \inference[\smallnrule{[T-SDel2]}]
    {{\tsentails{A} \sysS \colon \ptypeF}
    & 
    {\expand{\ptypeF}{\setenum{u}}\!(u)\ \text{honest}}}
    {\tsentails{A} (u)\sysS \colon \ptypeF\setenum{\bind{u}{\bot}}}
    \quad
    \inference[\smallnrule{[T-SPar2]}]
    {\tsentails{A} \sysS \colon \ptypeF \quad 
    \tsentails{A} \sysSi \tscompat \ptypeF}
    {\tsentails{A} \sysS \cocoPar \sysSi \colon \ptypeF}
    \\[12pt]
    \inference[\smallnrule{[T-SAFree0]}]
    {}
    {\tsentails{A} \emptysys \tscompat \ptypeF} 
    \qquad
    \inference[\smallnrule{[T-SAFree1]}]
    {\pmv B \neq \pmv A  & \fv{\procP} \cap \dom{\ptypeF} = \emptyset}
    {\tsentails{A} \sys {\pmv B} {\procP} \tscompat \ptypeF}
    \qquad
    \inference[\smallnrule{[T-SAFree2]}]
    {\pmv B \neq \pmv A & \ptypeF(x) = \bot}
    {\tsentails{A} {\setenum{\freeze x {\contrP}}_{\pmv B}}
      \tscompat \ptypeF}
    \\[12pt]
    \inference[\smallnrule{[T-SAFree3]}]
    {\sys s \gamma \text{ {\pmv A}-free} & \ptypeF(s) = \bot}
    {\tsentails{A} {\sys s \gamma} \tscompat \ptypeF}
    \quad
    \inference[\smallnrule{[T-SFz1]}]
    {(\contrP, \expand{\ptypeF}{\setenum{x}}\!(x)) \ \text{honest}}
    {\tsentails{A} {\setenum{\freeze x {\contrP}}_{\pmv A}} \tscompat \ptypeF}
    \quad
    \inference[\smallnrule{[T-SFuse]}]
    {(\cta{\gamma}, \expand{\ptypeF}{\setenum{s}}\!(s))\ \mbox{honest}}
    {\tsentails{A} {\sys s {\gamma}} \tscompat \ptypeF}
    \\[12pt]
    \inference[\smallnrule{[T-SFzS]}]
    {}
    {\tsentails{A} {\setenum{\freeze s {\contrP}}_{\pmvB}}
      \tscompat \ptypeF}
    \quad
    \inference[\smallnrule{[T-SDel1]}]
    {\tsentails{A} \sysS \tscompat \ptypeF\setenum{\bind{u}{\bot}}}
    {\tsentails{A} (u)\sysS \tscompat \ptypeF}
    \quad
    \inference[\smallnrule{[T-SPar1]}]
    {\tsentails{A} \sysS \tscompat \ptypeF \quad 
      \tsentails{A} \sysSi \tscompat \ptypeF}
    {\tsentails{A} \sysS \cocoPar \sysSi \tscompat \ptypeF} 
  \end{array}
  \]
  \caption{Typing rules for systems. Symmetric rules w.r.t.\ $\cocoPar$
    for \nrule{[T-SFuse]} and \nrule{[T-SPar2]} are omitted.}
  \label{fig:type-sys}
\end{figure}

The first three rules in~\Cref{fig:type-sys} deal with the typing judgements 
\mbox{$\tsentails{A}\colon$} for systems.
Rule~\nrule{[T-SA]} extends to $\sys {\pmvA} {\procP}$ a typing of $\procP$.
Rule~\nrule{[T-SDel2]} is similar to the rule~\nrule{[T-Del]} for processes.
Rule~\nrule{[T-SPar2]} types as $\ptypeF$ the parallel composition of two systems,
one of which must contain $\pmvA$ and be typeable with $\ptypeF$ 
(under the \mbox{$\tsentails{A}\colon$} typing), 
while the rest of the system must be compatible with $\ptypeF$ 
(using the $\tsentails{A}\tscompat$ typing).

All the other rules define the compatibility judgements 
\mbox{$\tsentails{A}\tscompat$}.
For instance, rules $\nrule{[T-SAFree*]}$ tell that
{\pmv A}-free systems are compatible with all types~$\ptypeF$
undefined on the channels in the conclusion of the rules.
For instance, in rule~\nrule{[T-SAFree1]} we forbid $\pmvB$
to use the free variables of $\pmvA$ (\ie, those in $\dom{\ptypeF}$),
to avoid potentially harmful name instantiations.
Similar preconditions are required by rules~\nrule{[T-SAFree2]}
and~\nrule{[T-SAFree3]}.
Rule~\nrule{[T-SFz1]} states that a latent contract 
$\latent[\pmvA]{x}{\contrP}$ is typeable with \mbox{$\tsentails{A}\tscompat$} only when
$\ptypeF(x)$ ``realizes’’ such contract.
Rule \nrule{[T-SFused]} is similar, 
except that a contract of $\pmvA$ occurs inside a session;
also in this case, $\pmvA$ must realize her contract.
Rule~\nrule{[T-SFzS]} deals with garbage latent contracts
$\latent[\pmvA]{s}{\contrP}$,
which cannot be fused in any sessions 
(because $s$ is already a session name, so it cannot be instantiated). 
Rule \nrule{[T-SDel1]} is symmetrical to~\nrule{[T-SDel2]}: 
for system $(u) \sysS$ to be compatible with $\ptypeF$, 
the use of $u$ in $\sysS$ has to be decoupled from the 
abstraction $\ptypeF(u)$. 
This is necessary to prevent confusion between
the channel $u$ occurring bound in $(u) \sysS$ and
\emph{another} channel named $u$ occurring in the process of $\pmvA$, 
hence found in $\dom{\ptypeF}$.
Although these two channels have the same name, 
their scope is different, so they must be treated as distinct.
For this reason, we need to check $\sysS$ to be compatible to
a type obtained by ignoring in $\ptypeF$ the presence of $u$ %
(see \Cref{ex:type-sys:honest-choice} below).
The last rule~\nrule{[T-SPar1]} is straightforward.

\begin{exa}[Honest choice] \label{ex:type-sys:honest-choice}
  Recall the process 
  $
  \procPi
  = 
  \tell{}{\freeze x {(\atomOut{a} \sumInt \atomOut{b}})} 
  \procseq \,
  \cocodo x {\atomOut a}
  $
  and its type $\ptypeFi$ from~\Cref{ex:type-proc:honest-choice}.
  We can type the system
  $
  \sysS
  = 
  \sys {\pmvA} {\procPi} \cocoPar (x) \, \sys {\pmvB} {\tell {} {\freeze x {\atomIn{a}}}}
  $
  as follows:
  \medskip
  \begin{center}
    \AXC{$\effentails \procPi \colon \ptypeFi$}
    \RL{\smallnrule{[T-SA]}}
    \UIC{$\tsentails{A} \sys {\pmvA} {\procPi} \colon \ptypeFi$}

    \AXC{$\pmvB \neq \pmvA$}
    \AXC{$\setenum{x} \cap \dom{\ptypeFi \setenum{\bind{x}{\bot}}} = \emptyset$}
    \RL{\smallnrule{[T-SAFree1]}}
    \BIC{$\tsentails{A} \sys {\pmvB} {\tell {} {\freeze x {\atomIn{a}}}} \tscompat \ptypeFi \setenum{\bind{x}{\bot}}$}
    \RL{\smallnrule{[T-SDel2]}}
    \UIC{$\tsentails{A} (x) \, \sys {\pmvB} {\tell {} {\freeze x {\atomIn{a}}}} \tscompat \ptypeFi$}

    \RL{\smallnrule{[T-SPar2]}}
    \BIC{$\tsentails{A} \sysS \colon \ptypeFi$}
    \DisplayProof
  \end{center}
  Note that the typing is possible because the delimited variable $x$
  in the process of $\pmvB$ does not interfere with the free variable $x$ in $\procPi$.
  This would be consistent with $\alpha$-converting $x$.
\end{exa}


\section{Basic properties of the type system}
\label{sec:type-basic-results}

In this~\namecref{sec:type-basic-results} we present some
basic properties of our type system for \coco;
we defer to the next~\namecref{sec:type-safety}
for subject reduction, progress, and type safety.

The type system assigns to $\effwildcard$ a pointed abstract process $\ptypeF(\effwildcard)$ which may only contain $\tau$ and $\efftauqm$ actions, hence $\ptypeF(\effwildcard)$ is always honest.

\begin{reslemma}[Honesty of $\ptypeF(\effwildcard)$]{lem:types:f-wildcard-honest}
  \begin{bartalign}
    \label{lem:types:proc@typing@effwildcard}
    \effentails \procP \colon \ptypeF
    & \implies 
    \ptypeF(\effwildcard)
    \text{ only contains $\tau$ and $\efftauqm$ actions}
    \\
    \label{lem:types:f-wildcard-honest:proc}
    \effentails \procP \colon \ptypeF
    & \implies
    \ptypeF(\effwildcard) \text{ honest}
    \\
    \label{lem:types:f-wildcard-honest:sys}
    \tsentails{A} \sysS \colon \ptypeF
    & \implies  
    \ptypeF(\effwildcard) \text{ honest}
  \end{bartalign}
\end{reslemma}

The following~\namecref{lem:types:dom-fnv} relates the free channels
of processes and systems with the domain of their type.
While these sets are the same for processes,
in the case of systems we only have inclusion.
For instance, for a system 
$
\sysS 
= 
\sys {\pmvA} {\tell {} {\freeze x \contrP}} \cocoPar \sys {\pmvB} {\tell {} {\freeze y \contrQ}}
$
with type $\ptypeF$, 
we have that $x$ and $y$ are free in $\sysS$,
but $y$ does not belong to $\dom{\ptypeF}$.

\begin{lem}[Free channels of typed processes/systems] 
  \label{lem:types:dom-fnv}
  \begin{bartalign}
    \label{lem:types:dom-fnv:proc}
    \effentails \procP \colon \ptypeF
    & \;\;\implies\;\;
    \dom \ptypeF = \fnv{\procP} \cup \setenum{\effwildcard}
    \\
    \label{lem:types:dom-fnv:sys}
    \effentails \sysS \colon \ptypeF
    & \;\;\implies\;\;
    \dom{\ptypeF} \subseteq \fnv{\sysS}\, \cup \setenum{\effwildcard}
  \end{bartalign}
\end{lem}
\begin{proof}
  Straightforward induction on the typing derivations of 
  $\effentails \procP \colon \ptypeF$ and $\effentails \sysS \colon \ptypeF$.
\end{proof}

Item~\eqref{lem:types:a-process-presence-typing:colon} of
the following~\namecref{lem:types:a-process-presence-typing} states that, 
when the typing relation $\tsentails{A} \sysS \colon \ptypeF$ holds,
then the process of $\pmvA$ occurs in $\sysS$.
Conversely, item~\eqref{lem:types:a-process-presence-typing:tscompat}
states that when the typing relation $\tsentails{A} \sysS \tscompat \ptypeF$
holds, then the process of $\pmvA$ does \emph{not} occur in $\sysS$.

\begin{lem}[Participants and system typing]
  \label{lem:types:a-process-presence-typing}
  For all systems $\sysS$ and process types $\ptypeF$:
  \begin{bartalign}
    \label{lem:types:a-process-presence-typing:colon}
    \tsentails{A} \sysS \colon \ptypeF
    & \;\implies\;
    \exists {\vec v}, \sysS[0], \procP \;\dotseq\;
    \sysS \equiv (\vec v)\left({\sys {\pmv A} \procP} \cocoPar \sysS[0]\right)
    \\
    \label{lem:types:a-process-presence-typing:tscompat}
    \tsentails{A} \sysS \tscompat \ptypeF
    & \;\implies\;
    \forall {\vec v}, \sysS[0], \procP \;\dotseq\;
    \sysS \not\equiv (\vec v)\left({\sys {\pmv A} \procP} \cocoPar \sysS[0]\right)
  \end{bartalign}
\end{lem}
\begin{proof}
  Easy induction on the typing derivation
  and inspection of the typing rules.
\end{proof}

Types are preserved by structural equivalence of processes and systems.

\begin{reslemma}[Type congruence]{lem:types:equiv}
  \begin{bartalign}
    \label{lem:types:P-equiv}
    \effentails \procP \colon \ptypeF
    \;\land \;\;
    \procP \equiv \procPi
    & \;\; \implies \;\;
    \effentails \procPi \colon \ptypeF
    \\
    \label{lem:types:S-equiv:colon}
    \tsentails{A} \sysS \colon \ptypeF
    \;\; \land \;\; 
    \sysS \equiv \sysSi
    & \;\; \implies \;\;
    \tsentails{A} \sysSi \colon \ptypeF 
    \\
    \label{lem:types:S-equiv:tscompat}
    \tsentails{A} \sysS \tscompat \ptypeF
    \;\; \land \;\; 
    \sysS \equiv \sysSi
    & \;\; \implies \;\;
    \tsentails{A} \sysSi \tscompat \ptypeF
  \end{bartalign}
\end{reslemma}

The following~\namecref{lem:types:uniqueness} states that
the type of a process is unique,
up-to structural equivalence of pointed abstract processes and systems.
The same holds for \mbox{$\tsentails{A}\colon$} typing of systems.
On the contrary, the type obtained by the judgements 
\mbox{$\tsentails{A}\tscompat$} is \emph{not} unique:
for instance, we have that $\tsentails{A} \sys \pmvB {\pnil} \tscompat \ptypeF$,
for all types $\ptypeF$.

\begin{lem}[Uniqueness of typing] \label{lem:types:uniqueness}
  \begin{bartalign}
    \label{lem:types:uniqueness:proc}
    \effentails \procP \colon \ptypeF
    \;\; \land \;\;\;\,
    \effentails \procP \colon \ptypeFi
    & \;\; \implies \;\;
    \ptypeF = \ptypeFi
    \\
    \label{lem:types:uniqueness:sys}
    \tsentails{A} \sysS \colon \ptypeF
    \;\; \land \;\;
    \tsentails{A} \sysS \colon \ptypeFi
    & \;\; \implies \;\;
    \ptypeF = \ptypeFi
  \end{bartalign}
\end{lem}
\begin{proof}
  Item~\eqref{lem:types:uniqueness:proc} follows
  by easy induction on the derivation of $\effentails \procP \colon \ptypeF$,
  by noting that each process can be typed with exactly one rule,
  and all the rules are deterministic. %
  
  \smallskip\noindent
  Likewise for item~\eqref{lem:types:uniqueness:sys}, where 
  in the case of rule~\nrule{[T-Sa]} we exploit~\Cref{lem:types:uniqueness:proc}\end{proof}

The following~\namecref{th:types:decidability} is a cornerstone
of our analysis technique,
since it establishes the decidability of type inference.
This gives us a terminating algorithm to statically analyse 
the honesty of a process $\procP$.
If we succeed in inferring the type of $\procP$,
then we know that $\procP$ is honest (by type safety, \Cref{th:type-safety});
otherwise, we cannot establish whether $\procP$ is honest or not.
Hence, our analysis safely over-approximates honesty.

\begin{thm}[Decidability of type inference]
  \label{th:types:decidability}
  Type inference for processes is decidable.
\end{thm}
\begin{proof}
  All the typing rules in~\Cref{fig:type-proc}
  follow the syntactic structure of processes.
  We can infer the type of $\procP$ by structural recursion, 
  inferring the types of the sub-processes, and composing
  them according to the typing rules.
  The only non-trivial rule is \nrule{[T-Del]}, which requires
  to check abstract honesty; however, this is decidable 
  by~\Cref{th:abs-honesty:decidable}.
\end{proof}



\section{Subject reduction and type safety} \label{sec:type-safety}

In this~\namecref{sec:type-safety} we establish the main result of this paper,
\ie type safety for \coco processes
(\Cref{th:type-safety}), ensuring that typeable processes are honest.
As usual, the proof of type safety builds upon subject reduction and progress,
hence we start by proving these results.

Subject reduction states that each step of the process of $\pmvA$ within a system
is matched by a step of its type.
Formalising this requires to define a transition system over types:
roughly, a type $\ptypeF$ takes a transition on a prefix $\pref$
when all its points $\ptypeF(u)$ agree to take a transition on
the abstract prefix~$[\pref]_u$.

\begin{defi}[Type transitions]
  \label{def:types:trans} 
  We write $\ptypeF \effmove{\pref} \ptypeFi$ when
  $\dom \ptypeFi \subseteq \dom \ptypeF$, and:
  \begin{align*}
    \forall u \in \dom \ptypeFi
    & \quad : \quad
    \ptypeF(u) \effmove{[\pref]_u} \ptypeFi(u)
    \\
    \forall u \in \dom \ptypeF \setminus \dom \ptypeFi
    & \quad : \quad
    \ptypeF(u) \effmove{[\pref]_u} \ptypeFi(\effwildcard)
  \end{align*}
\end{defi}

The second clause of \Cref{def:types:trans} accounts for the transitions of a system that discharge free channels: the transition system of types must also allow to restrict the domain of types accordingly, 
as per~\nameCref{lem:types:dom-fnv}~\ref{lem:types:dom-fnv:proc}. 
When a free channel $u$ is lost by a system transition, it is also lost from the domain of its type; further, when $u$ is no longer free in the system, at the type level it is represented by the dummy $\effwildcard$.

\begin{exa}
  \label{ex:types:trans}
  Consider the following transition, where $\pref = \cocodo s {\atomOut{a}}$:
  \begin{align*}
    \sysS 
    \; = \;\; 
    & \sys \pmvA {\cocodo s {\atomOut{a}}}
      \; \cocoPar \;
      \sys \pmvB \cdots 
      \; \cocoPar \;
      \sys s {\bic{\atomOut{a} \buffer{}}{\atomIn{a} \buffer{}}}
      \hspace{10pt}
      \sysmove A {\pref} {}
    \\
    & \sys \pmvA {\pnil}
      \hspace{23pt} \; \cocoPar \;
      \sys \pmvB \cdots 
      \; \cocoPar \;
      \sys s {\bic{\cnil \buffer{}\hspace{6pt}}{\atomIn{a} \buffer{\atomOut{a}}}}
      \; = \;
      \sysSi
  \end{align*}
  We have the following typings for $\sysS$ and its reduct $\sysSi$:
  \[
  \tsentails{A} \sysS  \colon 
  \setenum{\bind{s}{\atomOut{a}},\; \bind{\effwildcard}{\efftauqm}} \; = \; \ptypeF
  \hspace{50pt}
  \tsentails{A} \sysSi \colon 
  \setenum{\bind{\effwildcard}{\effempty}} \; = \; \ptypeFi  
  \]
  We have
  $\ptypeF(\effwildcard) \effmove{\efftauqm} \ptypeFi(\effwildcard)$,
  which satisfies the first clause of~\Cref{def:types:trans}, and 
  $\ptypeF(s) \effmove{\atomOut{a}} \ptypeFi(\effwildcard)$,  
  which satisfies also the second clause.
  Therefore, $\ptypeF \effmove{\pref} \ptypeFi$.
\end{exa}

To prove subject reduction, we need to cope with the fact that
rule~\nrule{[Fuse]} substitutes session names for variables. %
These substitutions affect its typing derivation,
as shown by the following~\namecref{ex:type-safety:fsubst}.

\begin{exa}\label{ex:type-safety:fsubst}
  Consider
  $
  \sysS = 
  \sys{\pmvA}{\cocodo x {\atomOut{a}}} \cocoPar 
  \latent{x}{\atomOut{a}} \cocoPar 
  \latent[\pmvB]{y}{\atomIn{a}}
  $, 
  And
  $
  \sysSi = 
  \sys{\pmvA}{\cocodo s {\atomOut{a}}} \cocoPar 
  \sys{s}{\bic{\atomOut{a} \buffer{}}{\atomIn{a} \buffer{}}}
  $.
  Then, by rule \nrule{[Fuse]} we have the transition 
  $(x,y) \, \sysS \sysmove{K}{\fuse{}}{} (s) \, \sysSi$.
  The typings of the (open) systems $\sysS, \sysSi$ are:
  \[
  \tsentails{A} \sysS \colon \ptypeF = \setenum{x \mapsto \atomOut{a},\; \effwildcard \mapsto \efftauqm}
  \hspace{50pt}
  \tsentails{A} \sysSi \colon \ptypeFi = \setenum{s \mapsto \atomOut{a},\; \effwildcard \mapsto \efftauqm}
  \]
  Note that variable $x$ in the domain of $\ptypeF$ has been 
  ``substituted'' with $s$ in $\ptypeFi$.
  Technically, such substitutions are obtained through the operator
  $\fsubst{}{}$ formalised below.
\end{exa}

\begin{defi}[Substitutions on types] \label{def:types:substitutions}
  \label{def:types:fsubst}
  We define substitutions on types as follows:
  \[
  \fsubst{\ptypeF}{\subs{s}{\vec x}} = \begin{cases}
    \ptypeF & \text{if ${\vec x} \cap \dom \ptypeF = \emptyset$} \\
    \ptypeF\setenum{\bind{y}{\bot}}\setenum{\bind{s}{\ptypeF(y)}}
    & \text{if ${\vec x} \cap \dom \ptypeF = \setenum{y}$} \\
    \text{undefined} & \text{otherwise}
  \end{cases}
  \]
\end{defi}

Substituting a (free) variable with a fresh session name does not
affect the typeability of a system or process, but requires adjusting
the type with operator $\fsubst{}{}$. 

\begin{reslemma}[Typing and substitution]{lem:types:substitution}
  \label{lem:types:sys-typing-renaming}
  For all processes $\procP$, systems $\sysS$, types $\ptypeF$, 
  and for all substitutions $\sigma = \subs{s}{\vec{x}}$
  such that $s \not\in \fnv{\procP} \cup \fnv{\sysS}$ 
  and $\fsubst{\ptypeF}{\sigma}$ is defined:
  \begin{bartalign}
    \label{lem:types:typing-renaming-proc}
    \effentails \procP \colon \ptypeF 
    & \;\;\implies\;\;
    \effentails \procP\sigma \colon \fsubst{\ptypeF}{\sigma}
    \\
    \label{lem:types:typing-renaming-colon}
    \tsentails{A} \sysS \colon \ptypeF 
    & \;\; \implies \;\;
    \tsentails{A} \sysS\sigma \colon \fsubst{\ptypeF}{\sigma} 
    \\
    \label{lem:types:typing-renaming-tscompat}
    \tsentails{A} \sysS \tscompat \ptypeF 
    & \;\; \implies \;\;
    \tsentails{A} \sysS\sigma \tscompat \fsubst{\ptypeF}{\sigma}
  \end{bartalign}
\end{reslemma}

We can now state subject reduction: 
typeability is preserved by system transitions. %
We need to consider a few cases, depending on which participant moves 
(either $\pmvA$ under typing, or any other participant $\pmvB$), 
and on which typing relation is used ($\colon$ or $\tscompat$).
Note that,
by~\nameCref{lem:types:a-process-presence-typing}~\ref{lem:types:a-process-presence-typing:tscompat}, 
when $\pmvA$ moves the typing relation $\tscompat$ cannot hold,
so we have only three cases.
When a system $\sysS$ takes a transition due to $\pmvA$, 
the $\colon$-type of the reduct cannot be the same as the type of $\sysS$,
because the action consumed in $\sysS$ has to be consumed also in the type.
Rather, such move of $\pmv A$ can be ``simulated'' by a corresponding move 
of the type (\Cref{def:types:trans}).
System transitions caused by $\pmvB \neq \pmvA$  
preserve the type (both for \mbox{$\colon$} and \mbox{$\tscompat$}).

Note that the hypothesis that $\ptypeF$ is honest always holds for 
\emph{closed} systems, as it is implied 
by~\nameCref{lem:types:f-wildcard-honest}~\ref{lem:types:f-wildcard-honest:sys}
and~\nameCref{lem:types:dom-fnv}~\ref{lem:types:dom-fnv:sys}.

\begin{thm}[Subject reduction]
  \label{th:types:subject-reduction}
  If $\sysS \sysmove{B}{\pref}{} \sysSi$ and $\ptypeF$ is honest:
  \begin{bartalign} 
    \label{th:types:subject-reduction-A-colon}  
    \tsentails{A}\sysS \colon \ptypeF
    \;\land\; 
    \pmvB = \pmvA
    &\;\implies\; \exists
    \ptypeFi\dotseq \ptypeF \effmove{\pref} \ptypeFi \;\wedge\; 
    \tsentails{A}\sysSi \colon \ptypeFi 
    \\
    \label{th:types:subject-reduction-B-colon}
    \tsentails{A}\sysS \colon \ptypeF
    \;\land\;
    \pmvB \neq \pmvA
    &\;\implies\;
    \;\tsentails{A} \sysSi \colon \ptypeF
    \\
    \label{th:types:subject-reduction-B-tscompat}
    \tsentails{A}\sysS \tscompat \ptypeF
    \;\land\;
    \pmvB \neq \pmvA 
    &\;\implies\;
    \;\tsentails{A} \sysSi \tscompat \ptypeF
  \end{bartalign}
\end{thm}
\begin{proof}
  See page~\pageref{proof:th:types:subject-reduction}.
\end{proof}

Progress is somehow dual to subject reduction: roughly,
it guarantees that type transitions are ``simulated'' by system transitions.
However, this does not necessarily holds for $\cocodo{}{\!}$ transitions,
since they may be enabled in the type but forbidden in the system.
For $\cocodo{}{\!}$ transitions, progress only guarantees that they
are \emph{ready} in the system.
For instance, in
\[
  \tsentails{A}
  \;
  \sys{\pmvA}{\cocodo{s}{\atomOut{a} \cocoPar \cocodo{s}{\atomOut{b}}}} \cocoPar 
  \sys{s} {\bic{\atomOut{b} \buffer{}}{\cdots}}
  \;\; \colon \;
  \setenum{s \mapsto \atomOut{a} \mid \atomOut{b},\; \effwildcard \mapsto \cdots}
\]
the type can perform both $\atomOut{a}$ and $\atomOut{b}$ at session $s$,
while the system can only perform $\atomOut{b}$ 
(the action $\atomOut{a}$ is ready but not fireable).

\begin{restheorem}[Progress]{th:progress}
  If $\,\tsentails{A}\sysS \colon \ptypeF$
  with $\ptypeF$ honest, 
  $\ptypeF \effmove{\pref} \ptypeFi$, 
  and $u \in \dom{\ptypeF}$, then:
  \begin{bartalign}
    \label{th:progress:tau}
    \pref \in \setenum{\tau, \; {\tell {\pmv B} {\freeze u \contrP}}}
    & \implies
    \exists \sysSi \dotseq
    \sysS \sysmove{A}{\pref}{} \sysSi
    \;\;\land\;\;
    \tsentails{A} \sysSi \colon \ptypeFi
    \\
    \label{th:progress:do}    
    \pref = {\cocodo u {\atomA}}
    & \implies
    \atomA \in \readydosys{u}{\pmv A}{\sysS}
  \end{bartalign}
\end{restheorem}

In order to prove type safety,
we first show that if $\ptypeF$ is the type associated to some process,
and $\ptypeF(u)$ takes a transition,
then the whole $\ptypeF$ can take a transition.

\begin{reslemma}[Self-concordance]{lem:self-concordance}
  \label{def:self-concordance}
  If $\ptypeF$ is inhabited, then for all $u \in \dom \ptypeF$:
  \[
  \ptypeF(u) \effmove{\alpha} \ctypeTi 
  \;\implies\; 
  \exists \pref,\ptypeFi \dotseq %
  [\pref]_u = \alpha 
  \;\land\; 
  \ptypeF \effmove{\pref} \ptypeFi
  \;\land\; 
  \expand{\ptypeFi}{\setenum{u}}(u) = \ctypeTi
  \]
\end{reslemma}

Type safety guarantees that typeable closed processes are honest.

\begin{thm}[Type safety]
  \label{th:type-safety}
  For all closed $\procP$, if
  $\;\effentails \procP \colon \ptypeF$ then $\procP$ is honest.
\end{thm}
\begin{proof}
  By~\Cref{def:honest}, we need to prove
  that for all {\pmv A}-free $\sysS$:
  \[
  \sys \pmvA \procP \cocoPar \sysS \rightarrow^{*} \sysSi
  \qquad
  \implies
  \qquad
  \pmvA \text{ ready in } \sysSi
  \]
  Since $\procP$ is closed, then by Lemma~\ref{lem:types:dom-fnv:proc}
  it follows that $\dom{\ptypeF} = \setenum{\effwildcard}$;
  together with the fact that $\sysS$ is $\pmvA$-free, then
  by a simple structural induction on $\sysS$ we can apply the rules in~\Cref{fig:type-sys}
  to obtain the typing
  $\tsentails{A} \sysS \tscompat \ptypeF$.
  Hence, we can reconstruct the following typing derivation:
  \[
    \inference[\smallnrule{[T-SPar2]}]
    {\inference[\smallnrule{[T-SA]}]
      {\effentails \procP \colon \ptypeF}
      {\tsentails{A} \sys \pmvA \procP \colon \ptypeF} &
    \tsentails{A} \sysS \tscompat \ptypeF}
    {\tsentails{A} \sys \pmvA \procP \cocoPar \sysS \colon \ptypeF}  
  \]
  Since $\dom{\ptypeF} = \setenum{\effwildcard}$,
  then by Lemma~\ref{lem:types:f-wildcard-honest:proc}
  it follows that $\ptypeF$ is honest.
  Since honesty is preserved by transitions of process types
  (Lemma~\ref{lem:types:chan-types-moves-honesty}),
  then by iterating~\Cref{th:types:subject-reduction} (Subject reduction)
  we have that:
  \[
  \tsentails{A} \sysSi \colon \ptypeFi  
  \qquad\qquad
  \text{for some $\ptypeFi$ honest }
  \]
  By~\Cref{def:ready}, we must prove that, whenever 
  $\sysSi \equiv (\vec v) \sysSi[0]$
  for some $\vec{v}$ and $\sysSi[0]$,
  then, for all $s$,
  $\sysSi[0] \in \RdyS{\pmv A}{s}$.
  This is equivalent to:
  \[
  \obbl{\pmv A}{s}{\sysSi[0]} \neq\emptyset
  \;\implies\;
  \obbl{\pmv A}{s}{\sysSi[0]} \cap \readydoweak{s}{\pmv A}{\sysSi[0]} 
  \; \neq \; \emptyset
  \]
  The above equivalence holds because %
  $\obbl{\pmv A}{s}{\sysSi[0]} \cap \AtomIn$ contains at most one
  element\footnote{
    This holds for both synchronous and asynchronous semantics of
    session types:
    see~\Cref{footnote:simplified-ready}.}.
  So, assume $\atomB \in \obbl{\pmv A}{s}{\sysSi[0]} \neq \emptyset$.
  Then, $\sysSi[0]$ must be structurally equivalent to:
  \[
  (\vec{u}) \left(
    \sys {\pmvA} \procPi \cocoPar 
    \sys s {\gamma} \cocoPar \sysSii[0] 
  \right)
  \tag*{with $\gamma = \bic{\contrP \buffer{\qmvA}}{\contrQ \buffer{}}$}
  \]
  for some $\vec{u}$, $\pmvB$, $\contrP, \contrQ$, $\procPi$, $\sysSii[0]$ 
  and $\qmvA$ such that $s \not\in \vec{u}$, and $\qmvA$ is either empty 
  or a singleton.
  Since 
  $
  \sysSi 
  \equiv 
  (\vec{v} \vec{u})
  \left(
    \sys {\pmvA} \procPi \cocoPar \sys s {\gamma} \cocoPar \sysSii[0] 
  \right)
  $
  and $\tsentails{A} \sysSi \colon \ptypeFi$,
  by~\Cref{lem:types:equiv} we have that:
  \[
  \tsentails{A} (\vec{v} \vec{u})
  \left(
    \sys {\pmvA} \procPi \cocoPar \sys s {\gamma} \cocoPar \sysSii[0] 
  \right)
  \colon \ptypeFi
  \]
  By inverting the typing derivation, we obtain some honest $\ptypeG$ such that:
  \[
  \tsentails{A} \sys {\pmvA} \procPi \colon \ptypeG
  \hspace{50pt}
  \tsentails{A} \sys s {\gamma} \tscompat \ptypeG
  \hspace{50pt}
  \tsentails{A} \sysSii[0]  \tscompat \ptypeG
  \]
  Since $\tsentails{A} \sys s {\gamma} \tscompat \ptypeG$
  can only by typed via rule~\nrule{[T-SFuse]}, then 
  $\expand{\ptypeG}{\setenum{s}}\!(s)$ must realize 
  $\abscontrP = \cta{\gamma}$.
  Let $\ctypeT = \expand{\ptypeG}{\setenum{s}}\!(s)$.
  Since $\ctypeT$ realizes $\abscontrP$, then 
  by~\Cref{def:abs-honesty} we know that
  $(\abscontrP,\ctypeT)$ is honest, 
  \ie for all $\abscontrQ,\ctypeQ$:
  \[
  (\abscontrP,\ctypeT)
  \;\ecmove^*\; (\abscontrQ,\ctypeQ) 
  \;\; \implies \;\;
  \ctypeQ \;\text{is abstractly ready for}\ \abscontrQ
  \]
  In particular, $\ctypeT$ is abstractly ready for $\abscontrP$,
  \ie by~\Cref{def:abs-honesty}:
  \[
  \abscontrP \;\abscmove{\atomB}
  \quad \implies \quad
  \exists \atomA \; \suchthat \;
  (\abscontrP,\ctypeT) \; {\ecmove[\tau]}{}^* \; \ecmove[\atomA]
  \]
  Since $\gamma \cmove{\pmvA \says \atomB}$,
  then by~\nameCref{lem:cabs:conc-to-abs}~\ref{item:lem:abs-contract:conc-to-abs-i} 
  we have that $\abscontrP \;\abscmove{\atomB}$,
  and so from the above implication:
  \[
  (\abscontrP,\ctypeT) 
  \ecmove[\tau]
  (\abscontrP,\ctypeT[1]) 
  \ecmove[\tau]
  \cdots
  \ecmove[\tau]
  (\abscontrP,\ctypeT[n]\,) 
  \ecmove[\atomA]
  (\abscontrPi,\ctypeTi\,) 
  \]
  By the rules in~\Cref{fig:abs-sys:semantics},
  we obtain a corresponding trace of $\ctypeT$:
  \[
  \ctypeT
  \effmove{\tau}
  \ctypeT[1]
  \effmove{\tau}
  \cdots
  \effmove{\tau}
  \ctypeT[n]
  \effmove{\atomA}
  \ctypeTi
  \]
  We now exploit~\Cref{lem:self-concordance} to prove that 
  there exist $\pref[1],\ldots,\pref[n],\prefi$,
  and $\ptypeG[1],\ldots,\ptypeG[n],\ptypeGi$
  such that $[\pref[i]] = \tau$ for all $i \in 1..n$, 
  $[\prefi] = a$, and:
  \[
  \ptypeG
  \effmove{\pref[1]}
  \ptypeG[1]
  \effmove{\pref[2]}
  \cdots
  \effmove{\pref[n]}
  \ptypeG[n]
  \effmove{\prefi}
  \ptypeGi
  \]
  To justify the first step, we observe that, by~\Cref{def:types:expand},
  it must either be 
  $\ctypeT = \ptypeG(s)$, or $\ctypeT = \ptypeG(\effwildcard)$.
  Since $\ptypeG$ is inhabited 
  (by $\sysS[0] = \sys {\pmvA} \procPi \cocoPar \sys s {\gamma} \cocoPar \sysSii[0]$), 
  by applying~\Cref{lem:self-concordance}
  on $\ctypeT \effmove{\tau} \ctypeT[1]$ 
  (with $u = s$ or $u = \effwildcard$, accordingly)
  we obtain that $\ptypeG \effmove{\pref[1]} \ptypeG[1]$,
  for some $\pref[1]$ such that $[\pref[1]] = \tau$
  (and so, we have either $\pref[1] = \tau$ or $\pref[1] = \tell{}{\!}$).
  Then, we have either 
  $\ptypeG[1](s) = \ctypeP[1]$ or $\ptypeG[1](\effwildcard) = \ctypeP[1]$.
  Moreover, since $\ptypeG$ is honest, 
  then by~\nameCref{th:progress}~\ref{th:progress:tau}
  it follows that $\ptypeG[1]$ is inhabited 
  by some $\sysS[1]$ such that $\sysS[0] \sysmove{A}{\pref[1]}{} \sysS[1]$.
  Hence, by Lemma~\ref{lem:types:chan-types-moves-honesty}
  also $\ptypeG[1]$ is honest.
  By iterating the same argument to $\ptypeG[1], \ldots, \ptypeG[n-1]$,
  we obtain that $\ptypeG[n]$ is honest, inhabited, 
  and either $\ptypeG[n](s) = \ctypeP[n]$ 
  or $\ptypeG[n](\effwildcard) = \ctypeP[n]$.
  Therefore, we can apply once again~\Cref{lem:self-concordance}, 
  from which we obtain some $\prefi$ and $\ptypeGi$
  such that $[\prefi] = \atomA$, and
  $
  \ptypeG[n]
  \effmove{\prefi}
  \ptypeGi
  $.
  Note that, in the meanwhile, we have constructed a trace:
  \[
  \sysS[0]
  \sysmove{A}{\pref[1]}{}
  \sysS[1]
  \sysmove{A}{\pref[2]}{}
  \cdots
  \sysmove{A}{\pref[n]}{}
  \sysS[n]
  \]
  Since $[\prefi] = \atomA$, 
  the abstraction cannot be done on $\effwildcard$,  
  and so $\prefi = \cocodo{s}{\atomA}$.
  Then, by~\nameCref{th:progress}~\ref{th:progress:do}
  it must be $\atomA \in \readydosys{s}{\pmv A}{\sysS[n]}$.
  Then, by~\Cref{def:readydo-weak},
  we also have $\atomA \in \readydoweak{s}{\pmv A}{\sysS[0]}$.
  Therefore, since $s \not\in \vec{u}$,
  then
  $\atomA \in \readydoweak{s}{\pmv A}{\sysSi[0]}$.
  To conclude, just note that,
  since $(\abscontrP,\ctypeT[n]\,) \ecmove[\atomA] (\abscontrPi,\ctypeTi\,)$, 
  then $\atomA \in \obbl{\pmv A}{s}{\sysSi[0]}$.
  We have then proved
  $\obbl{\pmv A}{s}{\sysSi[0]} \cap \readydoweak{s}{\pmv A}{\sysSi[0]} 
  \; \neq \; \emptyset$,
  which concludes the proof.
\end{proof}

The following~\namecref{ex:type-proc:incompleteness}
shows that our type system is incomplete,
\ie there exists an honest process which is not typeable.

\begin{exa} \label{ex:type-proc:incompleteness}
  We can type the process
  $
  \procPi 
  \; = \;
  \tell{}{\freeze x {\atomOut{a}}}
  \cocoSeq \,
  (
  \cocodo x {\atomOut a} \cocoPlus \cocodo y {\atomOut b}
  )
  $
  with:
  \[
  \ptypeF
  \; = \;
  \setenum{\;
    \bind{x}{\effcontract{\atomOut a} . (\atomOut{a} + \efftauqm)},\;\;
    \bind{y}{\tau . (\efftauqm + \atomOut{b}), \;\;}
    \bind{\effwildcard}{\tau . ( \efftauqm + \efftauqm) }
  \; }
  \]
  Since $\ptypeF(x)$ is \emph{not} abstractly honest,
  then $\procP = (x) (y) \procPi$ is \emph{not} typeable.
  However, $\procP$ is honest: indeed, 
  the branch $\cocodo y {\atomOut b}$ is immaterial, 
  since the session $y$ cannot be established.
\end{exa}



\section{Related work and conclusions}
\label{sec:related-work}

The concept of \emph{contract-oriented computing} 
(as surveyed in \Cref{sec:introduction}) %
has been introduced in~\cite{BZ10lics}, %
and \coco has been later proposed as a \emph{contract-agnostic}
calculus for contract-oriented computing in~\cite{BTZ12sacs}.
\coco has been instantiated with several contract models --- both
binary \cite{BTZ12coordination,BSTZ13forte,BMSZ15jlamp} 
and multiparty~\cite{BTZ12sacs,LangeChorSyntICE13,Bartoletti2015Wild}.
Here, similarly to~\cite{BMSZ15jlamp}, we consider bilateral
contracts, formalised as binary session types (\Cref{sec:contracts}).
A minor difference
w.r.t.~\cite{BTZ12coordination,BSTZ13forte,LangeChorSyntICE13} is that
in the present work we do not have $\fuse{}{}$ as a language
primitive: the creation of fresh sessions is performed
non-deterministically by the context (rule~\nrule{[Fuse]} in
\Cref{fig:co2:semantics}).  
This is equivalent to assume a contract
broker which collects all contracts, and may establish sessions when
compliant ones are found.
The notion of honesty used here is slightly different from the one in~\cite{BTZ12coordination}. 
There, {\pmv A} is considered \emph{culpable} in a session $s$ 
when she has enabled moves in $s$;
by performing such moves, $\pmvA$ can exculpate herself. 
Honesty in~\cite{BTZ12coordination} requires $\pmvA$ to be always able to exculpate herself, in all contexts and in all sessions.
This is a mild variation of the notion of honesty considered here: 
we believe that these two notions are equivalent, under a fair semantics.
A survey of other variants of honesty, and of their properties, 
is in~\cite{BZ15wsfm}.

A type system to safely over-approximate honesty in \coco has been
first proposed in~\cite{BSTZ13forte}.  
The present work improves those results in two main directions.
First, we have redesigned the type system so that now it 
has a decidable type inference (\Cref{th:types:decidability}).
This result relies on a (sound and complete) algorithm for deciding abstract honesty,
based on submarking reachability in Petri nets
(\Cref{th:abs-honesty:decidable}).
Second, we can safely over-approximate the honesty of processes 
which interact through \emph{asynchronous} session types 
(\Cref{th:honest-sync-async}).

The programming model envisioned by \coco has been implemented as a
\emph{contract-oriented middleware}~\cite{CO2} featuring
\emph{timed} session types~\cite{Bartoletti15forte} as contracts.
This middleware collects the contracts advertised by services, and
creates a session between two services when their contracts are compliant.
The middleware monitors all the intra-session communications
(similarly to~\cite{Neykova14beat}), 
also checking that services respect the time constraints specified in their
contracts.
When a participant is culpable of a contract violation 
its reputation is decreased, consequently reducing its chances of being involved in further sessions.

\medskip

The contract model in the present work (\Cref{sec:contracts}) is based
on an interpretation of session types as \emph{behavioural contracts}
equipped with an LTS semantics, which are also studied 
in~\cite{Barbanera10ppdp,BSZ14concur,Bernardi14concur,Barbanera15MSCS,Scalas15phd}.
Similarly to~\cite{BSZ14concur,Scalas15phd}, here we combine contracts with
buffers and define semantics and notions of compliance accounting for
asynchronous interactions.  
A novel contribution in this work is~\Cref{th:st:async-simulates-tm}, 
which proves the undecidability of
compliance between session types interacting via unbounded buffers.

The problem of ensuring safe interactions in session-based systems
has been addressed to a wide extent in the literature,
\eg in~\cite{DezaniY07TGC,Bettini08concur,Castagna09ppdp,Honda98esop,Kobayashi06concur,Bettini08concur,Honda16jacm,Vasconcelos12,Coppo13coordination,Coppo14mscs}. %
In many of these approaches (surveyed in~\cite{Huttel16acmsurvey}), 
deadlock-freedom in the presence
of interleaved sessions is not directly implied by typeability. %
For instance, the two (dishonest) processes:
\begin{align*}
  \procP
  & \; = \; {
    (x,y) \;
    \tell {} {\freeze {x} {\atomIn{a}}}
    \cocoSeq
    \tell {} {\freeze {y} {\atomIn{b}}}
    \cocoSeq
    \cocodo{x}{\atomIn{a}}
    \cocoSeq
    \cocodo{y}{\atomIn{b}}
  }
  \\
  \procQ
  & \; = \; {
    (x,y) \;
    \tell {} {\freeze {x} {\atomOut{a}}}
    \cocoSeq
    \tell {} {\freeze {y} {\atomOut{b}}}
    \cocoSeq
    \cocodo{y}{\atomOut{b}}
    \cocoSeq
    \cocodo{x}{\atomOut{a}}
  }
\end{align*}
would typically be well-typed. %
However, the composition %
$\sys{\pmvA}{\procP} \cocoPar \sys{\pmvB}{\procQ}$ 
reaches a deadlock after fusing the sessions: %
in fact, %
$\pmvA$ remains waiting on $x$ %
(while not being ready at $y$), %
and %
$\pmv{B}$ remains waiting on $y$ %
(while not being ready at $x$).

Multiple interleaved sessions has been tackled 
\eg in~\cite{DezaniY07TGC,Bettini08concur,Castagna09ppdp,Coppo13coordination,Coppo14mscs}. %
To guarantee deadlock freedom, these approaches usually
require that all the interactions on a session must
end before another session can be used. %
For instance, the system
$\sys{\pmvA}{\procP} \cocoPar \sys{\pmvB}{\procQ}$
above would \emph{not} be typeable in~\cite{Castagna09ppdp}, 
coherently with the fact that it is not deadlock-free. %
The resulting notions seem however quite different from honesty, %
because we do not necessarily classify as dishonest 
processes with interleaved sessions. %
For instance, the processes:
\begin{align*}
  & (x,y) \;
  \tell {} {{\freeze{x}{\atomIn{a}}}} \cocoSeq %
  \tell {} {{\freeze{y}{\atomOut{b}}}} \cocoSeq %
  \big(
  \cocodo{x}{\atomIn{a}} \cocoSeq \cocodo{y}{\atomOut{b}} 
  \cocoPlus
  \cocodo{y}{\atomOut{b}} \cocoSeq \cocodo{x}{\atomIn{a}}
  \big)
  \\
  & (x,y) \;
  \tell{}{\freeze x {\atomIn{a}}}
  \cocoSeq \,
  \tell{}{\freeze y {(\atomOut{b} \sumInt \atomOut{c})}}
  \cocoSeq \,
  \big(
  \cocodo x {\atomIn a} \cocoSeq \cocodo y {\atomOut b}
  \cocoPlus
  \tau \cocoSeq (\cocodo y {\atomOut c} \cocoPar \cocodo x {\atomIn a})
  \big)
\end{align*}
would not be typeable according to~\cite{Castagna09ppdp},
but they are honest in our theory (see~\Cref{ex:type-proc:honest-interleaving}). %
A further difference between these approaches and ours
is that we do \emph{not} assume the knowledge of the whole system,
but instead we focus on typing standalone participants. %
Once a participant $\pmvA$ is typed using the rules in~\Cref{fig:type-proc}, 
then $\pmvA$ will always be ready to make her sessions progress.
Our type discipline does not make assumptions on contexts other than
their $\pmvA$-freeness (which can be easily obtained via digital signatures). %
Deadlocks can only occur when the context starts behaving dishonestly.

The problem of checking if the abstract behaviour
of a service conforms to a role of a given choreography has
been investigated in~\cite{Bravetti07sc}.
Under suitable well-formed conditions, conformance is attained exploiting the
\emph{should testing} pre-order.
Similar techniques have been used  in~\cite{bz09} to define contract-based
composition of services.
A main difference between these approaches and ours is that we also consider
the contexts where some participants can be dishonest,
\ie we aim at establishing whether a process abides by
its own contract regardless of its execution context.

In the top-down approach to design a distributed application,
one specifies its overall communication behaviour through a \emph{choreography},
which validates some global properties of the application
(\eg safety, deadlock-freedom, \emph{etc.}). %
To ensure that the application enjoys such properties,
all the components forming the application have to be verified;
this can be done \eg by projecting the choreography to end-point views,
against which these components are verified~\cite{Aalst10cj,Honda16jacm}. %
This approach assumes that designers control the whole application,
\eg, they develop all the needed components. %
However, in many real-world scenarios 
several components are developed independently, without knowing
at design time which other components they will be integrated with. %
In these scenarios, 
the compositional verification pursued by the top-down approach 
is not immediately applicable, because
the choreography is usually unknown,
and even if it were known,
only a subset of the needed components is available for verification. %
The ideas pursued in this paper depart from the top-down approach,
because designers can advertise contracts to discover the needed components
(and so ours can be considered a \emph{bottom-up} approach). %
Coherently, the main property we are interested in is \emph{honesty}, 
which is a property of components, and not of global applications. %
Some works mixing top-down and bottom-up
composition have been proposed in the past few 
years~\cite{Denielou13icalp,Lange12concur,LangeChorSyntICE13,Bartoletti2015Wild}. %



\subsection*{Future works.}

An interesting direction for future research would be extending our type discipline to the version of \coco studied in~\cite{BMSZ15jlamp}, which features value-passing processes and conditionals.
The behaviour of conditional processes $\cocoifte{\expE}{\procP}{\procQ}$ 
is delicate to abstract in the presence of recursion. 
A na\"ive attempt could simply abstract it as $\tau . \ctypeP + \tau . \ctypeQ$, 
where $\ctypeP$ and $\ctypeQ$ are the abstractions of $\procP$ and $\procQ$, respectively.
However, this abstraction would not be safe. 
For instance, consider the process:
\[
\procR 
\; = \;
\tell {} {\freeze x {\atomOut{a}}}
\cocoSeq 
(\procRec {\procX()} {\cocoifte{\false}{\cocodo x {\atomOut a}}{\tau \cocoSeq \procX()}}) ()
\]
We have that $(x) \procR$ is \emph{not} honest, since the promised $\atomOut{a}$ action will never be performed.
However, the na\"ive abstraction of $\procPi$ on channel $x$ would be
$
\effcontract{\atomOut a} . 
(\effrec{\consttovar{\procX}}{\tau . \atomOut{a} + \tau . \tau . \consttovar{\procX}} )
$
which is abstractly honest, since after $\effcontract{\atomOut{a}}$ 
the action $\atomOut{a}$ is persistently weakly enabled. 

The issue is that, under recursion, 
$\tau . \ctypeP + \tau . \ctypeQ$ does not precisely represent the 
internal non-determinism caused by conditionals.
We could tackle this issue in two different ways:
either refining the notion of readiness as in~\cite{BMSZ15jlamp},
or --- more directly --- ruling out conditional processes wherein some recursive calls are not guarded by $\cocodo{}{}$-actions 
(as in $\tau \cocoSeq \procX()$ above).

Another possible extension of our type system is a more precise handling of recursion, 
so to type processes with non-empty lists of parameters in recursive calls $\procX(\vec{u})$. 
However this would be a non-trivial task, due to the presence of processes which can possibly extrude names through recursive invocations.
This is why the typing rule \nrule{[T-Rec]} in~\Cref{fig:type-proc} only allows recursion with no parameters. 
This forbids, \eg, to type the following honest process:
\[
(x) \,
\tell {} {\freeze x {\atomOut a}} 
\cocoSeq 
(\procRec {\procX(\mathit{y})} {\cocodo y {\atomOut a} \cocoSeq 
  (z) \, \tell {} {\freeze z {\atomOut a}} \cocoSeq \procX(z))(x)} 
\]
Note however that the current type system is precise enough to correctly 
establish the honesty of complex processes,
like \eg the travel agency case study in~\cite{BMSZ15jlamp}.

In our work, we focused mainly on the synchronous setting. 
Asynchrony exposes several technical challenges 
that are not present in the synchronous case.
For instance, the operational semantics of \coco becomes undecidable,
because compliance between asynchronous session types is undecidable 
(\Cref{th:st:compliance:sync-implies-async}).
As for honesty, the set of $\infty$-honest processes is larger than the set of $1$-honest one, 
and it is still undecidable (\Cref{th:honesty-undecidable}).
For instance, \Cref{ex:honesty:sync-async} provides a $\infty$-honest process which is not $1$-honest (and so, not typeable).
While~\Cref{th:honest-sync-async} allows us to lift $1$-honesty to $\infty$-honesty, making our type system sound even in the asynchronous setting, it would be desirable to extend our analysis so to cover a larger class of processes that includes some $\infty$-honest but not $1$-honest processes.
A possible refinement of our type system would be to modify 
the abstraction of prefixes to take into account the fact that, 
in the asynchronous setting,
enabled outputs on a session $u$ are never blocking once $u$ has been established.
Technically, this would require to improve the abstraction of prefixes
(\Cref{def:types:typing-prefixes}) 
so that the enabled output actions (but for the first action in $u$) 
are abstracted on $v \neq u$ as $\tau$, instead of $\efftauqm$.
For instance, the process in~\Cref{ex:honesty:sync-async}
would be typeable in this modified analysis.

Another research direction is the integration of contract-oriented primitives
within mainstream programming languages.
This can be done \eg as in~\cite{CO2}, where Java APIs are provided 
to interact with a middleware which handles contracts and sessions.
The problem of honesty of Java programs is analogous to that of \coco,
with the additional issue that contract violations are explicitly
sanctioned by the middleware in terms of reputation loss.
The suite of tools Diogenes~\cite{Atzei16forte}
supports programmers in writing honest Java code.
The tool translates honest \coco specifications into skeletal Java programs,
and checks that their honesty is preserved upon refinement.
To this purpose, the tool first 
infers a \coco process which approximates the behaviour 
of a Java program;
the honesty of this process is then verified
through the model checker in~\cite{BMSZ15jlamp}.



\section*{Acknowledgement}

This work has been partially supported by
Aut.\ Reg.\ of Sardinia P.I.A.\ 2013 ``NOMAD'',
 by EU COST Action IC1201
``Behavioural Types for Reliable Large-Scale Software Systems''
(BETTY), and by EU  COST Action IC1405 ``Reversible Computation -
extending horizons of computing''.
Alceste Scalas was partly supported by EPSRC grant EP/K011715/1. %
We thank the anonymous reviewers and Nicola Atzei 
for their insightful comments on a preliminary version of this paper.


\bibliographystyle{abbrv}
\bibliography{main}

\newpage

\begin{appendices}

\section{Proofs for~\Cref{sec:abs-honesty}} 
\label{sec:proofs-abs-honesty}

\begin{nota}[Realizes] \label{def:types:realizes}
  We say that: 
  \begin{itemize}
    
  \item $\ctypeT$ is \emph{ready for $\abscontrP$} whenever
    $(\abscontrP,\ctypeT)$ is ready;
    
  \item $\ctypeT$ \emph{realizes} $\abscontrP$ whenever
    $(\abscontrP,\ctypeT)$ is honest.
    
  \end{itemize}
\end{nota}

\begin{lem}
  \label{lem:types:channel-type-moves-realization}
  For all 
  $\alpha \in \setenum{\tau, \efftauqm, \effcontract{\contrPi}}$: \;
  $
  \ctypeT \effmove{\alpha} \ctypeTi
  \;\;\land\;\; 
  (\Gamma,\ctypeT) \text{ honest}
  \;\implies\;
  (\Gamma,\ctypeTi) \text{ honest}
  $
\end{lem}
\begin{proof}
  Assume $(\Gamma,\ctypeP)$ honest,
  $\alpha \in \setenum{\tau, \efftauqm, \effcontract{\contrPi}}$, and
  $\ctypeP \effmove{\alpha} \ctypePi$.
  To show $(\Gamma,\ctypePi)$ honest, assume that
  $(\Gamma,\ctypePi) \ecmove^* (\abscontrP,\ctypeTii)$.
  We have the following two cases:
  \begin{itemize}
    
  \item $\alpha \in \setenum{\tau,\efftauqm}$.
    Since $\ctypeP \effmove{\alpha} \ctypePi$, 
    then by rule \nrule{[A-Tau]} we have 
    $(\Gamma, \ctypeP) \ecmove[\alpha] (\Gamma, \ctypePi)$.
    Hence, 
    $
    (\Gamma, \ctypeP)
    \ecmove[\alpha] 
    (\Gamma, \ctypePi) 
    \ecmove^* 
    (\abscontrP, \ctypeTii)
    $.
    Since $(\Gamma,\ctypeP)$ is honest, then $(\abscontrP, \ctypeTii)$ is ready.

  \item $\alpha = \effcontract{\contrPi}$.
    We can modify the trace
    $(\Gamma, \ctypePi) \ecmove^* (\abscontrP,\ctypeTii)$
    by adding $\contrPi$ to all the sets of contracts in the trace.
    The result is still a valid trace, 
    because enlarging the set of contracts does not reduce the applicability 
    of the rules in~\Cref{fig:abs-sys:semantics}.
    In this way, we obtain the trace
    $(\Gammai, \ctypePi) \ecmove^* (\abscontrP, \ctypeTii)$,
    where $\Gammai$ is either $\Gamma \cup \setenum{\contrPi}$
    if $\Gamma$ is a set of contracts, or $\Gammai = \Gamma$ otherwise.
    Therefore, we have
    $
    (\Gammai, \ctypeP) 
    \effmove{\tau} 
    (\Gammai, \ctypePi) 
    \ecmove^* 
    (\abscontrP, \ctypeTii)
    $.
    Since $(\Gamma,\ctypeP)$ is honest, then $(\abscontrP, \ctypeTii)$ is ready.
    \qedhere
  \end{itemize}  
\end{proof}

\begin{proofoflem}{lem:types:chan-types-moves-honesty}
  Immediate consequence of~\Cref{lem:types:channel-type-moves-realization}.
\end{proofoflem}

\begin{proofoflem}[Abstract readiness and parallel composition]{lem:types:multiplicity}
  First, we note that if $\Gamma \not\abscmove{}$, 
  then both sides are trivially true by~\Cref{def:abs-honesty}, hence they are equivalent. 
  Therefore, we can assume that $\Gamma \abscmove{\atomA}$ for some $\atomA$, 
  and so $\Gamma = \abscontrP$ for some $\abscontrP$.

  For the $\Leftarrow$ direction, assume w.l.o.g.~that $(\abscontrP,\ctypeP)$ is ready.
  By~\Cref{def:abs-honesty}, this implies that 
  $(\abscontrP,\ctypeP) {\ecmove[\tau]}{}^* \ecmove[\atomB]$ for some $\atomB$. 
  Inverting the rules of the semantics of pointed abstract systems, 
  we find a corresponding trace for the pointed abstract process $\ctypeP$, namely
  $\ctypeP {\effmove{\tau}}{}^* \effmove{\atomB}$. 
  Consequently, by rule~\nrule{[C-ParL]},
  $\ctypeP \mid \ctypeQ {\effmove{\tau}}{}^* \effmove{\atomB}$, hence
  $(\abscontrP,\ctypeP \mid \ctypeQ) {\ecmove[\tau]}{}^* \ecmove[\atomB]$.

  For the $\Rightarrow$ direction, assume that $(\abscontrP,\ctypeP \mid \ctypeQ)$ is ready,
  hence $(\abscontrP,\ctypeP \mid \ctypeQ) {\ecmove[\tau]}{}^* \ecmove[\atomB]$.
  As above, inverting the rules we obtain $\ctypeP \mid \ctypeQ {\effmove{\tau}}{}^* \effmove{\atomB}$.
Therefore, we have that either of $\ctypeP {\effmove{\tau}}{}^* \effmove{\atomB}$ and
  $\ctypeQ {\effmove{\tau}}{}^* \effmove{\atomB}$ because parallel pointed abstract processes cannot interact with each other (as their semantics does not allow synchronization or communication among parallel processes).
  In both cases, we can then lift the trace on pointed abstract processes to a trace on pointed abstract systems, showing that $(\abscontrP,\ctypeP)$ or $(\abscontrP,\ctypeQ)$ is ready.
  \qed
\end{proofoflem}

\begin{lem}
  \label{lem:abs-honesty:tau}
  If $\ctypeQ$ only contains $\tau$ and $\efftauqm$ actions, then:
  \begin{bartalign}
    \label{lem:abs-honesty:tau:par}
    & \ctypeP \text{ honest }
    \iff
    \ctypeP \mid \ctypeQ \text{ honest }
    \\
    \label{lem:abs-honesty:tau:substitution}
    & \ctypeP \text{ honest }
    \implies
    \ctypeP\subs{\ctypeQ}{\ctypeX} \text{ honest}
  \end{bartalign}
\end{lem}
\begin{proof}
  For the $\Leftarrow$ direction of item~\eqref{lem:abs-honesty:tau:par},
  assume that 
  $(\emptyset,\ctypeP) \ecmove^* (\abscontrP,\ctypePi)$.
  By the semantics of pointed abstract systems, this implies that
  $(\emptyset,\ctypeP \mid \ctypeQ) \ecmove^* (\abscontrP,\ctypePi \mid \ctypeQ)$,
  which is ready because $\ctypeP \mid \ctypeQ$ is honest.
  By~\Cref{lem:types:multiplicity}, either 
  $(\abscontrP,\ctypePi)$ is ready or 
  $(\abscontrP,\ctypeQ)$ is ready.
  If $(\abscontrP,\ctypePi)$ is ready, we have the thesis.
  Otherwise, $\abscontrP$ has obligations, 
  and $(\abscontrP,\ctypeQ)$ is ready 
  --- contradicting $\ctypeQ$ performing only $\tau$ and $\efftauqm$ moves
  (indeed, by~\Cref{def:abs-honesty}, a pointed abstract systems 
  with only $\tau$ e $\efftauqm$ cannot fulfil obligations).

  \medskip\noindent
  For the $\Rightarrow$ direction of item~\eqref{lem:abs-honesty:tau:par},
  assume that 
  \begin{equation} \label{eq:abs-honesty:tau:par}
    (\emptyset,\ctypeP \mid \ctypeQ) \; \ecmove^* \; (\abscontrP,\ctypeR)
  \end{equation}
  Assuming that $\abscontrP$ has some obligations, we need to prove that 
  $(\abscontrP,\ctypeR)$ is ready, \ie 
  $(\abscontrP,\ctypeR) \ecmove[\tau]^* \ecmove[\atomA]$ for some $\atomA$.
  We must have that 
  $\ctypeR = \ctypePi \mid \ctypeQi$, where 
  $\ctypeP \effmove{}^* \ctypePi$ and 
  $\ctypeQ \effmove{}^* \ctypeQi$. 
  This is because two parallel components can only interact 
  by performing actions on the contract.
  Exploiting~\eqref{eq:abs-honesty:tau:par}, we can construct a trace
  $(\emptyset, \ctypeP) \ecmove^* (\abscontrP,\ctypePi)$.
  Indeed, in the transitions 
  $\ctypeQ \effmove{}^* \ctypeQi$ 
  there are only $\tau$ and $\efftauqm$ actions, 
  so the evolution of the contract in~\eqref{eq:abs-honesty:tau:par}  
  only depends on the transitions of $\ctypeP$.
  Since $\ctypeP$ is honest and $\abscontrP$ has some obligations, 
  we must have 
  $(\abscontrP,\ctypePi) \ecmove[\tau]^* \effmove{\atomA}$,
  hence by rule~\nrule{[C-ParL]} we conclude that 
  $(\abscontrP, \ctypePi \mid \ctypeQi) \ecmove[\tau]^* \ecmove[\atomA]$.

  \medskip\noindent
  For item~\eqref{lem:abs-honesty:tau:substitution}, note that
  the process $\ctypeP\subs{\ctypeQ}{\ctypeX}$ differs from $\ctypeP$ 
  in that the occurrences of the free variables $\ctypeX$ 
  have been replaced by $\ctypeQ$ 
  (with the usual assumption that the substitution is capture-avoiding).
  According to the transition semantics, $\ctypeX$ is a stuck process, 
  while by hypothesis $\ctypeQ$ only performs $\tau$ and $\efftauqm$ actions. 
  Intuitively, the substitution $\subs{\ctypeQ}{\ctypeX}$ is irrelevant 
  for the transitions of $\ctypeP\subs{\ctypeQ}{\ctypeX}$, 
  except for those cases where a trace of $\ctypeP$ would expose an $\ctypeX$ 
  at the top level. 
  Hence, a residual of $\ctypeP\subs{\ctypeQ}{\ctypeX}$ 
  is formed by a parallel component 
  where the substitution had no effect in the trace, 
  and the residuals of all the substituted top-level $\ctypeX$’s 
  --- which are residuals of $\ctypeQ$.
  More precisely, we have that in every trace 
  $(\emptyset, \ctypeP\subs{\ctypeQ}{\ctypeX}) \ecmove^* (\abscontrP, \ctypeR)$:
  \[
  \ctypeR \; = \; \ctypePi \subs{\ctypeQ}{\ctypeX} \mid \ctypeQ[1] \mid \cdots \mid \ctypeQ[k]
  \hspace{20pt}
  \text{where }
  \ctypeQ[i] \text{ are residuals of } \ctypeQ
  \text{ and } 
  \ctypeP \effmove{}^* 
  \ctypePi \mid \underbrace{\ctypeX \mid \cdots \mid \ctypeX}_{k \text{ times}}
  \]
  This fact relies on two properties of pointed abstract systems. 
  First, the residuals of $\ctypeQ$ never interact with 
  $\ctypePi\subs{\ctypeQ}{\ctypeX}$, 
  since $\ctypeQ$ can only perform $\tau$ and $\efftauqm$ actions. 
  Second, sums in pointed abstract processes are prefix-guarded
  (otherwise, substituting the honest 
  $(\atomOut{a}, \ctypeX + \atomOut{a})$ we would obtain 
  $(\atomOut{a}, \tau + \atomOut{a})$, which is no longer honest). 
  Roughly, prefix-guardedness ensures that the moves of $\ctypeQ$ in 
  $\ctypeP\subs{\ctypeQ}{\ctypeX}$ do not conflict with the moves of $\ctypeP$.

  Hence, from the trace 
  $(\emptyset, \ctypeP\subs{\ctypeQ}{\ctypeX}) \ecmove^* (\abscontrP, \ctypeR)$ 
  we can construct a trace 
  $(\emptyset, \ctypeP) \ecmove^*  (\abscontrP, \ctypePi \mid \ctypeX \mid \cdots \mid \ctypeX)$.
  Since $\ctypeP$ is honest, then 
  $(\abscontrP, \ctypePi \mid \ctypeX \mid \cdots \mid \ctypeX)$ is ready, 
  and since the components $\ctypeX$ are stuck, then also 
  $(\abscontrP, \ctypePi)$ must be ready.
  Then, 
  $(\abscontrP, \ctypePi \subs{\ctypeQ}{\ctypeX} \mid \ctypeQ[1] \mid \cdots \mid \ctypeQ[k])$ 
  is ready.  
\end{proof}


\section{Proofs for~\Cref{sec:type-system}} 
\label{sec:proofs-type-system}

\begin{proofoflem}[Honesty of $\ptypeF(\effwildcard)$]{lem:types:f-wildcard-honest}
  Item~\eqref{lem:types:proc@typing@effwildcard} follows
  by easy induction on the typing derivation of  
  $\effentails \procP \colon \ptypeF$ (\Cref{fig:type-proc}).
  Item~\eqref{lem:types:f-wildcard-honest:proc} is a 
  direct consequence of~\Cref{lem:types:proc@typing@effwildcard}.
  For item~\eqref{lem:types:f-wildcard-honest:sys},
  we proceed by induction on
  the typing derivation of $\tsentails{A} \sysS \colon \ptypeF$.
  \begin{itemize}

  \item \nrule{[T-SA]}. We have:
    \[
    \inference[\smallnrule{[T-SA]}]
    {\effentails \procP \colon \ptypeF}
    {\tsentails{A} \sysS = {\sys {\pmvA} {\procP}} \colon \ptypeF}
    \]
    Thus, $\ptypeF(\effwildcard)$ is honest 
    by item~\eqref{lem:types:f-wildcard-honest:proc}.

  \item \nrule{[T-SDel2]}, \nrule{[T-SPar2]}. 
    Straightforward, by applying the induction hypothesis on the rule premises.
    \qedhere

  \end{itemize}
\end{proofoflem}

\begin{lem}[Structural equivalence and substitutions]
  \label{lem:types:struct-equiv-subst}
  \label{lem:types:struct-equiv-subst-sys}
  For all processes $\procP, \procPi$, systems $\sysS, \sysSi$,
  and for all substitutions $\sigma$:
  \begin{align*}
    \procP \equiv \procPi 
    & \;\;\implies\;\; \procP\sigma \equiv \procPi\sigma
    \\
    \sysS \equiv \sysSi 
    & \;\;\implies\;\; \sysS\sigma \equiv \sysSi\sigma
  \end{align*}
\end{lem}
\begin{proof}
  Case analysis on all the different cases of structural equivalence.
\end{proof}

\begin{lem}[Substitution of delimited channels] 
  \label{lem:types:tscompat-bot}
  \label{lem:types:f-subst-notin}
  For all $\sysS$, $\ptypeF$, $u$, and $\ctypeT$:
  \[
  \effentails \sysS \tscompat \ptypeF
  \;\;\land\;\; u \not\in \fnv{\sysS}
  \;\;\;\implies\;\;\;
  \effentails \sysS \tscompat \ptypeF\setenum{\bind{u}{\ctypeT}}
  \;\text{ and }\;
  \effentails \sysS \tscompat \ptypeF\setenum{\bind{u}{\bot}}
  \]
  Furthermore, these typing derivations have the same depth as the original one.
\end{lem}
\begin{proof}
  Straightforward induction, 
  by inspection of the rules in~\Cref{fig:type-sys}.
\end{proof}

\begin{lem}[Substitution of recursion variables]
  \label{lem:types:type-subs-recvar}
  For all $\procP$, $\procQ$, $\ptypeF$, and $\ptypeG$:
  \[
  \effentails \procP \colon \ptypeF
  \;\; \land \;\;
  \effentails \procQ \colon \ptypeG  
  \;\implies\;
  \effentails 
  \procP \subs{\procQ}{\procX} 
  \colon 
  \lambda u \dotseq 
  \expand{\ptypeF}{A}(u) \subs{\expand{\ptypeG\,}{A}(u)}{\consttovar{\procX}}
  \tag*{ where $A = \fnv{\procP \subs{\procQ}{\procX}}$.}
  \]
\end{lem}
\begin{proof}
  Tedious induction on the typing derivation of 
  $\,\effentails \procP \colon \ptypeF$.
\end{proof}

\begin{lem}
  \label{lem:types:fsubst-honest}
  \label{lem:types:expand-honest}
  If $\ptypeF$ is honest, then
  $\fsubst{\ptypeF}{\sigma}$
  and 
  $\expand{\ptypeF}{A}$
  are honest,
  for all $\sigma$ and for all $A$.    
\end{lem}
\begin{proof}
  Assume that $\ptypeF$ is honest, 
  and that $u \in \dom{(\fsubst{\ptypeF}{\sigma})}$.
  We have that $(\fsubst{\ptypeF}{\sigma})(u) = \ptypeF(v)$,
  for some $v \in \dom{\ptypeF}$.
  Similarly, for all $u \in \dom{(\expand{\ptypeF}{A})}$,
  we have that $\expand{\ptypeF}{A}(u) = \ptypeF(v)$,
  for some $v \in \dom{\ptypeF}$. 
  The thesis follows by the assumption that $\ptypeF$ is honest.
\end{proof}

\begin{lem}
  \label{lem:types:tscompat-effmove}
  For all $\sysS$, $\ptypeF,\ptypeFi$, and $\pref$
  such that $\tsentails{A} \sysS \tscompat \ptypeF$
  and $\ptypeF \effmove{\pref} \ptypeFi$:
  \begin{align}
    \label{eq:lem:types:tscompat-effmove:do}
    \pref = \cocodo{s}{\atomA} \;\land\; \sys s {\cdots} \not\in \sysS  
    & \;\;\implies\;\; 
    \tsentails{A} \sysS \tscompat \ptypeFi
    \\
    \label{eq:lem:types:tscompat-effmove:not-do}
    (\forall s,\atomA \suchthat \pref \neq \cocodo{s}{\atomA})  
    & \;\;\implies\;\; 
    \tsentails{A} \sysS \tscompat \ptypeFi
  \end{align}
\end{lem}
\begin{proof}
  By induction on the typing derivation of $\tsentails{A} \sysS \tscompat \ptypeF$.
  All cases are straightforward, but the following ones:
  \begin{itemize}
  \item \nrule{[T-SFz1]}. We have:
    \[
    \inference[\smallnrule{[T-SFz1]}]
    {\expand{\ptypeF}{\setenum{x}}(x)\ \mbox{realizes $\contrP$}}
    {\tsentails{A} \sysS = {\setenum{\freeze x {\contrP}}_{\pmv A}} \tscompat \ptypeF}      
    \]
    We first note that, since $\ptypeF \effmove{\pref} \ptypeFi$, 
    then $\expand{\ptypeF}{\setenum{x}}(x) \effmove{[\pref]_x} \expand{\ptypeFi}{\setenum{x}}(x)$.
    By~\Cref{def:types:typing-prefixes} we have that
    for item~\eqref{eq:lem:types:tscompat-effmove:do} $[\pref]_x = \efftauqm$,
    while for item~\eqref{eq:lem:types:tscompat-effmove:not-do}
    $[\pref]_x \neq \atomA$, for all $\atomA$.
    Then, in both cases by~\Cref{lem:types:channel-type-moves-realization}
    we have that $\expand{\ptypeFi}{\setenum{x}}(x)$ realizes $\contrP$.
    Therefore, the thesis follows by using rule~\nrule{[T-SFz1]}.

  \item \nrule{[T-SFuse]}. We have:
    \[
    \inference[\smallnrule{[T-SFuse]}]
    {\expand{\ptypeF}{\setenum{t}}(t)\ \mbox{realizes $\cta{\gamma}$}}
    {\tsentails{A} \sysS = {\sys t {\gamma}} \tscompat \ptypeF}
    \]
    We first note that, since $\ptypeF \effmove{\pref} \ptypeFi$, 
    then $\expand{\ptypeF}{\setenum{t}}(t) \effmove{[\pref]_t} \expand{\ptypeFi}{\setenum{t}}(t)$.
    By~\Cref{def:types:typing-prefixes} we have that
    for item~\eqref{eq:lem:types:tscompat-effmove:do} $[\pref]_t = \efftauqm$
    (because the assumption $\sys s {\cdots} \not\in \sysS$ implies that $t \neq s$),
    while for item~\eqref{eq:lem:types:tscompat-effmove:not-do}
    $[\pref]_t \neq \atomA$, for all $\atomA$.
    Then, in both cases by~\Cref{lem:types:channel-type-moves-realization}
    we have that $\expand{\ptypeFi}{\setenum{t}}(t)$ realizes $\cta{\gamma}$.
    The thesis follows by using rule~\nrule{[T-SFuse]}.

  \item \nrule{[T-SDel1]}. We have:
    \[
    \inference[\smallnrule{[T-SDel1]}]
    {\tsentails{A} \sysS[0] \tscompat \ptypeF\setenum{\bind{u}{\bot}}}
    {\tsentails{A} \sysS = (u)\sysS[0] \tscompat \ptypeF}
    \]
    Since $\ptypeF \effmove{\pref} \ptypeFi$, then 
    $\ptypeF\setenum{\bind{u}{\bot}} \effmove{\pref} \ptypeFi\setenum{\bind{u}{\bot}}$.
    For item~\eqref{eq:lem:types:tscompat-effmove:not-do}
    we can apply the induction hypothesis, which gives
    $\tsentails{A} \sysS[0] \tscompat \ptypeFi\setenum{\bind{u}{\bot}}$;
    the thesis then follows by rule~\nrule{[T-SDel1]}.
    For item~\eqref{eq:lem:types:tscompat-effmove:do}
    there are two cases, according to whether $s = u$ or not.
    If $s \neq u$, then we can apply the induction hypothesis, 
    and proceed as above. 
    Otherwise, if $s = u$, then it might be the case that
    $\sys s {\cdots} \in \sysS[0]$,
    \ie $\sysS[0] = \sys s {\gamma} \cocoPar \cdots$.
    In this case, by~\Cref{def:types:trans} we also have that:
    \[
    \ptypeF\setenum{\bind{u}{\bot}} \effmove{\prefi} \ptypeFi\setenum{\bind{u}{\bot}}
    \]
    where $\prefi = \cocodo{t}{\atomA}$, with $\sys t {\cdots} \not\in \sysS[0]$.
    This holds because $[\pref]_v = \efftauqm = [\prefi]_v$, for all $v \in \dom{\ptypeF}$.
    Therefore, we can apply the induction hypothesis, which gives
    $\tsentails{A} \sysS[0] \tscompat \ptypeFi\setenum{\bind{u}{\bot}}$;
    we then conclude by rule~\nrule{[T-SDel1]}. 

  \item \nrule{[T-SPar1]}. Straightforward, by applying the induction hypothesis on both premises.

  \item \nrule{[T-SAFree2]}, \nrule{[T-SAFree3]}. 
    The thesis follows because $\dom{\ptypeFi} \subseteq \dom{\ptypeF}$.
    \qedhere

  \end{itemize}
\end{proof}



\begin{lem}     \label{lem:types:substitution:expand}
  \begin{bartalign}
    \label{lem:types:substitution:expand:ctype}
    & \expand{\ptypeF}{A} \setenum{\bind{u}{\ctypeT}} 
    \; = \; 
    \expand{(\ptypeF \setenum{\bind{u}{\ctypeT}})}{A}
    \\
    \label{lem:types:substitution:expand:bot}
    & \expand{\ptypeF}{A} \setenum{\bind{u}{\bot}} 
    \; = \; 
    \expand{(\ptypeF \setenum{\bind{u}{\bot}})}{A \setminus \setenum{u}}
  \end{bartalign}
\end{lem}
\begin{proof}
  Straightforward case analysis on $u \in A$ in~\Cref{def:types:expand}.
\end{proof}

\begin{nota} \label{notation:ptype:syntax}
  For all $\alpha$, and for $\circ \in \setenum{\mid,+}$, 
  we will use the compact notation 
  $\alpha.\ptypeF$ for $\lambda u.\; \alpha . \ptypeF(u)$, and
  $\ptypeF \circ \ptypeG$
  for $\lambda u.\; \ptypeF(u) \circ \ptypeG(u)$.
\end{nota}

\begin{proofoflem}[Structural equivalence and typing]{lem:types:equiv}
  By induction on the typing derivation.
  Most cases are straightforward; we only show the case of scope extrusion.

  For~\Cref{lem:types:P-equiv}, 
  consider the case where:
  \[
  \procP 
  \;\; = \;\;
  (u) (\procP[0] \cocoPar \procP[1])
  \;\;\equiv\;\;
  \procP[0] \cocoPar (u) \procP[1]
  \;\; = \;\;
  \procPi
  \tag*{with $u \not\in \fnv{\procP[0]}$}
  \]
  We have the following typing derivation for $\procP$,
  where $A = \dom{\ptypeG[0]} \cup \dom{\ptypeG[1]}$:
  \smallskip
  \begin{center}
    \AXC{$\effentails \procP[0] \colon \ptypeG[0]$}
    \AXC{$\effentails \procP[1] \colon \ptypeG[1]$}
    \RL{\nrule{[T-Par]}}
    \BIC{$\effentails \procP[0] \cocoPar \procP[1] \colon \expand{\ptypeG[0]}{A} \,\mid \expand{\ptypeG[1]}{A} \; = \; \ptypeG$}
    \AXC{$\expand{\ptypeG}{\setenum{u}}\!(u)$ honest}
    \RL{\nrule{[T-Del]}}
    \BIC{$\effentails \procP \colon \ptypeG \setenum{\bind{u}{\bot}} = \ptypeF$}
    \DisplayProof
  \end{center}
  \smallskip
  Since $\effentails \procP[0] \colon \ptypeG[0]$ and 
  $u \not\in \fnv{\procP[0]}$, then 
  by~\nameCref{lem:types:dom-fnv}~\ref{lem:types:dom-fnv:proc}
  it must be 
  $\expand{\ptypeG[0]}{\setenum{u}}\!(u) = \ptypeG[0](\effwildcard)$.
  Hence, 
  by~\nameCref{lem:types:f-wildcard-honest}~\ref{lem:types:proc@typing@effwildcard},
  $\expand{\ptypeG[0]}{\setenum{u}}\!(u)$
  only contains $\tau$ and $\efftauqm$ actions.
  Together with the fact that
  $\expand{\ptypeG}{\setenum{u}}\!(u) = \expand{\ptypeG[0]}{\setenum{u}}\!(u) \mid \expand{\ptypeG[1]}{\setenum{u}}\!(u)$
  is honest,
  by~~\nameCref{lem:abs-honesty:tau}~\ref{lem:abs-honesty:tau:par}
  we deduce that
  $\expand{\ptypeG[1]}{\setenum{u}}\!(u)$ is honest.
  We can then construct the following typing derivation for $\procPi$,
  where $A' = \dom{\ptypeG[0]} \cup (\dom{\ptypeG[1] \setminus \setenum{u}})$:
  \smallskip
  \begin{center}
    \AXC{$\effentails \procP[0] \colon \ptypeG[0]$}

    \AXC{$\effentails \procP[1] \colon \ptypeG[1]$}
    \AXC{$\expand{\ptypeG[1]}{\setenum{u}}\!(u)$ honest}
    \RL{\nrule{[T-Del]}}
    \BIC{$\effentails (u) \procP[1] \colon \ptypeG[1] \setenum{\bind{u}{\bot}}$}
    \RL{\nrule{[T-Par]}}
    \BIC{$\effentails \procPi \colon \expand{\ptypeG[0]}{A'} \,\mid \expand{\ptypeG[1] \setenum{\bind{u}{\bot}}}{A'} = \ptypeFi$}
    \DisplayProof
  \end{center}
  To conclude the proof, we need to prove that $\ptypeFi = \ptypeF$.
  Since $u \not \in \fnv{\procP[0]} \cup \setenum{\effwildcard} = \dom{\ptypeG[0]}$, 
  then $A' = (\dom{\ptypeG[0]} \cup \dom{\ptypeG[1]}) \setminus \setenum{u} = A \setminus \setenum{u}$, 
  and so $\ptypeG[0] = \ptypeG[0]\setenum{\bind{u}{\bot}}$.

  We then have:
  \begin{align*}
    \ptypeF 
    & = \ptypeG \setenum{\bind{u}{\bot}}
    && \text{by def.\ $\ptypeF$}
    \\
    & = (\expand{\ptypeG[0]}{A} \,\mid\, \expand{\ptypeG[1]}{A}) \setenum{\bind{u}{\bot}}
    && \text{by def.\ $\ptypeG$}
    \\
    & = \expand{\ptypeG[0]}{A} \setenum{\bind{u}{\bot}} \,\mid\, \expand{\ptypeG[1]}{A} \setenum{\bind{u}{\bot}}
    && 
    \\
    & = \expand{\ptypeG[0]\setenum{\bind{u}{\bot}}}{A'} \,\mid\, \expand{\ptypeG[1]\setenum{\bind{u}{\bot}}}{A'}
    && \text{by~\nameCref{lem:types:substitution:expand}~\ref{lem:types:substitution:expand:bot}}
    \\
    & = \expand{\ptypeG[0]}{A'} \,\mid\, \expand{\ptypeG[1]\setenum{\bind{u}{\bot}}}{A'}
    && \text{as shown before}
    \\
    & = \ptypeFi
    && \text{by def.\ $\ptypeFi$}
  \end{align*}


  The analogous case for systems (for~\ref{lem:types:S-equiv:colon}) is similar.
  \qed
\end{proofoflem}


\begin{proofoflem}[Typing and substitution]{lem:types:substitution}
  We start by proving item~\eqref{lem:types:typing-renaming-proc}.
  We proceed by induction on the typing derivation of 
  $\effentails \procP \colon \ptypeF$. %
  We have the following cases, according to the last rule
  used in the derivation:
  \begin{itemize}

  \item \nrule{[T-Nil]}, \nrule{[T-Var]}. 
    Trivial, since the substitution is vacuous
    (both on $\procP$ and on $\ptypeF$).

  \item \nrule{[T-Par]}. We have:
    \[
    \inference[\smallnrule{[T-Par]}]
    {\effentails \procP[0] \colon \ptypeF[0] & 
     \effentails \procP[1] \colon \ptypeF[1] & 
     A = \dom{\ptypeF[0]} \cup \dom{\ptypeF[1]}}
    {\effentails \procP = \procP[0] \cocoPar \procP[1] \colon
      \lambda u\dotseq \expand{\ptypeF[0]}{A}(u) \mid \expand{\ptypeF[1]}{A}(u)} 
    \]
    Since $\dom{\ptypeF[0]} \subseteq \dom{\ptypeF} \supseteq \dom{\ptypeF[1]}$ 
    and $\fsubst{\ptypeF}{\sigma}$ is defined, then also
    $\fsubst{\ptypeF[0]}{\sigma}$ and $\fsubst{\ptypeF[1]}{\sigma}$ are defined.
    Then, by applying the induction hypothesis on both premises:
    \[
    \effentails \procP[0]\sigma \colon \fsubst{\ptypeF[0]}{\sigma}
    \qquad
    \effentails \procP[1]\sigma \colon \fsubst{\ptypeF[1]}{\sigma}
    \]
    Then, by rule~\nrule{[T-Par]}:
    \[
    \inference[\smallnrule{[T-Par]}]
    {\effentails \procP[0]\sigma \colon \fsubst{\ptypeF[0]}{\sigma} & 
     \effentails \procP[1]\sigma \colon \fsubst{\ptypeF[1]}{\sigma} & 
     B = \dom{\fsubst{\ptypeF[0]}{\sigma}} \cup \dom{\fsubst{\ptypeF[1]}{\sigma}}}
    {\effentails \procP\sigma = \procP[0]\sigma \cocoPar \procP[1]\sigma \colon
      \lambda u\dotseq \expand{(\fsubst{\ptypeF[0]}{\sigma})}{B}(u) \mid \expand{\fsubst{(\ptypeF[1]}{\sigma})}{B}(u) 
      = \ptypeG} 
    \]
    To conclude, we need to prove that $\ptypeG = \fsubst{\ptypeF}{\sigma}$, \ie:
    \[
    \fsubst{(\expand{\ptypeF[i]}{A})}{\sigma}
    \;\; = \;\;
    \expand{(\fsubst{\ptypeF[i]}{\sigma})}{B} 
    \tag*{(for $i \in \setenum{0,1}$)}
    \]
    We proceed by cases on 
    $(\dom{\sigma} \cap \dom{\ptypeF[0]},\, \dom{\sigma} \cap \dom{\ptypeF[1]})$.
    Since $\fsubst{\ptypeF[0]}{\sigma}$ and $\fsubst{\ptypeF[1]}{\sigma}$
    are defined, we have the following cases:
    \begin{itemize}
      
      \item $(\emptyset,\emptyset)$.
        We have that $\fsubst{\ptypeF[i]}{\sigma} = \ptypeF[i]$ for $i \in \setenum{0,1}$,
        hence $B = A$.
        We have that:
        \begin{align*}
          \hspace{40pt}
          \fsubst{(\expand{\ptypeF[i]}{A})}{\sigma}
          & \; = \; \expand{\ptypeF[i]}{A}
          && (\dom \sigma \cap \dom{\expand{\ptypeF[i]}{A}} = \dom \sigma \cap A = \emptyset)
          \\
          & \; = \; \expand{(\fsubst{\ptypeF[i]}{\sigma})}{A}
          \\
          & \; = \; \expand{(\fsubst{\ptypeF[i]}{\sigma})}{B}
        \end{align*}
        
      \item $(\setenum{x},\emptyset)$.
        By~\Cref{def:types:fsubst}
        we have $B = (A \setminus \setenum{x}) \cup \setenum{s}$.
        For $i = 0$, we have:
        \begin{align*}
          \hspace{40pt}
          \fsubst{(\expand{\ptypeF[0]}{A})}{\sigma}
          & =
          \expand{\ptypeF[0]}{A} \setenum{\bind{x}{\bot}} \setenum{\bind{s}{\expand{\ptypeF[0]}{A}(x)}}
          && (\dom \sigma \cap \dom{\expand{\ptypeF[0]}{A}} = \setenum{x})
          \\
          & = 
          \expand{\ptypeF[0]}{A} \setenum{\bind{x}{\bot}} \setenum{\bind{s}{\ptypeF[0](x)}}
          && (x \in \dom{\ptypeF[0]})
          \\
          & = 
          \expand{(\ptypeF[0] \setenum{\bind{s}{\ptypeF[0](x)}})}{A} \setenum{\bind{x}{\bot}}
          && (\eqref{lem:types:substitution:expand:ctype})
          \\
          & = 
          \expand{(\ptypeF[0] \setenum{\bind{s}{\ptypeF[0](x)}} \setenum{\bind{x}{\bot}})}{A \setminus \setenum{x}}
          && (\eqref{lem:types:substitution:expand:bot})
          \\
          & = 
          \expand{(\ptypeF[0] \setenum{\bind{s}{\ptypeF[0](x)}} \setenum{\bind{x}{\bot}})}{(A \setminus \setenum{x}) \cup \setenum{s}}
          && (s \in \dom{\ptypeF[0] \setenum{\bind{s}{\ptypeF[0](x)}} \cdots})
          \\
          & = \expand{(\fsubst{\ptypeF[0]}{\sigma})}{B}
          && (B = (A \setminus \setenum{x}) \cup \setenum{s})
          \intertext{For $i = 1$, we have:}
          \fsubst{(\expand{\ptypeF[1]}{A})}{\sigma}
          & = 
          \expand{\ptypeF[1]}{A} \setenum{\bind{x}{\bot}} \setenum{\bind{s}{\expand{\ptypeF[1]}{A}(x)}}
          && (\dom \sigma \cap \dom{\expand{\ptypeF[1]}{A}} = \setenum{x})
          \\
          & = 
          \expand{\ptypeF[1]}{A} \setenum{\bind{x}{\bot}} \setenum{\bind{s}{\ptypeF[1](\effwildcard)}}
          && (x \not\in \dom{\ptypeF[1]})
          \\
          & = 
          \expand{\ptypeF[1]}{A \setminus \setenum{x}} \setenum{\bind{s}{\ptypeF[1](\effwildcard)}}
          && (x \not\in \dom{\ptypeF[1]})
          \\
          & = 
          \expand{\ptypeF[1]}{(A \setminus \setenum{x}) \cup \setenum{s}}
          && (s \not\in \dom{\ptypeF[1]})
          \\
          & =
          \expand{(\fsubst{\ptypeF[1]}{\sigma})}{B}
          && (B = (A \setminus \setenum{x}) \cup \setenum{s})
        \end{align*}

      \item $(\emptyset,\setenum{x})$. Symmetrical to the previous case.

      \item $(\setenum{x},\setenum{x})$.
        For $i = 0$, the proof is analogous to the corresponding subcase $i = 0$
        in the case $(\setenum{x},\emptyset)$.
        For $i = 1$, the proof is symmetric.
    
      \item $(\setenum{x},\setenum{y})$ with $x \neq y$.
        This case does not apply, because otherwise
        $\fsubst{\ptypeF}{\sigma}$ would be undefined.

    \end{itemize}

  \item \nrule{[T-Sum]}.
    Similar to the previous case.

  \item \nrule{[T-Del]}. We have:
    \[
    \inference[\smallnrule{[T-Del]}]
    {\effentails \procPi \colon \ptypeFi
      & \ptypeFi(u)\; \mbox{honest}}
    {\effentails \procP = (u)\procPi \colon
      \ptypeFi\setenum{\bind u {\bot}} = \ptypeF}
    \]
    Since $\alpha$-conversion does not change typing,
    we can assume that $s \neq u$.
    Now, let $\vec{z} = \dom \sigma \cap \dom {\ptypeF}$.
    We proceed by cases on the possible values of $\vec{z}$.
    Since $\fsubst{\ptypeF}{\sigma}$ is defined, we only have the following three cases:
    \begin{itemize}
      
    \item $\vec{z} = \emptyset$.
      By the induction hypothesis (with substitution $\sigma_{\neq u}$), we have:
      \[
      \effentails \procPi\sigma_{\neq u} \colon 
      \fsubst{\ptypeFi}{\sigma_{\neq u}}
      =
      \ptypeFi
      \]
      Hence, the thesis follows by rule~\nrule{[T-Del]}.
      
    \item $\vec{z} = \setenum{u}$. 
      This case does not apply, because $u \not\in \dom{\ptypeF}$.

    \item $\vec{z} = \setenum{x}$, with $x \neq u$. 
      By the induction hypothesis (with substitution $\sigma_{\neq u}$):
      \[
      \effentails \procPi\sigma_{\neq u} \colon 
      \fsubst{\ptypeFi}{\sigma_{\neq u}}
      =
      \ptypeFi \setenum{\bind{x}{\bot}} \setenum{\bind{s}{\ptypeFi(x)}}
      \]
      Let $\ptypeG = \ptypeFi \setenum{\bind{x}{\bot}} \setenum{\bind{s}{\ptypeFi(x)}}$.
      Since $\ptypeFi(u)$ is honest, then 
      $\ptypeG(u) = \ptypeFi(u)$ is honest as well.
      Then, by rule~\nrule{[T-Del]}:
      \[
      \inference[\smallnrule{[T-Del]}]
      {\effentails \procPi\sigma_{\neq u} \colon \ptypeG
        & \ptypeG(u)\; \mbox{honest}}
      {\effentails \procP\sigma = (u)\procPi\sigma_{\neq u} \colon
        \ptypeG\setenum{\bind u {\bot}}}
      \]
      The thesis follows because:
      \begin{align*}
        \hspace{60pt}
        \ptypeG \setenum{\bind u {\bot}} 
        & = \ptypeFi \setenum{\bind{x}{\bot}} \setenum{\bind{s}{\ptypeFi(x)}} \setenum{\bind u {\bot}} 
        && \text{(by def.\ of $\ptypeG$)}
        \\
        & = \ptypeFi \setenum{\bind{u}{\bot}}  \setenum{\bind x {\bot}} \setenum{\bind{s}{(\ptypeFi\setenum{\bind{u}{\bot}})(x)}}
        \\
        & = \fsubst{\ptypeFi{\setenum{\bind{u}{\bot}}}}{\sigma}
        && \text{(by~\Cref{def:types:fsubst})}
        \\
        & = \fsubst{\ptypeF}{\sigma}
        && \text{(by def.\ of $\ptypeF$)}
      \end{align*}
    \end{itemize}
    
  \item \nrule{[T-Rec]}. We have:
    \[
    \inference[\smallnrule{[T-Rec]}]
    { \effentails \procPi \colon \ptypeF}
    { \effentails \procP = (\procRec{\procX()}\procPi)() 
      \colon \lambda u.\ \effrec{\consttovar{\procX}} \ptypeF(u)}
    \]
    By the induction hypothesis we have
    $\effentails \procPi\sigma \colon \fsubst{\ptypeF}{\sigma}$.
    Then, by rule~\nrule{[T-Rec]}:
    \[
    \inference[\smallnrule{[T-Rec]}]
    { \effentails \procPi\sigma \colon \fsubst{\ptypeF}{\sigma} }
    { \effentails (\procRec{\procX()}\procPi\sigma)() 
      \colon \lambda u.\ \effrec{\consttovar{\procX}}{
        (\fsubst{\ptypeF}{\sigma})(u)}
    }
    \]
    The thesis is obtained from
    $(\procRec{\procX()}\procPi\sigma)() = 
    (\procRec{\procX()}\procPi)()\sigma$ and
    \[
    \lambda u.\ \effrec{\consttovar{\procX}}{
      (\fsubst{\ptypeF}{\sigma})(u) }
    =
    \fsubst{(\lambda u.\ \effrec{\consttovar{\procX}}{\ptypeF(u)})}{\sigma}
    \]
    which follows by case analysis on the argument: $x,s$ or something else.
  \end{itemize}
  
  \medskip\noindent
  We now prove item~\eqref{lem:types:typing-renaming-tscompat}
  (which is needed in order to prove item~\eqref{lem:types:typing-renaming-colon}).
  By induction on the typing derivation of $\tsentails{A}\sysS \tscompat \ptypeF$,
  we have the following cases, according to the last rule used in the typing derivation:
  \begin{itemize}

  \item \nrule{[T-SAFree0]}, \nrule{[T-SAFree1]}, \nrule{[T-SFzS]}.
    Trivial: the premise is not affected by $\sigma$.

  \item \nrule{[T-SAFree2]}.
    We have the following two cases:
    \begin{itemize}
      
      \item $x \in \dom{\sigma}$. 
        Then, $\sysS\sigma = \latent[\pmvB]{s}{\contrP}$,
        and so the thesis follows by rule~\nrule{[T-SFsZ]}.

      \item $x \not\in \dom{\sigma}$.
        Then, $\sysS\sigma = \sysS$, and
        since $\ptypeF(x) = \bot$, then $(\fsubst{\ptypeF}{\sigma})(x) = \bot$.
        Hence, the thesis follows by rule~\nrule{[T-SAFree2]}.

    \end{itemize}
    
  \item \nrule{[T-SAFree3]}. We have that:
    \[
    \inference[\smallnrule{[T-SAFree3]}]
    {\sys {s'} \gamma \text{ {\pmv A}-free} & \ptypeF(s') = \bot}
    {\tsentails{A} \sysS = {\sys {s'} \gamma} \tscompat \ptypeF}    
    \]
    Since $s \not\in \fnv{\sysS}$ by hypothesis, then it must be $s' \neq s$.
    The thesis follows because $(\fsubst{\ptypeF}{\sigma})(s') = \ptypeF(s')$.

  \item \nrule{[T-SFz1]}. We have:
    \[
    \inference[\smallnrule{[T-SFz1]}]
    {\expand{\ptypeF}{\setenum{x}}\!(x)\ \mbox{realizes $\contrP$}}
    {\tsentails{A} \sysS = {\setenum{\freeze x {\contrP}}_{\pmv A}} \tscompat \ptypeF}
    \]
    We have the following two cases:
    \begin{itemize}

      \item $x \in \dom{\sigma}$.
        Then, $\sysS\sigma = \latent[\pmvA]{s}{\contrP}$,
        and so the thesis follows by rule~\nrule{[T-SFsZ]}.

      \item $x \not\in \dom{\sigma}$.
        Then, 
        $\expand{(\fsubst{\ptypeF}{\sigma})}{\setenum{x}}\!(x) = \expand{\ptypeF}{\setenum{x}}\!(x)$,
        which realizes $\contrP$ by the premise.
        Hence, the thesis follows by rule~\nrule{[T-SFz1]}.

    \end{itemize}

  \item \nrule{[T-SFuse]}: 
    \[
    \inference[\smallnrule{[T-SFuse]}]
    {\expand{\ptypeF}{\setenum{s'}}(s')\ \mbox{realizes $\cta{\gamma}$}}
    {\tsentails{A} \sysS = {\sys {s'} {\gamma}} \tscompat \ptypeF}
    \]
    Since $s \not\in \fnv{\sysS}$ by hypothesis, then it must be $s' \neq s$.
    The thesis follows because $(\fsubst{\ptypeF}{\sigma})(s') = \ptypeF(s')$,
    which preserves the truth of the premise.

  \item \nrule{[T-SDel1]}. We have:
    \[
    \inference[\smallnrule{[T-SDel1]}]
    {\tsentails{A} \sysS[0] \tscompat \ptypeF\setenum{\bind{u}{\bot}}}
    {\tsentails{A} (u)\sysS[0] \tscompat \ptypeF}    
    \]
    W.l.o.g.\ we can assume $u \neq s$: otherwise we can $\alpha$-convert
    the typing derivation and obtain the same typing.
    Now, let $\vec{z} = \dom \sigma \cap \dom {\ptypeF}$.
    We proceed by cases on the possible values of $\vec{z}$.
    Since $\fsubst{\ptypeF}{\sigma}$ is defined, we only have the following three cases:
    \begin{itemize}
      
    \item $\vec{z} = \emptyset$.
      By the induction hypothesis (with substitution $\sigma$), we have:
      \[
      \tsentails{A} \sysS[0]\sigma
      \tscompat \fsubst{\ptypeF\setenum{\bind{u}{\bot}}}{\sigma}
      \]
      By~\Cref{def:types:fsubst} we have that:
      \begin{align*}
        \fsubst{\ptypeF\setenum{\bind{u}{\bot}}}{\sigma}
        & =
        \ptypeF \setenum{\bind{u}{\bot}}
        = 
        (\fsubst{\ptypeF}{\sigma}) \setenum{\bind{u}{\bot}}
      \end{align*}
      Hence, the thesis follows by rule~\nrule{[T-SDel1]}.
    
    \item $\vec{z} = \setenum{u}$. 
      Since $s \not\in \fnv{\sysS[0]}$,
      by applying~\Cref{lem:types:f-subst-notin} we obtain:
      \[
      \hspace{40pt}
      \tsentails{A} \sysS[0] \tscompat 
      \ptypeF \setenum{\bind{u}{\bot}} \setenum{\bind{s}{\ptypeF(u)}}
      \]
      with a typing derivation of the same depth as the original one.
      By the induction hypothesis (with substitution $\sigma_{\neq u}$), we have:
      \[
      \tsentails{A} \sysS[0]\sigma_{\neq u}
      \tscompat \fsubst{(\ptypeF \setenum{\bind{u}{\bot}} \setenum{\bind{s}{\ptypeF(u)}})}{\sigma_{\neq u}}
      \]
      By~\Cref{def:types:fsubst} we have that:
      \begin{align*}
        \fsubst{(\ptypeF \setenum{\bind{u}{\bot}} \setenum{\bind{s}{\ptypeF(u)}})}{\sigma_{\neq u}}
        & \; = \;
        \ptypeF \setenum{\bind{u}{\bot}} \setenum{\bind{s}{\ptypeF(u)}}
        \\
        & \; = \;
        \ptypeF \setenum{\bind{s}{\ptypeF(u)}} \setenum{\bind{u}{\bot}} 
      \end{align*}
      Then, by rule~\nrule{[T-SDel1]}:
      \[
      \inference[\smallnrule{[T-SDel1]}]
      {\tsentails{A} \sysS[0]\sigma_{\neq u} \tscompat \ptypeF \setenum{\bind{s}{\ptypeF(u)}} \setenum{\bind{u}{\bot}}}
      {\tsentails{A} \sysS\sigma = (u) \sysS[0]\sigma_{\neq u} \tscompat \ptypeF \setenum{\bind{s}{\ptypeF(u)}}}
      \]
      Since $u \not\in \fnv{\sysS\sigma}$,
      by applying~\Cref{lem:types:f-subst-notin} again we conclude that:
      \[
      \tsentails{A} \sysS\sigma \tscompat 
      \ptypeF \setenum{\bind{s}{\ptypeF(u)}} \setenum{\bind{u}{\bot}} 
      \; = \;
      \fsubst{\ptypeF}{\sigma}
      \]

    \item $\vec{z} = \setenum{x}$, with $x \neq u$. 
      By applying the induction hypothesis on the premise, we have:
      $
      \tsentails{A} \sysS[0]\sigma
      \tscompat \fsubst{\ptypeF\setenum{\bind{u}{\bot}}}{\sigma}
      $.
      We have that:
      \begin{align*}
        \fsubst{\ptypeF\setenum{\bind{u}{\bot}}}{\sigma}
        & =
        \ptypeF \setenum{\bind{u}{\bot}} \setenum{\bind{x}{\bot}} \setenum{\bind{s}{\ptypeF\setenum{\bind{u}{\bot}}(x)}}
        \\
        & =
        \ptypeF \setenum{\bind{x}{\bot}} \setenum{\bind{s}{\ptypeF(x)}} \setenum{\bind{u}{\bot}} 
        \\
        & =
        (\fsubst{\ptypeF}{\sigma}) \setenum{\bind{u}{\bot}}
      \end{align*}
      Then, by rule~\nrule{[T-SDel1]} we conclude:
      \[
      \inference[\smallnrule{[T-SDel1]}]
      {\tsentails{A} \sysS[0]\sigma \tscompat (\fsubst{\ptypeF}{\sigma}) \setenum{\bind{u}{\bot}}}
      {\tsentails{A} \sysS\sigma = (u) \sysS[0]\sigma \tscompat \fsubst{\ptypeF}{\sigma}}
      \]

    \end{itemize}
    
  \item \nrule{[T-SPar1]}. 
    Straightforward, by applying the induction hypothesis on both premises.

  \end{itemize}
  
  \medskip\noindent
  To prove item~\eqref{lem:types:typing-renaming-colon},
  we proceed by  induction on the typing derivation of 
  $\tsentails{A}\sysS \colon \ptypeF$.
  We have the following cases, according to the last rule used
  in the typing derivation:
  \begin{itemize}

  \item \nrule{[T-SA]}. 
    We have:
    \[
    \inference[\smallnrule{[T-SA]}]
    {\effentails \procP \colon \ptypeF}
    {\tsentails{A} \sysS = {\sys {\pmv A} {\procP}} \colon \ptypeF}    
    \]
    By applying~\Cref{lem:types:substitution} on the rule premise, 
    we obtain $\effentails \procP\sigma \colon \fsubst{\ptypeF}{\sigma}$.
    The thesis follows by rule~\nrule{[T-SA]}.

  \item \nrule{[T-SDel2]}. We have:
    \[
    \inference[\smallnrule{[T-SDel2]}]
    {\tsentails{A} \sysS[0] \colon \ptypeF[0] & \expand{\ptypeF[0]}{\setenum{u}}\!(u)\ \mbox{honest}}
    {\tsentails{A} \sysS = (u)\sysS[0] \colon \ptypeF[0]\setenum{\bind{u}{\bot}} = \ptypeF}
    \]
    W.l.o.g.\ we can assume $u \neq s$: otherwise we can $\alpha$-convert
    the typing derivation and obtain the same typing.
    Since $u \neq s$, then $s \not\in \fnv{\sysS[0]}$,
    hence by~\Cref{lem:types:dom-fnv} $s \not\in \dom{\ptypeF[0]}$.
    Since $\fsubst{\ptypeF}{\sigma}$ is defined and
    $\dom{\ptypeF[0]} \cap \dom{\sigma_{\neq u}} \subseteq \dom{\ptypeF} \cap \dom{\sigma}$,
    then $\fsubst{\ptypeF[0]}{\sigma_{\neq u}}$ is defined.
    Then, by the induction hypothesis we obtain
    $\effentails \sysS[0] \, \sigma_{\neq u} \colon \fsubst{\ptypeF[0]}{\sigma_{\neq u}}$.
    To prove that 
    $\expand{\ptypeF[0]}{\setenum{u}}\!(u)$ honest
    implies $\expand{(\fsubst{\ptypeF[0]}{\sigma_{\neq u}})}{\setenum{u}}\!(u)$ honest,
    we consider the following two cases:
    \begin{itemize}

    \item $u \in \dom{\ptypeF[0]}$. Then:
      \[
      \expand{(\fsubst{\ptypeF[0]}{\sigma_{\neq u}})}{\setenum{u}}\!(u)
      =
      \expand{\ptypeF[0]}{\setenum{u}}\!(u)
      =
      \ptypeF[0](u)
      \quad
      \text{honest}
      \]

    \item $u \not\in \dom{\ptypeF[0]}$. Then: 
      \[
      \expand{(\fsubst{\ptypeF[0]}{\sigma_{\neq u}})}{\setenum{u}}\!(u)
      =
      \expand{\ptypeF[0]}{\setenum{u}}\!(u)
      =
      \ptypeF[0](\effwildcard)
      \quad
      \text{honest}
      \]

    \end{itemize}
    Then we can construct the following typing derivation:
    \[
    \inference[\smallnrule{[T-SDel2]}]
    {\tsentails{A} \sysS[0] \, \sigma_{\neq u} \colon \fsubst{\ptypeF[0]}{\sigma_{\neq u}}
      & 
      \expand{(\fsubst{\ptypeF[0]}{\sigma_{\neq u}})}{\setenum{u}}\!(u)\ \mbox{honest}}
    {\tsentails{A} \sysS \, \sigma_{\neq u} = (u)(\sysS[0] \, \sigma_{\neq u}) \colon 
      (\fsubst{\ptypeF[0]}{\sigma_{\neq u}}) \setenum{\bind{u}{\bot}}}
    \]
    Now, let $\vec{z} = \dom \sigma \cap \dom {\ptypeF}$.
    We proceed by cases on the possible values of $\vec{z}$.
    Since $\fsubst{\ptypeF}{\sigma}$ is defined, we only have the following three cases:
    \begin{itemize}
      
    \item $\vec{z} = \emptyset$.
      We have that:
      \begin{align*}
        (\fsubst{\ptypeF[0]}{\sigma_{\neq u}})\setenum{\bind{u}{\bot}}
        & =
        \ptypeF[0] \setenum{\bind{u}{\bot}}
        = 
        \ptypeF
        =           
        \fsubst{\ptypeF}{\sigma}
      \end{align*}
      
    \item $\vec{z} = \setenum{u}$. 
      This case does not apply, because $u \not\in \dom{\ptypeF}$.

    \item $\vec{z} = \setenum{x}$, with $x \neq u$. 
      We have that:
      \begin{align*}
        \hspace{40pt}
        (\fsubst{\ptypeF[0]}{\sigma_{\neq u}})\setenum{\bind{u}{\bot}}
        & =
        \ptypeF[0] \setenum{\bind{x}{\bot}} \setenum{\bind{s}{\ptypeF[0](x)}} \setenum{\bind{u}{\bot}}
        && \text{($\dom{\sigma_{\neq u}} \cap \dom{\ptypeF[0] = \setenum{x}}$)}
        \\
        & = 
        \ptypeF[0] \setenum{\bind{x}{\bot}} \setenum{\bind{s}{\ptypeF(x)}} \setenum{\bind{u}{\bot}}
        && \text{($\ptypeF[0](x) = \ptypeF(x)$)}
        \\
        & = 
        \ptypeF \setenum{\bind{x}{\bot}} \setenum{\bind{s}{\ptypeF(x)}} 
        && \text{($\ptypeF = \ptypeF[0] \setenum{\bind{u}{\bot}}$)}
        \\
        & = 
        \fsubst{\ptypeF}{\sigma}
        && \text{($\dom{\sigma} \cap \dom{\ptypeF} = \setenum{x}$)}
      \end{align*}

    \end{itemize}
    
  \item \nrule{[T-SPar2]}. We have:
    \[
    \inference[\smallnrule{[T-SPar2]}]
    {\tsentails{A} \sysS[0] \colon \ptypeF \quad 
      \tsentails{A} \sysS[1] \tscompat \ptypeF}
    {\tsentails{A} \sysS = \sysS[0] \cocoPar \sysS[1] \colon \ptypeF}
    \]
    By the induction hypothesis of item~\eqref{lem:types:typing-renaming-colon},
    we have $\tsentails{A} \sysS[0]\sigma \colon \fsubst{\ptypeF}{\sigma}$.
    By item~\eqref{lem:types:typing-renaming-tscompat},
    we have $\tsentails{A} \sysS[1]\sigma \tscompat \fsubst{\ptypeF}{\sigma}$.
    Then, by applying rule~\nrule{[T-SPar2]}:
    \[
    \inference[\smallnrule{[T-SPar2]}]
    {\tsentails{A} \sysS[0]\sigma \colon \fsubst{\ptypeF}{\sigma} \quad 
      \tsentails{A} \sysS[1]\sigma \tscompat \fsubst{\ptypeF}{\sigma}}
    {\tsentails{A} \sysS \sigma = \sysS[0]\sigma \cocoPar \sysS[1]\sigma \colon \fsubst{\ptypeF}{\sigma}}
    \]

  \end{itemize}
\end{proofoflem}


\section{Proof of \Cref{th:types:subject-reduction} (Subject Reduction)}
\label{proof:th:types:subject-reduction}

Assume that $\sysS \sysmove{B}{\pref}{} \sysSi$, for some $\pmvB$.
We proceed by induction on the proof of the reduction 
from $\sysS$ to $\sysSi$.
We have the following cases, according to the last rule used
in the reduction.

\subsection*{Rule~\nrule{[Tau]}}

We have: 
\[
\sysS 
\; = \;
\sys {\pmvB} {\tau \cocoSeq \procP \cocoPlus \procPi \cocoPar \procQ}
\sysmove{B}{\tau}{} 
{\sys {\pmvB} {\procP \cocoPar \procQ}}
\; = \;
\sysSi
\]
\begin{itemize}
\item Item~\eqref{th:types:subject-reduction-A-colon}.
  Let $\procP[0] = \tau \cocoSeq \procP \cocoPlus \procPi \cocoPar \procQ$.
  By hypothesis we have that $\pmv B = \pmv A$, and
  $\tsentails{A} \sysS \colon \ptypeF$.
  The only typing rule applicable for $\sysS$ is \nrule{[T-SA]},
  which requires $\tsentails{A} \procP[0] \colon \ptypeF$.
  Let $A = \dom \ptypeF = \fnv{\procP[0]} \cup \setenum{\effwildcard}$, 
  as implied by~\Cref{lem:types:dom-fnv}. 
  We have:
  \[
  \tsentails{A} \procP[0] \colon \ptypeF 
  \; = \;
  \lambda u \dotseq [\tau]_u \dotseq \expand{\ptypeF^{\procP}}{A}(u) \cocoPlus \expand{\ptypeF^{\procPi}}{A}(u) \cocoPar \expand{\ptypeF^{\procQ}}{A}(u)
  \]
  where $\ptypeF^{\procP}$, $\ptypeF^{\procPi}$ and $\ptypeF^{\procQ}$ 
  are, respectively, the types of $\procP$, $\procPi$ and $\procQ$. %
  Let $A' = \fnv{\procP} \cup \fnv{\procQ}$, and let:
  \[
  \ptypeFi \; = \; \lambda u \dotseq \expand{\ptypeF^{\procP}}{A'}(u) \mid \expand{\ptypeF^{\procQ}}{A'}(u)
  \]

  We now prove that $\ptypeF \effmove{\tau} \ptypeFi$, 
  according to~\Cref{def:types:trans}.
  There are the following two cases:
  \begin{itemize}
  \item $v \in \dom \ptypeFi$. 
    We must show that $\ptypeF(v) \effmove{\tau} \ptypeFi(v)$.
    We have that:
    \[
    \hspace{60pt}
    \small
    \inference[\smallnrule{[C-ParL]}]
    {\inference[\smallnrule{[C-SumL]}]
      {\inference[\smallnrule{[C-Pref]}]
        {}
        {\tau \effseq \expand{\ptypeF^{\procP}}{A}(v) \effmove{\tau} \expand{\ptypeF^{\procP}}{A}(v)}}
      {\tau \effseq \expand{\ptypeF^{\procP}}{A}(v) + \expand{\ptypeF^{\procPi}}{A}(v) \effmove{\tau} \expand{\ptypeF^{\procP}}{A}(v)}}
    {\tau \effseq \expand{\ptypeF^{\procP}}{A}(v) + \expand{\ptypeF^{\procPi}}{A}(v) \mid \expand{\ptypeF^{\procQ}}{A}(v)
      \effmove{\tau} \expand{\ptypeF^{\procP}}{A}(v) \mid \expand{\ptypeF^{\procQ}}{A}(v)}
    \]
    and since $\dom \ptypeFi \subseteq A$, then
    $\ptypeFi(v) = \expand{\ptypeF^{\procP}}{A}(v) \mid \expand{\ptypeF^{\procQ}}{A}(v)$.
    Hence, we conclude that $\ptypeF(v) \effmove{\tau} \ptypeFi(v)$.

  \item $v \in \dom \ptypeF \setminus \dom \ptypeFi$. 
    We must show that $\ptypeF(v) \effmove{\tau} \ptypeFi(\effwildcard)$.
    We have that $v \not\in \fnv{\procP} \cup \fnv{\procQ}$, thus 
    $\expand{\ptypeF^{\procP}}{A}(v) \mid \expand{\ptypeF^{\procQ}}{A}(v) = 
    \ptypeF^{\procP}(\effwildcard) \mid \ptypeF^{\procQ}(\effwildcard) = \ptypeFi(\effwildcard)$.
    The thesis follows by using the same derivation of the previous case.

  \end{itemize}
  
  \medskip
  We conclude the proof by showing that $\tsentails{A} \sysSi \colon \ptypeFi$:
  \[
  \inference[\smallnrule{[T-SA]}]
  {\inference[\smallnrule{[T-Par]}]
    {\effentails \procP \colon \ptypeF^{\procP}
      & \effentails \procQ \colon \ptypeF^{\procQ}}
    {\effentails \procP \cocoPar \procQ \colon
      \lambda u \dotseq \expand{\ptypeF^{\procP}}{A'}(u) \mid \expand{\ptypeF^{\procQ}}{A'}(u)
      = \ptypeFi}}
  {\tsentails{A} \sysSi \colon \ptypeFi}
  \]

\item Item~\eqref{th:types:subject-reduction-B-colon}.
  The thesis holds vacuously,
  since, for any $\ptypeF$, it is not possible to have a typing
  $\tsentails{A} \sys {\pmvB} {\cdots} \colon \ptypeF$ 
  when $\pmvB \neq \pmvA$
  (see~\Cref{lem:types:a-process-presence-typing}).
  
\item Item~\eqref{th:types:subject-reduction-B-tscompat}.
  The typing $\tsentails{A} {\sysS} \tscompat \ptypeF$
  must have been obtained by rule \nrule{[T-SAFree1]}, %
  whose premise requires that
  $\fv{\tau \cocoSeq \procP \cocoPlus \procPi \cocoPar \procQ}$.
  With the same rule we can derive
  $\tsentails{A} \sysSi \tscompat \ptypeF$, since
  $
  \fv{\procP \cocoPar \procQ}
  \subseteq
  \fv{\tau \cocoSeq \procP \cocoPlus \procPi \cocoPar \procQ}
  $.

\end{itemize}

\subsection*{Rule~\nrule{[Tell]}}

We have:
\[
\sysS 
\; = \;
{\sys {\pmvB} {{\tell {} {\freeze w \contrP}} \cocoSeq \procP \cocoPlus \procPi \cocoPar \procQ}} 
\sysmove{B}{\tell {} {\freeze w \contrP}}{} 
{\sys {\pmvB} {\procP \cocoPar \procQ}} \cocoPar
\setenum{\freeze w \contrP}_{\pmvB} 
\; = \; 
\sysSi
\]

\begin{itemize}
\item Item~\eqref{th:types:subject-reduction-A-colon}.
  Let $\procP[0] = {\tell {} {\freeze w \contrP}} \cocoSeq \procP \cocoPlus \procPi \cocoPar \procQ$.
  By hypothesis we have that $\pmv B = \pmv A$, and
  $\tsentails{A} \sysS \colon \ptypeF$.
  The only typing rule applicable for $\sysS$ is \nrule{[T-SA]},
  which requires $\tsentails{A} \procP[0] \colon \ptypeF$.
  Let $A = \dom \ptypeF = \fnv{\procP[0]} \cup \setenum{\effwildcard}$, 
  as implied by~\Cref{lem:types:dom-fnv}. 
  We have:
  \[
  \tsentails{A} \procP[0] \colon \ptypeF 
  \; = \;
  \lambda u \dotseq [{\tell {} {\freeze w \contrP}}]_u \dotseq \expand{\ptypeF^{\procP}}{A}(u) \cocoPlus \expand{\ptypeF^{\procPi}}{A}(u) \cocoPar \expand{\ptypeF^{\procQ}}{A}(u)
  \]
  where $\ptypeF^{\procP}$, $\ptypeF^{\procPi}$ and $\ptypeF^{\procQ}$ 
  are, respectively, the types of $\procP$, $\procPi$ and $\procQ$. %
  Also, by~\Cref{def:types:typing-prefixes} 
  we have that 
  $[\tell {} {\freeze w \contrP}]_u = \wif{w=u}{\effcontract{\contrP}}{\tau}$.
  Let $A' = \fnv{\procP} \cup \fnv{\procQ}$, and let:
  \[
  \ptypeFi \; = \; \lambda u \dotseq \expand{\ptypeF^{\procP}}{A'}(u) \mid \expand{\ptypeF^{\procQ}}{A'}(u)
  \]

  We now prove that $\ptypeF \effmove{\tell {} {\freeze w \contrP}} \ptypeFi$, 
  according to~\Cref{def:types:trans}.
  There are the following two cases:
  \begin{itemize}
  \item $v \in \dom \ptypeFi$. 
    We must show that $\ptypeF(v) \effmove{[\tell {} {\freeze w \contrP}]_v} \ptypeFi(v)$.
    We have that:
    \[
    \hspace{60pt}
    \small
    \inference[\smallnrule{[C-ParL]}]
    {\inference[\smallnrule{[C-SumL]}]
      {\inference[\smallnrule{[C-Pref]}]
        {}
        {[\tell {} {\freeze w \contrP}]_v \effseq \expand{\ptypeF^{\procP}}{A}(v) \effmove{[\tell {} {\freeze w \contrP}]_v} \expand{\ptypeF^{\procP}}{A}(v)}}
      {[\tell {} {\freeze w \contrP}]_v \effseq \expand{\ptypeF^{\procP}}{A}(v) + \expand{\ptypeF^{\procPi}}{A}(v) \effmove{[\tell {} {\freeze w \contrP}]_v} \expand{\ptypeF^{\procP}}{A}(v)}}
    {[\tell {} {\freeze w \contrP}]_v \effseq \expand{\ptypeF^{\procP}}{A}(v) + \expand{\ptypeF^{\procPi}}{A}(v) \mid \expand{\ptypeF^{\procQ}}{A}(v)
      \effmove{[\tell {} {\freeze w \contrP}]_v} \expand{\ptypeF^{\procP}}{A}(v) \mid \expand{\ptypeF^{\procQ}}{A}(v)}
    \]
    and since $\dom \ptypeFi \subseteq A$, then
    $\ptypeFi(v) = \expand{\ptypeF^{\procP}}{A}(v) \mid \expand{\ptypeF^{\procQ}}{A}(v)$.
    Hence, we conclude that $\ptypeF(v) \effmove{[\tell {} {\freeze w \contrP}]_v} \ptypeFi(v)$.

  \item $v \in \dom \ptypeF \setminus \dom \ptypeFi$. 
    We must show that $\ptypeF(v) \effmove{[\tell {} {\freeze w \contrP}]_v} \ptypeFi(\effwildcard)$.
    We have that $v \not\in \fnv{\procP} \cup \fnv{\procQ}$, thus 
    $\expand{\ptypeF^{\procP}}{A}(v) \mid \expand{\ptypeF^{\procQ}}{A}(v) = 
    \ptypeF^{\procP}(\effwildcard) \mid \ptypeF^{\procQ}(\effwildcard) = \ptypeFi(\effwildcard)$.
    The thesis follows by using the same derivation of the previous case.

  \end{itemize}
  
  \medskip
  We conclude the proof by showing that $\tsentails{A} \sysSi \colon \ptypeFi$.
  We start by proving that $\expand{\ptypeFi}{\setenum{w}}\!(w)$ realizes $\contrP$.
  To do that, we use the assumption of~\Cref{th:types:subject-reduction} 
  which says that $\ptypeF$ is honest.
  By~\Cref{def:abs-honesty} and~\Cref{def:types:f}, 
  since $w \in \fnv{\procP[0]} = \dom \ptypeF$,
  this implies that  $(\emptyset,\ptypeF(w))$ is honest.
  Let $\ctypeT = \expand{\ptypeF^{\procP}}{A}(w) \cocoPar \expand{\ptypeF^{\procQ}}{A}(w)$.
  We have that:
  \[
  \hspace{40pt}
  (\emptyset,\ptypeF(w))        
  \;\ecmove\;   
  (\setenum{\contrP},\ctypeT) 
  \;\ecmove\;   
  (\contrP,\ctypeT) 
  \]
  By~\Cref{def:abs-honesty} we know that $\ctypeT$
  realizes $\contrP$. We are left with proving that 
  $\expand{\ptypeFi}{\setenum{w}}\!(w) = \ctypeT$. For this
  we consider two cases, depending on whether $w \in A'$.

  If $w \in A' = \dom{\ptypeFi}$, then we have
  $\expand{\ptypeFi}{\setenum{w}}(w) =
  \ptypeFi(w) =
  \expand{\ptypeF^{\procP}}{A'}(w) \mid \expand{\ptypeF^{\procQ}}{A'}(w)$.
  From this, and $A' \subseteq A$, 
  we obtain
  $\expand{\ptypeF^{\procP}}{A'}(w) \mid \expand{\ptypeF^{\procQ}}{A'}(w)
  =\expand{\ptypeF^{\procP}}{A}(w) \cocoPar \expand{\ptypeF^{\procQ}}{A}(w)
  = \ctypeT$, hence $\expand{\ptypeFi}{\setenum{w}}\!(w)$ 
  realizes $\contrP$.

  Otherwise, if $w \not \in A' = \dom{\ptypeFi} = 
  \fnv{\procP \cocoPar \procQ}$. Here we have
  \begin{align*}
    \expand{\ptypeFi}{\setenum{w}}(w) & =
                                        \ptypeFi(\effwildcard) =
                                        \expand{\ptypeF^{\procP}}{A'}(\effwildcard) \mid \expand{\ptypeF^{\procQ}}{A'}(\effwildcard) 
    \\
                                      & = \expand{\ptypeF^{\procP}}{A}(\effwildcard) \mid \expand{\ptypeF^{\procQ}}{A}(\effwildcard) =
                                        \expand{\ptypeF^{\procP}}{A}(w) \mid \expand{\ptypeF^{\procQ}}{A}(w) = \ctypeT
  \end{align*}
  Therefore, $\expand{\ptypeFi}{\setenum{w}}\!(w)$ realizes $\contrP$,
  and we have the following typing derivation:
  \[
  \small
  \inference[\smallnrule{[T-SPar2]}]
  {
    \inference[\smallnrule{[T-SA]}]
    {\inference[\smallnrule{[T-Par]}]
      {\effentails \procP \colon \ptypeF^{\procP}
        & \effentails \procQ \colon \ptypeF^{\procQ}}
      {\effentails \procP \cocoPar \procQ \colon
        \lambda u \dotseq \expand{\ptypeF^{\procP}}{A'}(u) \mid \expand{\ptypeF^{\procQ}}{A'}(u)
        = \ptypeFi}}
    {\tsentails{A} \sys{\pmvA}{\procP \cocoPar \procQ} \colon \ptypeFi}
    &
    \inference[\smallnrule{[T-SFz1]}]
    {\expand{\ptypeFi}{\setenum{w}}\!(w) \ \mbox{realizes $\contrP$}}
    {\tsentails{A} \latent{w}{\contrP} \tscompat \ptypeFi}
  }
  {\tsentails{A} \sys{\pmvA}{\procP \cocoPar \procQ} \cocoPar \latent{w}{\contrP} \colon \ptypeFi}
  \]

\item Item~\eqref{th:types:subject-reduction-B-colon}.
  The thesis holds vacuously,
  since, for any $\ptypeF$, it is not possible to have a typing
  $\tsentails{A} \sys {\pmvB} {\cdots} \colon \ptypeF$ 
  when $\pmvB \neq \pmvA$
  (see~\Cref{lem:types:a-process-presence-typing}).
  
\item Item~\eqref{th:types:subject-reduction-B-tscompat}.
  Let $\procP[0] = \tell {} {\freeze w \contrP} \cocoSeq \procP \cocoPlus \procPi \cocoPar \procQ$.
  The typing $\tsentails{A} {\sysS} \tscompat \ptypeF$
  must have been obtained by rule \nrule{[T-SAFree1]} as follows: %
  \[
  \inference[\smallnrule{[T-SAFree1]}]
  {\pmvB \neq \pmvA & \fv{\procP[0]} \cap \dom{\ptypeF} = \emptyset}
  {\tsentails{A} \sysS = {\sys {\pmvB} {\procP[0]}} \tscompat \ptypeF}
  \]
  Since $\fv{\procP \cocoPar \procQ} \subseteq \fv{\procP[0]}$,
  then $\fv{\procP \cocoPar \procQ} \cap \dom{\ptypeF} = \emptyset$.
  Further, since $w \in \fv{\procP[0]}$ 
  and $\fv{\procP[0]} \cap \dom{\ptypeF} = \emptyset$,
  then $\ptypeF(w) = \bot$.
  Then, we can construct the following typing derivation for $\sysSi$:
  \[
  \inference[\smallnrule{[T-SPar1]}]
  {\inference[\smallnrule{[T-SAFree1]}]
    {\pmvB \neq \pmvA & \fv{\procP \cocoPar \procQ} \cap \dom{\ptypeF} = \emptyset}
    {\tsentails{A} {\sys {\pmvB} {\procP \cocoPar \procQ}} \tscompat \ptypeF}
    & 
    \inference[\smallnrule{[T-SAFree2]}]
    {\pmvB \neq \pmvA & \ptypeF(w) = \bot}
    {\tsentails{A} \setenum{\freeze w \contrP}_{\pmvB} \tscompat \ptypeF}
  }
  {\tsentails{A} \sysSi \tscompat \ptypeF}
  \]

\end{itemize}

\subsection*{Rule~\nrule{[Fuse]}}

We have $\pmvB = \pmv K \neq \pmvA$, and:
\[
\inference[\smallnrule{[Fuse]}]
{\contrP \compliant \contrQ &
  \gamma = \pbic{C}{\contrP\buffer{}}{D}{\contrQ\buffer{}} &
  \sigma = \subs s {x,y} &
  s \not\in \fnv{\sysS[0]}
}
{\sysS \; = \; 
  (x,y)(\sysS[0] \cocoPar 
  \setenum{\freeze x \contrP}_{\pmv C} \cocoPar
  \setenum{\freeze y \contrQ}_{\pmv D})
  \;\sysmove{K}{\fuse{}}{} \;
  (s)(\sysS[0]\sigma \cocoPar \sys s {\gamma}) 
  \; = \; 
  \sysSi
}
\]
where $\pmv C \neq \pmv D$.
We prove the following items:
\begin{itemize}

\item Item~\eqref{th:types:subject-reduction-A-colon}
  is trivial, because $\pmvB = \pmv K \neq \pmvA$.

\item Item~\eqref{th:types:subject-reduction-B-colon}.
  The redex $\sysS$ can only be typed as follows (up-to associativity):
  \[
  \small
  \inference[\smallnrule{[T-SDel2]}]
  {
    \inference[\smallnrule{[T-SDel2]}]
    {
      \inference[\smallnrule{[T-SPar2]}]
      {\tsentails{A} \sysS[0] \colon \ptypeF[0]
        & 
        \inference[\smallnrule{[T-SPar1]}]
        {\fbox{$D_1$} & \fbox{$D_2$}}
        {
          \tsentails{A} 
          \setenum{\freeze x \contrP}_{\pmv C} \cocoPar 
          \setenum{\freeze y \contrQ}_{\pmv D}
          \tscompat \ptypeF[0]
        }
      }
      {\tsentails{A} \sysS[0] \cocoPar 
        \setenum{\freeze x \contrP}_{\pmv C} \cocoPar 
        \setenum{\freeze y \contrQ}_{\pmv D}
        \colon \ptypeF[0]}
      & \begin{array}{c} \expand{\ptypeF[0]}{\setenum{y}}\!(y) \\ \text{ honest} \end{array}
    }
    {\tsentails{A} (y)(\sysS[0] \cocoPar 
      \setenum{\freeze x \contrP}_{\pmv C} \cocoPar
      \setenum{\freeze y \contrQ}_{\pmv D})
      \colon \ptypeF[0]\setenum{\bind{y}{\bot}}}
    & \begin{array}{c} \expand{\ptypeF[0]\setenum{\bind{y}{\bot}}}{\setenum{x}}\!(x) \\ \text{ honest} \end{array}
  }
  {\tsentails{A} \sysS \colon \ptypeF[0]\setenum{\bind{y}{\bot}}\setenum{\bind{x}{\bot}} 
    = \ptypeF}
  \]
  where the typing derivations $D_1$ and $D_2$ for the latent contracts
  depend on whether $\pmvA \in \setenum{\pmv C, \pmv D}$ or not.
  By~\Cref{lem:types:dom-fnv}
  we have $\dom{\ptypeF[0]} \subseteq \fnv{\sysS[0]} \cup \setenum{\effwildcard}$,
  so from the premise $s \not\in \fnv{\sysS[0]}$ it follows that
  $s \not\in \dom{\ptypeF[0]}$.
  Since $\tsentails{A} \sysS[0] \colon \ptypeF[0]$, 
  by~\Cref{lem:types:sys-typing-renaming} we have that
  $\tsentails{A} \sysS[0]\sigma \colon \fsubst{\ptypeF[0]}{\sigma}$.
  There are the following two cases:
  \begin{itemize}

  \item $\pmvA \not\in \setenum{\pmv C, \pmv D}$.
    Then: 
    \[
    \begin{array}{l}
      \vspace{10pt}\fbox{$D_1$} 
      \; =
      \begin{array}{c}
        \inference[\smallnrule{[T-SAFree2]}] 
        {\ptypeF[0](x) = \bot}
        {\tsentails{A} \setenum{\freeze x \contrP}_{\pmv C} \tscompat \ptypeF[0]}
        \\[-3pt]
      \end{array}
      \\[10pt]
      \vspace{10pt}\fbox{$D_2$} 
      \; =
      \begin{array}{c}
        \inference[\smallnrule{[T-SAFree2]}] 
        {\ptypeF[0](y) = \bot}
        {\tsentails{A} \setenum{\freeze y \contrQ}_{\pmv D} \tscompat \ptypeF[0]}
        \\[-3pt]
      \end{array}
    \end{array}
    \]
    Since $x,y \not\in \dom{\ptypeF[0]}$, 
    by~\Cref{def:types:fsubst} we have that
    $\fsubst{\ptypeF[0]}{\sigma} = \ptypeF[0]$.
    Since $s \not\in \dom{\ptypeF[0]}$, we have that
    $(\fsubst{\ptypeF[0]}{\sigma})(s) = \ptypeF[0](s) = \bot$.
    Then, 
    \[
    \expand{(\fsubst{\ptypeF[0]}{\sigma})}{\setenum{s}}(s) 
    \; = \;
    \ptypeF[0](\effwildcard) 
    \; = \;
    \ptypeF(\effwildcard)
    \]
    which is honest by the assumption that $\ptypeF$ is honest.
    Then, we can construct the following typing derivation:
    \[
    \small
    \hspace{50pt}
    \inference[\smallnrule{[T-SDel2]}]
    {
      \inference[\smallnrule{[T-SPar2]}]
      {\tsentails{A} \sysS[0]\sigma \colon \fsubst{\ptypeF[0]}{\sigma}
        & 
        \inference[\smallnrule{[T-SAFree3]}]
        { \sys s \gamma \text{ $\pmvA$-free} &
          (\fsubst{\ptypeF[0]}{\sigma})(s) = \bot
        }
        {
          \tsentails{A} \sys {s} {\gamma} 
          \tscompat \fsubst{\ptypeF[0]}{\sigma}
        }
      }
      {\tsentails{A} \sysS[0]\sigma \cocoPar \sys {s} {\gamma} 
        \colon \fsubst{\ptypeF[0]}{\sigma}}
      & \begin{array}{c} \expand{(\fsubst{\ptypeF[0]}{\sigma})}{\setenum{s}}\!(s) \\ \text{ honest} \end{array}
    }
    {\tsentails{A} \sysSi \colon (\fsubst{\ptypeF[0]}{\sigma})\setenum{\bind{s}{\bot}}}
    \]
    Since $x,y,s \not\in \dom{\ptypeF[0]}$, 
    we conclude that:
    \[
    \hspace{80pt}
    (\fsubst{\ptypeF[0]}{\sigma})\setenum{\bind{s}{\bot}}
    \; = \;
    \ptypeF[0]\setenum{\bind{s}{\bot}}
    \; = \;
    \ptypeF[0]
    \; = \;
    \ptypeF[0]\setenum{\bind{y}{\bot}}\setenum{\bind{x}{\bot}}
    \; = \;
    \ptypeF 
    \]

  \item $\pmvA = \pmv D$. 
    Then, $\pmvA \neq \pmv C$, and:
    \[
    \begin{array}{l}
      \vspace{10pt}\fbox{$D_1$} 
      \; =
      \begin{array}{c}
        \inference[\smallnrule{[T-SAFree2]}] 
        {\ptypeF[0](x) = \bot}
        {\tsentails{A} \setenum{\freeze x \contrP}_{\pmv C} \tscompat \ptypeF[0]}
        \\[-3pt]
      \end{array}
      \\[10pt]
      \vspace{10pt}\fbox{$D_2$} 
      \; =
      \begin{array}{c}
        \inference[\smallnrule{[T-SFz1]}] 
        {\expand{\ptypeF[0]}{\setenum{y}}\!(y) \text{ realizes } \contrQ}
        {\tsentails{A} \setenum{\freeze y \contrQ}_{\pmv D} \tscompat \ptypeF[0]}
        \\[-3pt]
      \end{array}
    \end{array}
    \]
    Therefore, $x \not\in \dom{\ptypeF[0]}$, 
    and we have two cases, according to whether 
    $y \in \dom{\ptypeF[0]}$ or not.

    If $y \not\in \dom{\ptypeF[0]}$, 
    then $\fsubst{\ptypeF[0]}{\sigma} = \ptypeF[0]$.
    Since 
    $\expand{\ptypeF[0]}{\setenum{y}}\!(y) = \ptypeF[0](\effwildcard)$ 
    realizes $\contrQ$,
    and since $\contrQ = \cta{\gamma}$ by~\Cref{def:abs-contracts},
    we have that
    $\expand{\ptypeF[0]}{\setenum{s}}\!(s) = \ptypeF[0](\effwildcard)$
    realizes $\cta{\gamma}$,
    and it is honest since $\ptypeF$ is honest.
    Then, we can construct the following typing derivation:
    \[
    \small
    \hspace{80pt}
    \inference[\smallnrule{[T-SDel2]}]
    {
      \inference[\smallnrule{[T-SPar2]}]
      {\tsentails{A} \sysS[0]\sigma \colon \ptypeF[0]
        & 
        \inference[\smallnrule{[T-SFuse]}]
        {\expand{\ptypeF[0]}{\setenum{s}}\!(s) \text{ realizes } \cta{\gamma}}
        {
          \tsentails{A} \sys {s} {\gamma} 
          \tscompat \ptypeF[0]
        }
      }
      {\tsentails{A} \sysS[0]\sigma \cocoPar \sys {s} {\gamma} 
        \colon \ptypeF[0]}
      & \expand{\ptypeF[0]}{\setenum{s}}\!(s) \text{ honest}
    }
    {\tsentails{A} \sysSi \colon \ptypeF[0]\setenum{\bind{s}{\bot}}}
    \]
    Since $x,y,s \not\in \dom{\ptypeF[0]}$, 
    we conclude that:
    \[
    \hspace{80pt}
    \ptypeF[0]\setenum{\bind{s}{\bot}}
    \; = \;
    \ptypeF[0]\setenum{\bind{y}{\bot}}\setenum{\bind{x}{\bot}}
    \; = \;
    \ptypeF
    \]

    Otherwise, 
    if $y \in \dom{\ptypeF[0]}$, the premise of~\nrule{[T-SFz1]} 
    implies that
    $\ptypeF[0](y) \text{ realizes } \contrQ$.
    By~\Cref{def:types:fsubst} we have that
    \[
    \fsubst{\ptypeF[0]}{\sigma} 
    \; = \;
    \ptypeF[0]\setenum{\bind{y}{\bot}}\setenum{\bind{s}{\ptypeF[0](y)}}
    \]
    from which we can compute:
    \[
    \expand{(\fsubst{\ptypeF[0]}{\sigma})}{\setenum{s}}(s) 
    \; = \;
    (\fsubst{\ptypeF[0]}{\sigma})(s) 
    \; = \;
    \ptypeF[0](y)
    \]
    Since $\ptypeF[0](y)$ realizes $\contrQ$,
    and since $\contrQ = \cta{\gamma}$ by~\Cref{def:abs-contracts},
    we have that
    $\expand{(\fsubst{\ptypeF[0]}{\sigma})}{\setenum{s}}\!(s)$
    realizes $\cta{\gamma}$.
    Furthermore, since $y \in \dom{\ptypeF[0]}$, then:
    \[
    \expand{(\fsubst{\ptypeF[0]}{\sigma})}{\setenum{s}}\!(s)
    \; = \;
    \ptypeF[0](y)
    \; = \;
    \expand{\ptypeF[0]}{\setenum{y}}\!(y)        
    \]
    which is honest by the premise of~\nrule{[T-SDel2]} 
    in the typing derivation of~$\sysS$.
    Then, we can construct the following typing derivation:
    \[
    \hspace{80pt}
    \inference[\smallnrule{[T-SDel2]}]
    {
      \inference[\smallnrule{[T-SPar2]}]
      {\tsentails{A} \sysS[0]\sigma \colon \fsubst{\ptypeF[0]}{\sigma}
        & 
        \inference[\smallnrule{[T-SFuse]}]
        {\begin{array}{c} \expand{(\fsubst{\ptypeF[0]}{\sigma})}{\setenum{s}}\!(s) \\ \text{ realizes } \cta{\gamma} \end{array}}
        {
          \tsentails{A} \sys {s} {\gamma} 
          \tscompat \fsubst{\ptypeF[0]}{\sigma}
        }
      }
      {\tsentails{A} \sysS[0]\sigma \cocoPar \sys {s} {\gamma} 
        \colon \fsubst{\ptypeF[0]}{\sigma}}
      & \begin{array}{c} \expand{(\fsubst{\ptypeF[0]}{\sigma})}{\setenum{s}}\!(s) \\ \text{ honest} \end{array}
    }
    {\tsentails{A} \sysSi \colon (\fsubst{\ptypeF[0]}{\sigma})\setenum{\bind{s}{\bot}}}
    \]
    Since $x,s \not\in \dom{\ptypeF[0]}$, 
    we conclude that:
    \[
    \hspace{80pt}
    (\fsubst{\ptypeF[0]}{\sigma})\setenum{\bind{s}{\bot}}
    \; = \;
    \ptypeF[0]\setenum{\bind{y}{\bot}}\setenum{\bind{s}{\bot}}
    \; = \;
    \ptypeF[0]\setenum{\bind{y}{\bot}}\setenum{\bind{x}{\bot}}
    \; = \;
    \ptypeF
    \]

  \end{itemize}

\item Item~\eqref{th:types:subject-reduction-B-tscompat}
  We have the following typing derivation for $\sysS$:
  \[
  \inference[\smallnrule{[T-SDel1]}]
  {
    \inference[\smallnrule{[T-SDel1]}]
    {
      \inference[\smallnrule{[T-SPar1]}]
      {\tsentails{A} \sysS[0] \tscompat \ptypeF[0]
        &
        \inference[\smallnrule{[T-SPar1]}]
        {\fbox{$D_1$} & \fbox{$D_2$}}
        {\tsentails{A} \latent[{\pmv C}]{x}{\contrP} \cocoPar \latent[{\pmv D}]{y}{\contrQ} \tscompat \ptypeF[0]}}
      {\tsentails{A} \sysS[0] \cocoPar \latent[{\pmv C}]{x}{\contrP} \cocoPar \latent[{\pmv D}]{y}{\contrQ}
        \tscompat \ptypeF\setenum{\bind{y}{\bot}}\setenum{\bind{x}{\bot}}
        = \ptypeF[0]}
    }
    {\tsentails{A} (y) (\sysS[0] \cocoPar \latent[{\pmv C}]{x}{\contrP} \cocoPar \latent[{\pmv D}]{y}{\contrQ}) 
      \tscompat \ptypeF\setenum{\bind{x}{\bot}}}
  }
  {\tsentails{A} \sysS \tscompat \ptypeF}
  \]
  where the typing derivations $D_1$ and $D_2$ for the latent contracts
  depend on whether $\pmvA \in \setenum{\pmv C, \pmv D}$ or not.
  There are the following two cases:
  \begin{itemize}
    
  \item $\pmvA \not\in \setenum{\pmv C, \pmv D}$.
    Then: 
    \[
    \begin{array}{l}
      \vspace{10pt}\fbox{$D_1$} 
      \; =
      \begin{array}{c}
        \inference[\smallnrule{[T-SAFree2]}] 
        {\ptypeF[0](x) = \bot}
        {\tsentails{A} \setenum{\freeze x \contrP}_{\pmv C} \tscompat \ptypeF[0]}
        \\[-3pt]
      \end{array}
      \\[10pt]
      \vspace{10pt}\fbox{$D_2$} 
      \; =
      \begin{array}{c}
        \inference[\smallnrule{[T-SAFree2]}] 
        {\ptypeF[0](y) = \bot}
        {\tsentails{A} \setenum{\freeze y \contrQ}_{\pmv D} \tscompat \ptypeF[0]}
        \\[-3pt]
      \end{array}
    \end{array}
    \]
    Since $x,y \not\in \dom{\ptypeF[0]}$, 
    by~\Cref{def:types:fsubst} we have that:
    $\fsubst{\ptypeF[0]}{\sigma} = \ptypeF[0]$.
    Since $\tsentails{A} \sysS[0] \tscompat \ptypeF[0]$ and
    $s \not\in \fnv{\sysS[0]}$, then by~\Cref{lem:types:tscompat-bot}
    it follows that:
    \[
    \tsentails{A} \sysS[0] \tscompat \ptypeF[0]\setenum{\bind{s}{\bot}}
    \]
    and so by~\Cref{lem:types:sys-typing-renaming},
    and since $x,y \not\in \dom{\ptypeF[0]}$ we have that:
    \[
    \tsentails{A} \sysS[0]\sigma \tscompat 
    \fsubst{(\ptypeF[0]\setenum{\bind{s}{\bot}})}{\sigma}
    \; = \;
    \ptypeF[0]\setenum{\bind{s}{\bot}}
    \]
    Therefore, we can construct the following typing derivation
    for $\sysSi$:
    \[
    \hspace{80pt}
    \inference[\smallnrule{[T-SDel1]}]
    {
      \inference[\smallnrule{[T-SPar1]}]
      { \tsentails{A} \sysS[0]\sigma \tscompat \ptypeF[0]\setenum{\bind{s}{\bot}}
        & 
        \inference[\smallnrule{[T-SAFree3]}]
        {\sys s \gamma \text{ $\pmvA$-free} & \ptypeF[0]\setenum{\bind{s}{\bot}}(s) = \bot}
        {\tsentails{A} \sys s \gamma \tscompat \ptypeF[0]\setenum{\bind{s}{\bot}}}
      }
      {\tsentails{A} \sysS[0]\sigma \cocoPar \sys s \gamma \tscompat \ptypeF[0]\setenum{\bind{s}{\bot}}}
    }
    {\tsentails{A} \sysSi \tscompat \ptypeF[0]}
    \]
    Since $\sysSi = (s) (\sysS[0]\sigma \cocoPar \sys s \gamma)$
    and $\dom{\sigma} = \setenum{x,y}$,
    then $x,y \not\in \fnv{\sysSi}$.
    Then, from the typing 
    $\tsentails{A} \sysSi \tscompat \ptypeF[0]$
    by~\Cref{lem:types:tscompat-bot} we obtain the typing:
    \[
    \tsentails{A} \sysSi \tscompat 
    \ptypeF[0]\setenum{\bind{y}{\ptypeF(y)}}\setenum{\bind{x}{\ptypeF(x)}}
    \; = \;
    \ptypeF
    \]

  \item $\pmvA = \pmv D$. 
    Then, $\pmvA \neq \pmv C$, and:
    \[
    \begin{array}{l}
      \vspace{10pt}\fbox{$D_1$} 
      \; =
      \begin{array}{c}
        \inference[\smallnrule{[T-SAFree2]}] 
        {\ptypeF[0](x) = \bot}
        {\tsentails{A} \setenum{\freeze x \contrP}_{\pmv C} \tscompat \ptypeF[0]}
        \\[-3pt]
      \end{array}
      \\[10pt]
      \vspace{10pt}\fbox{$D_2$} 
      \; =
      \begin{array}{c}
        \inference[\smallnrule{[T-SFz1]}] 
        {\expand{\ptypeF[0]}{\setenum{y}}\!(y) \text{ realizes } \contrQ}
        {\tsentails{A} \setenum{\freeze y \contrQ}_{\pmv D} \tscompat \ptypeF[0]}
        \\[-3pt]
      \end{array}
    \end{array}
    \]
    By the typing derivation of $\sysS$, we have that
    $x,y \not\in \dom{\ptypeF[0]}$, hence
    by~\Cref{def:types:fsubst} we have that
    $\fsubst{\ptypeF[0]}{\sigma} = \ptypeF[0]$.
    Since $\tsentails{A} \sysS[0] \tscompat \ptypeF[0]$ and
    $s \not\in \fnv{\sysS[0]}$, then by~\Cref{lem:types:tscompat-bot}
    it follows that:
    \[
    \tsentails{A} \sysS[0] \tscompat \ptypeF[0]\setenum{\bind{s}{\bot}}
    \]
    and so by~\Cref{lem:types:sys-typing-renaming},
    using the fact that $x,y \not\in \dom{\ptypeF[0]}$, we have that:
    \[
    \tsentails{A} \sysS[0]\sigma \tscompat 
    \fsubst{(\ptypeF[0]\setenum{\bind{s}{\bot}})}{\sigma}
    \; = \;
    \ptypeF[0]\setenum{\bind{s}{\bot}}
    \]
    Since $y \not\in \dom{\ptypeF[0]}$, 
    the premise of~\nrule{[T-SFz1]} 
    implies that
    $\ptypeF[0](\effwildcard) \text{ realizes } \contrQ$.
    Then, also
    $
    \expand{(\ptypeF[0]\setenum{\bind{s}{\bot}})}{\setenum{s}}\!(s)
    =
    \ptypeF[0](\effwildcard)
    $
    realizes $\contrQ$.
    By~\Cref{def:abs-contracts} we have $\contrQ = \cta{\gamma}$,
    so
    $\expand{(\ptypeF[0]\setenum{\bind{s}{\bot}})}{\setenum{s}}\!(s)$
    realizes $\cta{\gamma}$.
    Therefore, we can construct the following typing derivation
    for $\sysSi$:
    \[
    \hspace{80pt}
    \inference[\smallnrule{[T-SDel1]}]
    {
      \inference[\smallnrule{[T-SPar1]}]
      { \tsentails{A} \sysS[0]\sigma \tscompat \ptypeF[0]\setenum{\bind{s}{\bot}}
        & 
        \inference[\smallnrule{[T-SFuse]}]
        {\expand{(\ptypeF[0]\setenum{\bind{s}{\bot}})}{\setenum{s}}\!(s)
          \text{ realizes } \cta{\gamma}}
        {\tsentails{A} \sys s \gamma \tscompat \ptypeF[0]\setenum{\bind{s}{\bot}}}
      }
      {\tsentails{A} \sysS[0]\sigma \cocoPar \sys s \gamma \tscompat \ptypeF[0]\setenum{\bind{s}{\bot}}}
    }
    {\tsentails{A} \sysSi \tscompat \ptypeF[0]}
    \]
    Since $\sysSi = (s) (\sysS[0]\sigma \cocoPar \sys s \gamma)$
    and $\dom{\sigma} = \setenum{x,y}$,
    then $x,y \not\in \fnv{\sysSi}$.
    Then, from the typing 
    $\tsentails{A} \sysSi \tscompat \ptypeF[0]$
    by~\Cref{lem:types:tscompat-bot} we obtain the typing:
    \[
    \tsentails{A} \sysSi \tscompat 
    \ptypeF[0]\setenum{\bind{y}{\ptypeF(y)}}\setenum{\bind{x}{\ptypeF(x)}}
    \; = \;
    \ptypeF
    \]

  \end{itemize}      

\end{itemize}

\subsection*{Rule~\nrule{[Do]}}

We have:
\[
\inference[\smallnrule{[Do]}]
{\gamma \cmove{{\pmvB} \says {\atomA}} \gammai}
{\sysS 
  \; = \;      
  \sys {\pmvB} {\cocodo s {\atomA} \cocoSeq \procP \cocoPlus \procPi \cocoPar \procQ}
  \cocoPar  
  \sys s {\gamma} 
  \;\sysmove{B}{\cocodo s {\atomA}}{}\; 
  \sys {\pmvB} {\procP \cocoPar \procQ}
  \cocoPar 
  \sys s {\gammai} 
  \; = \;
  \sysSi
}
\]

\begin{itemize}

\item Item~\eqref{th:types:subject-reduction-A-colon}. 
  We have $\pmvB = \pmvA$,
  and the typing derivation for $\sysS$ must be of the following form,
  where $\gamma = \pbic{A}{\contrP\buffer{\qmvA}}{C}{\contrQ\buffer{\qmv[{\pmv C}]}}$:   
  \[
  \inference[\smallnrule{[T-SPar2]}]
  {\inference[\smallnrule{[T-SA]}]
    {\inference[\smallnrule{[T-Par]}]
      {\vdots}
      {\effentails {\cocodo s {\atomA} \cocoSeq \procP
          \cocoPlus \procPi \cocoPar \procQ} \colon \ptypeF}}
    {\tsentails{A} {\sys {\pmv A} {\cocodo s {\atomA} \cocoSeq \procP
          \cocoPlus \procPi \cocoPar \procQ}} \colon \ptypeF}
    &
    \inference[\smallnrule{[T-SFuse]}]
    {\expand{\ptypeF}{\setenum{s}}\!(s)\text{ realizes $\cta{\gamma}$}}
    {\tsentails{A} {\sys s {\gamma}} \tscompat \ptypeF}}
  {\tsentails{A} \sysS \colon \ptypeF}
  \]
  Let $A = \fnv{\sysS} \cup \setenum{\effwildcard} = \dom{\ptypeF}$, 
  as implied by~\Cref{lem:types:dom-fnv}.
  Hence:
  \[
  \ptypeF 
  \; = \; 
  \lambda u \dotseq [{\cocodo s
    {\atomA}}]_u \dotseq \expand{\ptypeF^{\procP}}{A}(u) + \expand{\ptypeF^{\procPi}}{A}(u) \mid \expand{\ptypeF^{\procQ}}{A}(u)
  \]
  where $\ptypeF^{\procP}$, $\ptypeF^{\procPi}$ and $\ptypeF^{\procQ}$ 
  are, respectively, the types of $\procP$, $\procPi$ and $\procQ$. %
  Let $A' = \fnv{\procP} \cup \fnv{\procQ}$.
  Hence, if we choose:
  \[
  \ptypeFi
  \; = \; 
  \lambda u \dotseq 
  \expand{\ptypeF^{\procP}}{A'}(u) \mid \expand{\ptypeF^{\procQ}}{A'}(u)
  \]
  we have a typing judgement 
  $\effentails {\procP \cocoPar \procQ} \colon \ptypeFi$.
  Since $A = \dom \ptypeF$ and $s \in \fnv{\sysS}$, 
  then $s \in \dom \ptypeF$.
  Hence, $\ptypeF(s)$ realizes $\cta{\gamma}$.
  By using the similar arguments to those used in case~\nrule{[Tau]},
  we can deduce that 
  $
  \ptypeF \effmove{\cocodo s {\atomA}} \ptypeFi
  $.
  Then, 
  \[
  \expand{\ptypeF}{\setenum{s}}(s) 
  \; = \;
  \ptypeF(s) 
  \effmove{[\cocodo s \atomA]_s}
  \expand{\ptypeFi}{\setenum{s}}(s) 
  \]
  Note that $[\cocodo s \atomA]_s = \atomA$.
  Since $\gamma \cmove{\pmv A \says {\absatomA}} \gammai$
  then by~\Cref{lem:cabs:conc-to-abs} it follows that
  $\cta{\gamma} \abscmove{\absatomA} \cta{\gammai}$.
  Then, by rule~\nrule{[A-Do]} in~\Cref{fig:abs-sys:semantics}
  and by~\Cref{def:abs-honesty}
  we deduce that
  $\expand{\ptypeFi}{\setenum{s}}(s)$ realizes $\cta{\gammai}$.
  Thus, we have the following typing derivation:
  \[
  \inference[\smallnrule{[T-SPar2]}]
  {\inference[\smallnrule{[T-SA]}]
    {\inference[\smallnrule{[T-Par]}]
      {\vdots}
      {\effentails {\procP \cocoPar \procQ} \colon \ptypeFi}}
    {\tsentails{A} {\sys {\pmv A} {\procP \cocoPar \procQ}} \colon \ptypeFi}
    &
    \inference[\smallnrule{[T-SFuse]}]
    {\expand{\ptypeFi}{\setenum{s}}\!(s)\text{ realizes $\cta{\gammai}$}}
    {\tsentails{A} {\sys s {\gammai}} \tscompat \ptypeFi}}
  {\tsentails{A} \sysSi \colon \ptypeFi}
  \]

\item Item~\eqref{th:types:subject-reduction-B-colon}
  is vacuous, because $\sysS$ is not typeable with ``$\colon$''.

\item Item~\eqref{th:types:subject-reduction-B-tscompat}.
  Let $\procP[0] = \cocodo s {\atomA} \cocoSeq \procP \cocoPlus \procPi \cocoPar \procQ$.
  There are two possible typing derivations for
  $\tsentails{A} \sysS \tscompat \ptypeF$, 
  according to whether $\pmvA$ occurs in $\gamma$ or not.

  If $\pmvA$ does \emph{not} occur in $\gamma$, then we
  have the following typing derivation:
  \[
  \inference[\smallnrule{[T-SPar1]}]
  {\inference[\smallnrule{[T-SAFree1]}]
    {\pmvB \neq \pmvA & \fv{\procP[0]} \cap \dom{\ptypeF} = \emptyset}
    {\tsentails{A} \sys {\pmvB} {\procP[0]} \tscompat \ptypeF}
    &
    \inference[\smallnrule{[T-SAFree3]}] 
    {\sys s \gamma \; \text{$\pmvA$-free} & \ptypeF(s) = \bot}
    {\tsentails{A} \sys s {\gamma} \tscompat \ptypeF}}
  {\tsentails{A}
    \sys {\pmvB} {\procP[0]} \cocoPar \sys s {\gamma} 
    \tscompat \ptypeF}
  \]
  The thesis follows by a similar typing derivation,
  where rule~\nrule{[T-SAFree1]} can be used because 
  $\gammai$ is still $\pmvA$-free and
  $\fv{\procP \cocoPar \procQ} \subseteq \fv{\procP[0]}$.

  Otherwise, if $\pmvA$ occurs in $\gamma$, then we
  have the following typing derivation:
  \[
  \inference[\smallnrule{[T-SPar1]}]
  {\inference[\smallnrule{[T-SAFree1]}]
    {\pmvB \neq \pmvA & \fv{\procP[0]} \cap \dom{\ptypeF} = \emptyset}
    {\tsentails{A} \sys {\pmvB} {\procP[0]} \tscompat \ptypeF}
    &
    \inference[\smallnrule{[T-SFuse]}] 
    {\expand{\ptypeF}{\setenum{s}}(s) \text{ realizes } \cta{\gamma}}
    {\tsentails{A} \sys s {\gamma} \tscompat \ptypeF}}
  {\tsentails{A}
    \sys {\pmvB} {\procP[0]} \cocoPar \sys s {\gamma} 
    \tscompat \ptypeF}
  \]
  Since $\gamma \cmove{\pmvB \says {\absatomA}} \gammai$
  then by~\Cref{lem:cabs:conc-to-abs} it follows that
  $\cta{\gamma} \abscmove{\ctx \says \absatomA} \cta{\gammai}$.
  Then, by rule~\nrule{[A-Ctx]} in~\Cref{fig:abs-sys:semantics}
  and by~\Cref{def:abs-honesty}
  we deduce that
  $\expand{\ptypeF}{\setenum{s}}(s)$ realizes $\cta{\gammai}$.
  Thus, we have the following typing derivation:
  \[
  \hspace{20pt}
  \inference[\smallnrule{[T-SPar1]}]
  {
    \inference[\smallnrule{[T-SAFree1]}]
    {\pmvB \neq \pmvA & \fv{\procP \cocoPar \procQ} \cap \dom{\ptypeF} = \emptyset}
    {\tsentails{A} {\sys {\pmvB} {\procP \cocoPar \procQ}} \tscompat \ptypeF}
    &
    \inference[\smallnrule{[T-SFuse]}]
    {\expand{\ptypeF}{\setenum{s}}\!(s)\text{ realizes $\cta{\gammai}$}}
    {\tsentails{A} {\sys s {\gammai}} \tscompat \ptypeF}}
  {\tsentails{A} \sysSi \tscompat \ptypeF}
  \]

\end{itemize}

\subsection*{Rule~\nrule{[Del]}}

We have $\pref = \del{u}{\prefi}$, and:
\[
\inference[\smallnrule{[Del]}]
{\sysS[0] \sysmove{B}{\prefi}{} \sysSi[0]}
{\sysS \; = \;
  (u)\sysS[0] \sysmove{B}{\del{u}{\prefi}}{} (u)\sysSi[0]
  \; = \; \sysSi}
\]

\begin{itemize}
\item Item~\eqref{th:types:subject-reduction-A-colon}.
  The only possible typing derivation for $\sysS$ is the following:
  \[
  \inference[\smallnrule{[T-SDel2]}]
  {\tsentails{A} \sysS[0] \colon \ptypeF[0]
    & \expand{\ptypeF[0]}{\setenum{u}}\!(u)\text{ honest}}
  {\tsentails{A} \sysS = (u)\sysS[0] \colon
    \ptypeF[0]\setenum{\bind{u}{\bot}} = \ptypeF}
  \]
  We have that $\ptypeF[0]$ is honest.
  Indeed, for all $v \in \dom {\ptypeF[0]}$ we have that:
  if $v \neq u$, then $\ptypeF[0](v) = \ptypeF(v)$, 
  which is honest by the hypothesis that $\ptypeF$ is honest;
  instead, if $v = u$ then 
  $\expand{\ptypeF[0]}{\setenum{u}}(u) = \ptypeF[0](u)$,
  which is honest by the rightmost premise of~\nrule{[T-SDel2]}.
  Then, by the induction hypothesis we have that 
  there exists some $\ptypeFi[0]$ such that 
  $\ptypeF[0] \effmove{\prefi} \ptypeFi[0]$ and
  $\tsentails{A} \sysSi[0] \colon \ptypeFi[0]$.
  Since $\ptypeF[0]$ is honest, 
  then by~\Cref{lem:types:chan-types-moves-honesty} also
  its reduct $\ptypeFi[0]$ is honest.
  By~\Cref{lem:types:expand-honest}, also
  $\expand{\ptypeFi[0]}{\setenum{u}}$ is honest.
  Then, we have the following typing derivation:
  \[
  \inference[\smallnrule{[T-SDel2]}]
  {\tsentails{A} \sysSi[0] \colon \ptypeFi[0]
    & \expand{\ptypeFi[0]}{\setenum{u}}\!(u)\text{ honest}}
  {\tsentails{A} \sysSi = (u)\sysSi[0] \colon
    \ptypeFi[0]\setenum{\bind{u}{\bot}}}
  \]

  Let $\ptypeFi = \ptypeFi[0]\setenum{\bind{u}{\bot}}$.
  Since $\ptypeF[0] \effmove{\prefi} \ptypeFi[0]$,
  then by~\Cref{def:types:trans} we have that
  $\dom{\ptypeFi[0]} \subseteq \dom{\ptypeF[0]}$, and:
  \begin{align}
    \label{eq:proof-subject-reduction:del:1}
    \forall v \in \dom{\ptypeFi[0]}
    & \quad : \quad
      \ptypeF[0](v) \effmove{[\prefi]_v} \ptypeFi[0](v)
    \\
    \label{eq:proof-subject-reduction:del:2}
    \forall v \in \dom{\ptypeF[0]} \setminus \dom{\ptypeFi[0]}
    & \quad : \quad
      \ptypeF[0](v) \effmove{[\prefi]_v} \ptypeFi[0](\effwildcard)
  \end{align}      
  We have that:
  \[
  \dom{\ptypeFi} 
  \; = \; 
  \dom{\ptypeFi[0]} \setminus \setenum{u}
  \; \subseteq \; 
  \dom{\ptypeF[0]} \setminus \setenum{u}
  \; = \; 
  \dom{\ptypeF}
  \]
  To prove that $\ptypeF \effmove{\pref} \ptypeFi$, 
  we need to consider the following two cases.
  \begin{itemize}
  \item $v \in \dom{\ptypeFi}$.
    Then, $v \in \dom{\ptypeFi[0]}$, 
    so from~\eqref{eq:proof-subject-reduction:del:1} we have
    $\ptypeF[0](v) \effmove{[\prefi]_v} \ptypeFi[0](v)$.
    Since $u \neq v$, then
    $\ptypeF(v) \effmove{[\prefi]_v} \ptypeFi(v)$.
    There are three further subcases:
    \begin{itemize}
    \item $\prefi = \cocodo{u}{\atomA}$.
      In this case we have 
      $[\prefi]_v = \efftauqm = \del{u}{\prefi} = \pref = [\pref]_v$.

    \item $\prefi = \tell{}{\freeze u \contrP}$.
      In this case we have 
      $[\prefi]_v = \tau = \del{u}{\prefi} = \pref = [\pref]_v$.

    \item otherwise, 
      $\prefi = \del{u}{\prefi} = \pref$, hence
      $[\prefi]_v = [\pref]_v$.
    \end{itemize}

  \item 
    $v \in \dom{\ptypeF} \setminus \dom{\ptypeFi}$, then
    $v \in \dom{\ptypeF[0]} \setminus \dom{\ptypeFi[0]}$,
    so from~\eqref{eq:proof-subject-reduction:del:2} we have
    $\ptypeF[0](v) \effmove{[\prefi]_v} \ptypeFi[0](\effwildcard)$.
    Since $u \neq v$, we conclude that
    $\ptypeF(v) \effmove{[\prefi]_v} \ptypeFi(\effwildcard)$.
    The thesis follows because $[\prefi]_v = [\pref]_v$,
    as in the previous case.
  \end{itemize}
  
\item Item~\eqref{th:types:subject-reduction-B-colon}.
  The only possible typing derivation for $\sysS$ is the following:
  \[
  \inference[\smallnrule{[T-SDel2]}]
  {\tsentails{A} \sysS[0] \colon \ptypeF[0]
    & \expand{\ptypeF[0]}{\setenum{u}}\!(u)\text{ honest}}
  {\tsentails{A} \sysS = (u)\sysS[0] \colon
    \ptypeF[0]\setenum{\bind{u}{\bot}} = \ptypeF}
  \]
  We have that $\ptypeF[0]$ is honest.
  Indeed, for all $v \in \dom {\ptypeF[0]}$ we have that:
  if $v \neq u$, then $\ptypeF[0](v) = \ptypeF(v)$, 
  which is honest by the hypothesis that $\ptypeF$ is honest;
  instead, if $v = u$ then 
  $\ptypeF[0](v) = \ptypeF[0](u) = \expand{\ptypeF[0]}{\setenum{u}}\!(u)$,
  which is honest by the rightmost premise of~\nrule{[T-SDel2]}.
  Hence, by the induction hypothesis we have that 
  $\tsentails{A} \sysSi[0] \colon \ptypeF[0]$.
  Since $\ptypeF[0]$ is honest,
  then by~\Cref{lem:types:fsubst-honest} it follows that   
  $\expand{\ptypeF[0]}{\setenum{u}}$ is honest.
  Then, we have the following typing derivation:
  \[
  \inference[\smallnrule{[T-SDel2]}]
  {\tsentails{A} \sysSi[0] \colon \ptypeF[0]
    & \expand{\ptypeF[0]}{\setenum{u}}\!(u) \text{ honest}}
  {\tsentails{A} \sysSi = (u)\sysSi[0] \colon
    \ptypeF[0]\setenum{\bind{u}{\bot}}}
  \]
  The thesis follows by choosing 
  $\ptypeFi[0]\setenum{\bind{u}{\bot}} = \ptypeF$.

\item Item~\eqref{th:types:subject-reduction-B-tscompat}.
  The only possible typing derivation for $\sysS$ is the following:
  \[
  \inference[\smallnrule{[T-SDel1]}]
  {\tsentails{A} \sysS[0] \tscompat \ptypeF\setenum{\bind{u}{\bot}}}
  {\tsentails{A} \sysS = (u)\sysS[0] \tscompat \ptypeF}
  \]
  Since $\ptypeF\setenum{\bind{u}{\bot}}$ is honest and
  $\tsentails{A} \sysS[0] \tscompat \ptypeF\setenum{\bind{u}{\bot}}$,
  then by the induction hypothesis it follows that
  $\tsentails{A} \sysSi[0] \tscompat \ptypeF\setenum{\bind{u}{\bot}}$.
  Then, we can reconstruct the following derivation:
  \[
  \inference[\smallnrule{[T-SDel1]}]
  {\tsentails{A} \sysSi[0] \tscompat 
    \ptypeF\setenum{\bind{u}{\bot}}}
  {\tsentails{A} \sysSi = (u)\sysSi[0] \tscompat \ptypeF}
  \]

\end{itemize}

\subsection*{Rule~\nrule{[Par]}}

We have:  
\[
\inference[\smallnrule{[Par]}]
{\sysS[0] \sysmove{B}{\pref}{} \sysSi[0]}
{\sysS \; = \;
  \sysS[0] \cocoPar \sysS[1] \sysmove{B}{\pref}{} \sysSi[0] \cocoPar \sysS[1]
  \; = \; \sysSi}
\]
\begin{itemize}
\item Item~\eqref{th:types:subject-reduction-A-colon}. 
  We have $\pmvB = \pmvA$, and
  since $\sysS[0]$ reduces through $\pmvA$,
  then a process of $\pmvA$ cannot appear in $\sysS[1]$.
  Hence, the only typing derivation for $\sysS$ is the following:
  \[
  \inference[\smallnrule{[T-SPar2]}]
  {\tsentails{A} \sysS[0] \colon \ptypeF
    & 
    \tsentails{A} \sysS[1] \tscompat \ptypeF}
  {\tsentails{A} \sysS \colon \ptypeF}
  \]
  
  By applying the induction hypothesis on the leftmost premise
  of~\nrule{[T-SPar2]}, 
  we have some $\ptypeFi$ such that:
  \[
  \ptypeF \effmove{\pref} \ptypeFi
  \qquad \text{and} \qquad
  \tsentails{A} \sysSi[0] \colon \ptypeFi
  \]
  Note that if $\pref = \cocodo{s}{\atomA}$, then by rule~\nrule{[Do]}
  it must be $s[\gamma] \in \sysS[0]$ (and so, $s[\gamma] \not\in \sysS[1]$).
  Hence, by~\Cref{lem:types:tscompat-effmove}
  we infer that $\tsentails{A} \sysS[1] \tscompat \ptypeFi$.
  Then, we can construct the following typing derivation for $\sysSi$:
  \[
  \inference[\smallnrule{[T-SPar2]}]
  {\tsentails{A} \sysSi[0] \colon \ptypeFi
    & 
    \tsentails{A} \sysS[1] \tscompat \ptypeFi}
  {\tsentails{A} \sysSi \colon \ptypeFi}
  \]

\item Item~\eqref{th:types:subject-reduction-B-colon}.
  We have $\pmvB \neq \pmvA$. 
  There are two possible
  typing derivations for $\sysS$:
  \[
  \hspace{40pt}
  \inference[\smallnrule{[T-SPar2]}]
  {\tsentails{A} \sysS[0] \colon \ptypeF
    & 
    \tsentails{A} \sysS[1] \tscompat \ptypeF}
  {\tsentails{A} \sysS \colon \ptypeF}
  \hspace{40pt}
  \inference[\smallnrule{[T-SPar2]}]
  {\tsentails{A} \sysS[0] \tscompat \ptypeF
    & 
    \tsentails{A} \sysS[1] \colon \ptypeF}
  {\tsentails{A} \sysS \colon \ptypeF}
  \]
  If the leftmost typing derivation has been used,
  by applying the induction hypothesis on its leftmost premise
  we obtain  $\tsentails{A} \sysSi[0] \colon \ptypeF$.
  So, we can construct the following typing derivation for $\sysSi$:
  \[
  \inference[\smallnrule{[T-SPar2]}]
  {\tsentails{A} \sysSi[0] \colon \ptypeF
    & 
    \tsentails{A} \sysS[1] \tscompat \ptypeF}
  {\tsentails{A} \sysSi \colon \ptypeF}
  \]
  If the rightmost typing derivation has been used,
  by applying the induction hypothesis of item~\eqref{th:types:subject-reduction-B-tscompat}
  on its leftmost premise we obtain $\tsentails{A} \sysSi[0] \tscompat \ptypeF$.
  So, we can construct the following typing derivation for $\sysSi$:
  \[
  \inference[\smallnrule{[T-SPar2]}]
  {\tsentails{A} \sysSi[0] \tscompat \ptypeF
    & 
    \tsentails{A} \sysS[1] \colon \ptypeF}
  {\tsentails{A} \sysSi \colon \ptypeF}
  \]

\item Item~\eqref{th:types:subject-reduction-B-tscompat}.
  We have the following typing derivation for $\sysS$:
  \[
  \inference[\smallnrule{[T-SPar1]}]
  {\tsentails{A} \sysS[0] \tscompat \ptypeF
    & \tsentails{A} \sysS[1] \tscompat \ptypeF}
  {\tsentails{A} \sysS \tscompat \ptypeF}
  \]
  By applying the induction hypothesis on the leftmost premise, 
  we have that $\tsentails{A} \sysSi[0] \tscompat \ptypeF$.
  So, we can construct the following typing derivation for $\sysSi$:
  \[
  \inference[\smallnrule{[T-SPar1]}]
  {\tsentails{A} \sysSi[0] \tscompat \ptypeF
    & \tsentails{A} \sysS[1] \tscompat \ptypeF}
  {\tsentails{A} \sysSi \tscompat \ptypeF}
  \]

\end{itemize}

\subsection*{Rule~\nrule{[Rec]}}

We have:
\[
\inference[\smallnrule{[Rec]}]
{\sys {\pmvB} {\procP\subs{\procRec{\procX(\vec y)} \procP}{\procX}\subs{\vec u}{\vec y}
    \cocoPar \procQ} \cocoPar \sysS[0] 
  \sysmove{B}{\pref}{} \sysSi}
{\sysS \; = \; 
  \sys {\pmvB} {(\procRec{\procX(\vec y)} \procP)(\vec u) \cocoPar \procQ} \cocoPar \sysS[0] \sysmove{B}{\pref}{} \sysSi}
\]
Let $\procP[0] = (\procRec{\procX(\vec y)} \procP)(\vec u) \cocoPar \procQ$.

\begin{itemize}

\item Item~\eqref{th:types:subject-reduction-A-colon}.
  We have $\pmvB = \pmvA$, $\vec y = \vec u = \emptyset$.
  The only typing derivation for $\sysS$ is the following, 
  where $\ptypeF = \lambda u.\, \expand{\ptypeF^{\procX}}{A}(u) \,\mid\, \expand{\ptypeF^{\procQ}}{A}(u)$,
  $\ptypeF^{\procX} = \lambda u .\, \effrec{\consttovar{\procX}}{\ptypeF^{\procP}(u)}$, and
  $A = \dom{\ptypeF^{\procX}} \cup \dom{\ptypeF^{\procQ}} = \dom{\ptypeF^{\procP}} \cup \dom{\ptypeF^{\procQ}}$:
  \[
  \hspace{40pt}
  \inference[\smallnrule{[T-SPar2]}]
  {\inference[\smallnrule{[T-SA]}]
    {\inference[\smallnrule{[T-Par]}]
      {\inference[\smallnrule{[T-Rec]}]
        {\effentails \procP \colon \ptypeF^{\procP}}
        {\effentails (\procRec{\procX()} \procP)() \colon \ptypeF^{\procX}}
        & \effentails \procQ \colon \ptypeF^{\procQ}}
      {\effentails {(\procRec{\procX()} \procP)() \cocoPar \procQ \colon \ptypeF}}}
    {\tsentails{A} \sys {\pmv A} {(\procRec{\procX()} \procP)() \cocoPar \procQ} \colon \ptypeF}
    & \tsentails{A} \sysS[0] \tscompat \ptypeF}
  {\tsentails{A} \sysS \colon \ptypeF}
  \]
  Since $\effentails \procP \colon \ptypeF^{\procP}$ 
  and $\effentails (\procRec{\procX()} \procP)() \colon \ptypeF^{\procX}$,
  then by~\Cref{lem:types:type-subs-recvar} we have:
  \[
  \effentails 
  \procP \subs{(\procRec{\procX()} \procP)()}{\procX} 
  \colon 
  \lambda u \dotseq 
  \expand{\ptypeF^{\procP}}{B}(u) \subs{\expand{\ptypeF^{\procX}\,}{B}(u)}{\consttovar{\procX}}
  \; = \;
  \ptypeG
  \]
  where $B = \dom{\ptypeF^{\procP}} \cup \dom{\ptypeF^{\procX}} = \dom{\ptypeF^{\procP}}$.
  Let: 
  \[
  \ptypeGi \; = \; \lambda u.\; \expand{\ptypeG}{C}(u) \mid \expand{\ptypeF^{\procQ}}{C}(u)
  \]
  where $C = \dom{\ptypeG} \cup \dom{\ptypeF^{\procQ}} = \dom{\ptypeF^{\procP}} \cup \dom{\ptypeF^{\procQ}} = A$.
  Note that $\ptypeGi$ can be obtained from $\ptypeF$, 
  by unfolding the recursion therein:
  however, the unfolding does not affect the typing
  $\tsentails{A} \sysS[0] \tscompat \ptypeF$,
  which can then be re-typed as 
  $\tsentails{A} \sysS[0] \tscompat \ptypeGi$.
  Therefore, we have the following typing for the premise of rule~\nrule{[Rec]}:
  \[
  \inference[\smallnrule{[T-SPar2]}]
  { \inference[\smallnrule{[T-SA]}]
    {
      \inference[\smallnrule{[T-Par]}]
      {\effentails \procP \subs{\procRec{\procX()} \procP}{\procX} \colon \ptypeG
        & 
        \effentails \procQ \colon \ptypeF^{\procQ}}
      {\effentails \procP \subs{\procRec{\procX()} \procP}{\procX} \cocoPar \procQ \colon \ptypeGi}}
    {\tsentails{A} \sys {\pmvA} {\procP\subs{\procRec{\procX()} \procP}{\procX} \cocoPar \procQ} \colon \ptypeGi}
    & 
    \tsentails{A} \sysS[0] \tscompat \ptypeGi}
  {\tsentails{A} \sys {\pmvA} {\procP\subs{\procRec{\procX()} \procP}{\procX} \cocoPar \procQ} \cocoPar \sysS[0] \colon \ptypeGi}
  \]
  Now, since $\ptypeF$ is honest, 
  then also its unfolding $\ptypeGi$ is honest as well.
  Therefore, by applying the induction hypothesis on the premise of rule~\nrule{[Rec]},
  we obtain some $\ptypeFi$ such that:
  \[
  \ptypeGi \effmove{\pref} \ptypeFi
  \hspace{40pt}
  \tsentails{A} \sysSi \colon \ptypeFi
  \]
  To conclude, note that $\ptypeF \effmove{\pref} \ptypeFi$ 
  follows by rule~\nrule{[C-Rec]} in~\Cref{fig:abs-sys:semantics}.

\item Item~\eqref{th:types:subject-reduction-B-colon}.
  We have $\pmvB \neq \pmvA$, so the only 
  typing derivation for $\sysS$ is the following: 
  \[
  \inference[\smallnrule{[T-SPar2]}]
  {\inference[\smallnrule{[T-SAFree1]}]
    {\pmvA \neq \pmvB & \fv{\procP[0]} \cap \dom{\ptypeF} = \emptyset} 
    {\tsentails{A} \sys {\pmv B} {\procP[0]} \tscompat \ptypeF}
    & \tsentails{A} \sysS[0] \colon \ptypeF}
  {\tsentails{A} \sysS \colon \ptypeF}
  \]
  Note that the typing with rule~\nrule{[T-SAFree1]} 
  is not affected by the unfolding of $\procP[0]$,
  because the unfolding can only decrease the set of free variables.
  Hence, we can use the same rule to obtain
  $\tsentails{A} \sys {\pmv B} {\procPi} \tscompat \ptypeF$,
  where $\procPi$ is the process within $\sys{\pmvB}{\cdots}$ 
  in the premise of rule~\nrule{[Rec]}.
  By applying~\nrule{[T-SPar2]}, we then type as $\ptypeF$
  the whole premise of rule~\nrule{[Rec]}.
  Then, by the induction hypothesis of item~\eqref{th:types:subject-reduction-B-colon},
  we obtain the thesis.

\item Item~\eqref{th:types:subject-reduction-B-tscompat}.
  The only typing derivation for $\sysS$ is the following:
  \[
  \inference[\smallnrule{[T-SPar1]}]
  {\inference[\smallnrule{[T-SAFree1]}]
    {{\pmv A} \neq {\pmv B} & \fv{\procP[0]} \cap \dom{\ptypeF} = \emptyset}
    {\tsentails{A} \sys {\pmvB} {\procP[0]} \tscompat \ptypeF}
    & \tsentails{A} \sysS[0] \tscompat \ptypeF}
  {\tsentails{A} \sysS \tscompat \ptypeF}
  \]
  Note that the typing with rule~\nrule{[T-SAFree1]} 
  is not affected by the unfolding of $\procP[0]$,
  because the unfolding can only decrease the set of free variables.
  Hence, we can use the same rule to obtain
  $\tsentails{A} \sys {\pmv B} {\procPi} \tscompat \ptypeF$,
  where $\procPi$ is the process within $\sys{\pmvB}{\cdots}$ 
  in the premise of rule~\nrule{[Rec]}.
  By applying~\nrule{[T-SPar1]}, we then type as $\ptypeF$
  the premise of rule~\nrule{[Rec]}.
  By the induction hypothesis of item~\eqref{th:types:subject-reduction-B-tscompat},
  we conclude.
  \qed

\end{itemize}



\section{Proof of~\Cref{th:progress} (Progress)} 
\label{sec:proofs-progress}

\begin{lem}[Process progress]
  \label{lem:types:process-progress}
  If $\,\effentails \procP \colon \ptypeF \effPar \ptypeG$ 
  with $\ptypeF \effPar \ptypeG$ honest,
  $\ptypeF \effmove{\pref} \ptypeFi$, and $u \in \dom{\ptypeF} = \dom{\ptypeG}$,
  then:
  \begin{bartalign}
    \label{lem:types:process-progress-item-tau}
    \pref \in \setenum{\tau,\tell {\pmv B} {\freeze u \contrP}}
    & \implies
    \exists \vec{x},\procPi,\sysSi \dotseq\;
    \begin{array}{rl}
      \sys \pmvA \procP \sysmove{A}{\pref}{} & (\vec{x}) (\sys {\pmvA} {\procPi} \cocoPar \sysSi)
      \\[5pt]
      \;\;\land\; \tsentails{A} & (\vec{x}) (\sys {\pmvA} {\procPi} \cocoPar \sysSi) \colon \ptypeFi \effPar \ptypeG
    \end{array}
    \\
    \label{lem:types:process-progress-item-fact}
    \pref = {\cocodo u {\atomA}}
    & \implies
    \atomA \in \readydosys{u}{\pmvA}{\sys {\pmvA} \procP}
  \end{bartalign}
\end{lem}
\begin{proof}
  We prove item~\eqref{lem:types:process-progress-item-tau}
  by induction on the proof of $\ptypeF \effmove{\pref} \ptypeFi$,
  by using the syntactic notation for process types introduced 
  in~\Cref{notation:ptype:syntax},
  and by universally quantifying $\procP$ and $\ptypeG$ 
  in the inductive statement.
  To reduce the technical overhead,
  we will omit the domain expansions $\expand{\ptypeF}{A}$
  throughout the proof.
  We have the following cases:
  \begin{itemize}

  \item \nrule{[C-Pref]}.
    Then, $\ptypeF = \alpha \effseq \ptypeFi$, and $\alpha = [\pref]$.
    Since $\effentails \procP \colon \ptypeF \effPar \ptypeG$, then $\procP$ has the form
    $(\vec{x}) (\prefi \cocoSeq \procP[0] \cocoPar \procQ)$, and
    $\del{\vec{x}}{\prefi} = \pref$.
    We proceed by cases on the form of~$\prefi$.
    Since~\eqref{lem:types:process-progress-item-tau} assumes that
    $\pref$ is either a $\tau$ or a $\tell{}{}$, 
    we only have the following two cases:
    \begin{itemize}

    \item $\prefi = \tau$.
      Let $\sysSi = \sysNil$. Then:
      \[
      \sys \pmvA {(\vec{x}) (\prefi \cocoSeq \procP[0] \cocoPar \procQ)} 
      \;\; \sysmove{A}{\tau}{} \;\; 
      (\vec{x}) (\sys {\pmvA} {\procP[0]  \cocoPar \procQ} \cocoPar \sysSi)
      \; = \;
      \sys {\pmvA} {(\vec{x}) (\procP[0] \cocoPar \procQ)}
      \]
      Let $\procPi = \procP[0] \cocoPar \procQ$.
      To prove that 
      $\tsentails{A} (\vec{x}) (\sys {\pmvA} {\procPi} \cocoPar \sysSi) \colon \ptypeFi \effPar \ptypeG$,
      we first deconstruct the typing $\effentails \procP \colon \ptypeF \effPar \ptypeG$ as follows:
      \[
      \small
      \hspace{50pt}
      \inference[{\smallnrule{[T-Del${}^*$]}}]
      {\inference[{\smallnrule{[T-Par]}}]
        {
          \inference[{\smallnrule{[T-Sum]}}]
          {\effentails \procP[0] \colon \ptypeFi[0]}
          {\effentails \prefi \cocoSeq \procP[0] \colon \ptypeF[0] = \lambda u . [\tau]_u \effseq \ptypeFi[0](u) = \tau \effseq \ptypeFi[0]}
        & \effentails \procQ \colon \ptypeG[0]}
        {\effentails \prefi \cocoSeq \procP[0] \cocoPar \procQ \colon \ptypeF[0] \effPar \ptypeG[0]}
          & (\ptypeF[0] \effPar \ptypeG[0])(\vec{x}) \text{ honest}}
      {\effentails (\vec{x}) (\prefi \cocoSeq \procP[0] \cocoPar \procQ) \colon \ptypeF \effPar \ptypeG = (\ptypeF[0] \effPar \ptypeG[0]) \setenum{\bind{\vec{x}}{\bot}}}
      \]
      By Lemma~\ref{lem:types:chan-types-moves-honesty},
      we have that $(\ptypeFi[0] \effPar \ptypeG[0])(\vec{x})$ is honest.
      We can then reconstruct the typing for $(\vec{x}) (\sys {\pmvA} {\procPi} \cocoPar \sysSi)$:
      \[
      \inference[\smallnrule{[T-SA]}]
      {
        \inference[{\smallnrule{[T-Del${}^*$]}}]
        {\effentails \procPi \colon \ptypeFi[0] \effPar \ptypeG[0] & (\ptypeFi[0] \effPar \ptypeG[0])(\vec{x}) \text{ honest}}
        {\effentails (\vec{x}) \procPi \colon (\ptypeFi[0]  \effPar \ptypeG[0]) \setenum{\bind{\vec{x}}{\bot}}}
      }
      {\tsentails{A} \sys {\pmvA} {(\vec{x}) \procPi} \colon (\ptypeFi[0] \effPar \ptypeG[0]) \setenum{\bind{\vec{x}}{\bot}}}
      \]
      Since $\ptypeF[0] \effmove{\tau} \ptypeFi[0]$ 
      and $\ptypeF \effPar \ptypeG = \ptypeF[0]\setenum{\bind{\vec{x}}{\bot}} \effPar \ptypeG[0]\setenum{\bind{\vec{x}}{\bot}}$,
      then we have the thesis: 
      \[
      \ptypeFi \effPar \ptypeG \; = \; \ptypeFi[0]\setenum{\bind{\vec{x}}{\bot}} \effPar \ptypeG[0]\setenum{\bind{\vec{x}}{\bot}}
      \]

    \item $\prefi = \tell {} {\freeze w \contrP}$.
      Let $\sysSi = \latent{w}{\contrP}$. Then:
      \[
      \sys \pmvA {(\vec{x}) (\prefi \cocoSeq \procP[0] \cocoPar \procQ)} 
      \;\; \sysmove{A}{\pref}{} \;\; 
      (\vec{x}) (\sys {\pmvA} {\procP[0] \cocoPar \procQ} \cocoPar \sysSi)
      \]
      Let $\procPi = \procP[0] \cocoPar \procQ$.
      To prove that 
      $\tsentails{A} (\vec{x}) (\sys {\pmvA} {\procPi} \cocoPar \sysSi) \colon \ptypeFi \effPar \ptypeG$,
      we first deconstruct the typing $\effentails \procP \colon \ptypeF \effPar \ptypeG$ 
      as in the previous case:
      \[
      \small
      \hspace{50pt}
      \inference[{\smallnrule{[T-Del${}^*$]}}]
      {\inference[{\smallnrule{[T-Par]}}]
        {
          \inference[{\smallnrule{[T-Sum]}}]
          {\effentails \procP[0] \colon \ptypeFi[0]}
          {\effentails \prefi \cocoSeq \procP[0] \colon \ptypeF[0] = \lambda u . [\prefi]_u \effseq \ptypeFi[0](u) = [\prefi] \effseq \ptypeFi[0]}
        & \effentails \procQ \colon \ptypeG[0]}
        {\effentails \prefi \cocoSeq \procP[0] \cocoPar \procQ \colon \ptypeF[0] \effPar \ptypeG[0]}
          & (\ptypeF[0] \effPar \ptypeG[0])(\vec{x}) \text{ honest}}
      {\effentails (\vec{x}) (\prefi \cocoSeq \procP[0] \cocoPar \procQ) \colon \ptypeF \effPar \ptypeG = (\ptypeF[0] \effPar \ptypeG[0]) \setenum{\bind{\vec{x}}{\bot}}}
      \]
      By Lemma~\ref{lem:types:chan-types-moves-honesty},
      we have that $(\ptypeFi[0] \effPar \ptypeG[0])(\vec{x})$ is honest.
      We now prove that 
      $\expand{(\ptypeFi[0] \effPar \ptypeG[0])}{w}(w)$ realizes $\contrP$.
      There are two cases, according to whether $w \in \vec{x}$ or not.
      If $w \in \vec{x}$, then 
      since $(\ptypeF[0] \effPar \ptypeG[0])(\vec{x})$ honest,
      so in particular $(\ptypeF[0] \effPar \ptypeG[0])(w)$ is honest.
      Hence, by 
      $
      (\ptypeF[0] \effPar \ptypeG[0])(w) \effmove{\effcontract{\contrP}} (\ptypeFi[0] \effPar \ptypeG[0])(w)
      $
      we infer that $(\ptypeFi[0] \effPar \ptypeG[0])(w)$
      realizes $\contrP$.
      If $w \not\in \vec{x}$, since 
      $\ptypeF \effPar \ptypeG$ is honest and
      $(\ptypeF \effPar \ptypeG)(w) = (\ptypeF[0] \effPar \ptypeG[0])(w)$,
      then $(\ptypeF[0] \effPar \ptypeG[0])(w)$ is honest;
      we infer that $(\ptypeFi[0] \effPar \ptypeG[0])(w)$
      realizes $\contrP$ as in the other case.
      
      We can then reconstruct the typing for $(\vec{x}) (\sys {\pmvA} {\procPi} \cocoPar \sysSi)$
      as follows:
      \[
      \hspace{40pt}
      \small
        \inference[{\smallnrule{[T-SDel2${}^*$]}}]
        {
          \inference[{\smallnrule{[T-SPar2]}}]
          {
            \inference[{\smallnrule{[T-SA]}}]
            {\effentails \procPi \colon \ptypeFi[0] \effPar \ptypeG[0]}
            {\tsentails{A} \sys {\pmvA} {\procPi} \colon \ptypeFi[0] \effPar \ptypeG[0]}
            &
            \inference[\smallnrule{[T-SFz1]}]
            {\expand{(\ptypeFi[0] \effPar \ptypeG[0])}{\setenum{w}}(w) \text{ realizes } \contrP}
            {\tsentails{A} \latent{w}{\contrP} \tscompat \ptypeFi[0] \effPar \ptypeG[0]}}
          {\tsentails{A} \sys {\pmvA} {\procPi} \cocoPar \sysSi \colon \ptypeFi[0] \effPar \ptypeG[0]}
          & \expand{(\ptypeFi[0] \effPar \ptypeG[0])}{\vec{x}}(\vec{x}) \text{ honest}}
        {\tsentails{A} (\vec{x}) (\sys {\pmvA} {\procPi} \cocoPar \sysSi) \colon (\ptypeFi[0] \effPar \ptypeG[0])\setenum{\bind{\vec{x}}{\bot}}}
      \]
      Since $\ptypeF[0] \effmove{\prefi} \ptypeFi[0]$,
      then $\ptypeF[0]\setenum{\bind{\vec{x}}{\bot}} \effmove{\pref} \ptypeFi[0]\setenum{\bind{\vec{x}}{\bot}}$.
      Therefore:
      \[
      \hspace{40pt}
      \ptypeF \effPar \ptypeG 
      \; = \; 
      \ptypeF[0]\setenum{\bind{\vec{x}}{\bot}} \effPar \ptypeG[0]\setenum{\bind{\vec{x}}{\bot}}
      \; \effmove{\pref} \;
      \ptypeFi \effPar \ptypeG 
      \; = \; 
      \ptypeFi[0]\setenum{\bind{\vec{x}}{\bot}} \effPar \ptypeG[0]\setenum{\bind{\vec{x}}{\bot}}
      \]

    \end{itemize}
    
    \medskip
    For item~\eqref{lem:types:process-progress-item-fact}
    we have $\pref = \cocodo {u} {\atomA}$.
    Since $u \in \dom{\ptypeF}$, 
    then by~\Cref{lem:types:dom-fnv} we have $u \in \fv{\procP}$, 
    and so $u \not\in \vec{x}$ and $\prefi = \cocodo {u} {\atomA}$.
    The thesis follows by~\Cref{def:readydo}.
   
  \item \nrule{[C-SumL]}.
    Easy generalisation of case \nrule{[C-Pref]}, since all sums are guarded.

  \item \nrule{[C-ParL]}.
    Then, $\ptypeF = \ptypeF[0] \effPar \ptypeF[1]$
    and $\ptypeF[0] \effmove{\pref} \ptypeFi[0]$
    for some $\ptypeF[0]$, $\ptypeF[1]$ and $\ptypeFi[0]$.
    Therefore, $\ptypeFi = \ptypeFi[0] \effPar \ptypeF[1]$,
    and $u \in \dom{\ptypeF[0]} = \dom{\ptypeF[1] \effPar \ptypeG}$.
    Both items~\eqref{lem:types:process-progress-item-tau} 
    and~\eqref{lem:types:process-progress-item-fact}
    follow by the induction hypothesis, 
    instantiating the parallel component $\ptypeG$ of the inductive statement
    as $\ptypeF[1] \effPar \ptypeG$.

  \item \nrule{[C-Rec]}.
    We have that $\ptypeF = \effrec{\ctypeX}{\ptypeF[0]}$, and:
    \[
    \inference
    {\ptypeF[0]\subs{\effrec{\ctypeX}{\ptypeF[0]}}{\ctypeX} \effmove{\pref} \ptypeFi}
    {\effrec{\ctypeX}{\ptypeF[0]} \effmove{\pref} \ptypeFi}    
    \]
    Therefore, 
    $
    \procP = (\vec{x}) (\procRec{\procX}{\procP[0]}) 
    $,
    and
    $
    \effentails (\vec{x}) (\procP[0]\subs{\procP}{\procX}) \colon \ptypeF[0]\subs{\ptypeF}{\ctypeX}
    $
    by~\Cref{lem:types:type-subs-recvar}.
    Since $\ptypeF \effPar \ptypeG$ is honest, then
    $\ptypeF[0]\subs{\effrec{\ctypeX}{\ptypeF[0]}}{\ctypeX} \effPar \ptypeG$
    is honest as well, since unfolding preserves the semantics.
    Further, 
    $u \in \dom{\ptypeF[0]\subs{\effrec{\ctypeX}{\ptypeF[0]}}{\ctypeX}} = \dom{\ptypeG}$.
    Both items~\eqref{lem:types:process-progress-item-tau} 
    and~\eqref{lem:types:process-progress-item-fact}
    follow by the induction hypothesis, 
    instantiating the process $\procP$ of the inductive statement
    as $(\vec{x}) (\procP[0]\subs{\procP}{\procX})$.
    \qedhere

  \end{itemize}
  
\end{proof}

\begin{proofofthm}{th:progress}
  By~\Cref{lem:types:a-process-presence-typing},
  there exist ${\vec v}, \sysS[0], \procP$ such that:
  \[
  \sysS \equiv (\vec v)\left({\sys {\pmv A} \procP} \cocoPar \sysS[0]\right)
  \]
  By inverting the typing derivation of $\tsentails{A} \sysS \colon \ptypeF$, 
  we must have
  $\ptypeF = \ptypeF[0] \setenum{\bind{\vec{v}}{\bot}}$,
  for some $\ptypeF[0]$ such that $\effentails \procP \colon \ptypeF[0]$
  and $\expand{\ptypeF[0]}{\vec{v}}(\vec{v})$ is honest;
  Together with the fact that $\ptypeF$ is honest,
  we obtain that $\ptypeF[0]$ is honest.
  Further, $\tsentails{A} \sysS[0] \tscompat \ptypeF[0]$.
  We have the following cases, according to the form of $\pref$:
  \begin{itemize}

  \item $\pref = \tau$.
    By~\Cref{def:types:typing-prefixes},
    the hypothesis $\ptypeF \effmove{\pref} \ptypeFi$ implies 
    $\ptypeF[0] \effmove{\prefi} \ptypeFi[0]$,
    where $\prefi \in \setenum{\tau,\tell {}{\freeze w \contrP}}$ 
    (with $w \in \vec{v}$), and
    $\ptypeFi = \ptypeFi[0] \setenum{\bind{\vec{v}}{\bot}}$.
    Note that if $\prefi = \tell {}{\freeze w \contrP}$ but
    $w \not\in \dom{\ptypeF[0]}$, then we also have
    $\ptypeF[0] \effmove{\tau} \ptypeFi[0]$, so in this
    case we could also choose $\prefi = \tau$.
    Consequently, w.l.o.g.\ we can assume $w \in \dom{\ptypeF[0]}$.
    By~\Cref{lem:types:process-progress} 
    (by choosing $\ptypeG = \effempty$)
    there exist $\vec{x}, \procPi, \sysS[0]$ such that:
    \[
    \sys {\pmvA} {\procP}
    \; \sysmove{A}{\prefi}{} \;
    (\vec{x}) (\sys {\pmvA} {\procPi} \cocoPar \sysS[0]) 
    \; = \; \sysSi[0]
    \qquad\qquad
    \tsentails{A} \sysSi[0] \colon \ptypeFi[0]
    \]
    Hence we have the following derivation:
    \[
    \inference[\smallnrule{[Del${}^*$]}]
    {
      \inference[\smallnrule{[Par]}]
      {\sys {\pmvA} {\procP} \sysmove{A}{\prefi}{} \sysSi[0]}
      {\sys {\pmvA} {\procP} \cocoPar \sysS[0] \sysmove{A}{\prefi}{} \sysSi[0] \cocoPar \sysS[0]}}
    {\sysS \sysmove{A}{\del{\vec{v}}{\prefi}}{} (\vec{v}) (\sysSi[0] \cocoPar \sysS[0]) = \sysSi}
    \]
    By definition of $\prefi$ we have that
    $\del{\vec{v}}{\prefi} = \pref$.
    Since $\tsentails{A} \sysS[0] \tscompat \ptypeF[0]$ and
    $\ptypeF[0] \effmove{\prefi} \ptypeFi[0]$
    with $\prefi \neq \cocodo{}{\cdots}$, 
    then by~\Cref{lem:types:tscompat-effmove} we also have
    $\tsentails{A} \sysS[0] \tscompat \ptypeFi[0]$.
    Let $\sysSi = (\vec{v}) (\sysSi[0] \cocoPar \sysS[0])$.
    Then, we have the following typing derivation for $\sysSi$:
    \[
    \inference[\smallnrule{[T-SDel2${}^*$]}]
    {
      \inference[\smallnrule{[T-SPar2]}]
      {\tsentails{A} \sysSi[0] \colon \ptypeFi[0] 
        &
        \tsentails{A} \sysS[0] \tscompat \ptypeFi[0]
      }
      {\tsentails{A} \sysSi[0] \cocoPar \sysS[0] \colon \ptypeFi[0]}
      & 
      \expand{\ptypeFi[0]}{\vec{v}}(\vec{v}) \text{ honest}}
    {\tsentails{A} \sysSi \colon \ptypeFi[0]\setenum{\bind{\vec{v}}{\bot}} = \ptypeFi}
    \]

  \item $\pref = {\tell {\pmv B} {\freeze u \contrP}}$.
    By~\Cref{def:types:typing-prefixes},
    the hypothesis $\ptypeF \effmove{\pref} \ptypeFi$ implies 
    $\ptypeF[0] \effmove{\pref} \ptypeFi[0]$,
    using the assumption $u \in \dom{\ptypeF} \subseteq \dom{\ptypeF[0]}$.
    The proof then proceeds similarly to the previous case,
    by exploiting~\Cref{lem:types:process-progress}.

  \item $\pref = {\cocodo u {\atomA}}$. 
    By~\Cref{def:types:typing-prefixes},
    the hypothesis $\ptypeF \effmove{\pref} \ptypeFi$ implies 
    $\ptypeF[0] \effmove{\pref} \ptypeFi[0]$,
    using the assumption $u \in \dom{\ptypeF} \subseteq \dom{\ptypeF[0]}$.
    By~\Cref{lem:types:process-progress},
    $\atomA \in \readydosys{u}{\pmvA}{\sys {\pmvA} \procP}$.
    Since $u \not\in \vec{v}$, then we have the thesis
    $\atomA \in \readydosys{u}{\pmvA}{\sysS}$.
    \qed

  \end{itemize}
\end{proofofthm}


\section{Other proofs for~\Cref{sec:type-safety} (Type safety)}
\label{sec:proofs-type-safety}

\begin{proofoflem}{lem:self-concordance}
  We prove the following stronger statement. %
  If $\Gamma \effentails \procP \colon \ptypeF$,
  for some process $\procP$ and some self-concordant $\Gamma$,
  then $\ptypeF$ is self-concordant. %
  We proceed by induction on this typing derivation. %
  We have the following cases, according to the 
  last rule applied in the derivation:
  \begin{itemize}
    
  \item \nrule{[T-Sum]}.
    The rule specifies a set of prefixes $\pref[i]$;
    then, $\alpha = [\pref[i]]_u$ for some of these $i$. %
    Taking $\pref = \pref[i]$ and $\ptypeFi = \ptypeF[i]$ 
    gives the thesis. %

  \item \nrule{[T-Par]}. The thesis follows directly by the induction hypothesis. %

  \item \nrule{[T-Def]}. %
    We have $\procP = \procX(\vec{v})$, and:
    \[
    \inference[\smallnrule{[T-Def]}]
    {\procX(\vec{u}) \mmdef \procPi &
      \procX(\vec{v}) \not\in \dom{\Gamma} &
      \Gamma\setenum{\bind{\procX(\vec{v})}{\lambda v .\, \ctypeX}} \effentails 
      \procPi\setenum{\bind{\vec{u}}{\vec{v}}}
      \colon \ptypeG}
    {\Gamma \effentails \procX(\vec{v}) \colon \lambda v . \, \effrec{\ctypeX}{\ptypeG(v)}}
    \]
    with $\ptypeF = \lambda v . \, \effrec{\ctypeX}{\ptypeG(v)}$. %
    Since $\effrec{\ctypeX}{\ptypeG(u)} \effmove{\alpha} \ctypeTi$,
    then by inverting rule~\nrule{[C-Rec]} it must be
    $\ptypeG(u)\setenum{\bind{\ctypeX}{\effrec{\ctypeX}{\ptypeG(v)}}} \effmove{\alpha} \ctypeTi$. %
    From this, we can prove that
    $\ptypeG(u) \effmove{\alpha} \ctypeTii$, %
    for some $\ctypeTii$ such that 
    $\ctypeTi = \ctypeTii\setenum{\bind{\ctypeX}{\effrec{\ctypeX}{\ptypeG(u)}}}$. %
    Note that
    $\Gamma\setenum{\bind{\procX(\vec{v})}{\lambda v .\, \ctypeX}}$
    is self-concordant. %
    By the induction hypothesis, there exist $\pref$ and $\ptypeGi$
    such that
    $[\pref]_u = \alpha$, 
    $\ptypeGi(u) = \ctypeTii$, and 
    $\ptypeG \effmove{\pref} \ptypeGi$. %
    The thesis follows by taking 
    $\ptypeFi = \lambda v. \ptypeGi(v)\setenum{\bind{\ctypeX}{\effrec{\ctypeX}{\ptypeG(v)}}}$.
    
  \item \nrule{[T-Var]}. Trivial, since $\Gamma$ is self-concordant. %

  \item \nrule{[T-Del]}. We have $\procP = (v)\procPi$, and:
    \[
    \inference[\smallnrule{[T-Del]}]
    {\Gamma_{\neq v} \effentails \procPi \colon \ptypeG & \ptypeG(v)\; \mbox{honest}}
    {\Gamma \effentails (v)\procPi \colon
      \ptypeG\setenum{\bind{v}{\ptypeG(\effwildcard)}}}
    \]
    where $\ptypeF = \ptypeG\setenum{\bind{v}{\ptypeG(\effwildcard)}}$. %
    There are the following two subcases:
    \begin{itemize}
    \item $v \neq u$. %
      By the induction hypothesis, there exist $\pref$, $\ptypeGi$ such that
      $[\pref]_u = \alpha$, 
      $\ptypeGi(u) = \ctypeTi$, and 
      $\ptypeG \effmove{\pref} \ptypeGi$. %
      The thesis follows by taking
      $\ptypeFi = \ptypeGi\setenum{\bind{v}{\ptypeGi(\effwildcard)}}$. %

    \item $v = u$. %
      Since $\ptypeF(u) \effmove{\alpha} \ctypeTi$, then
      $\ptypeG(\effwildcard) \effmove{\alpha} \ctypeTi$. %
      By the induction hypothesis (applied on $\effwildcard$),
      there exist $\pref$, $\ptypeGi$ such that
      $[\pref]_{\effwildcard} = \alpha$, 
      $\ptypeGi(\effwildcard) = \ctypeTi$, and 
      $\ptypeG \effmove{\pref} \ptypeGi$. %
      The thesis follows by taking
      $\ptypeFi = \ptypeGi\setenum{\bind{v}{\ptypeGi(\effwildcard)}}$. %
      \qedhere
    \end{itemize}

  \end{itemize}
\end{proofoflem}

\end{appendices}
\vspace{-20 pt}

\end{document}